\documentclass[a4paper,11pt]{article}
\pdfoutput=1 

\usepackage{jheppub} 

\usepackage{amsmath,amsfonts}
\usepackage{color}
\usepackage{graphicx, tikz}
\usepackage{float}
\usepackage{amsthm}
\usepackage{subcaption}
\usepackage{amssymb}
\usetikzlibrary{decorations.pathreplacing,decorations.pathmorphing}

\newcommand*{\cE}{\mathcal{E}}

\newcommand*{\stile}{\mathcal{T}_{\mathcal{S}}}
\newcommand*{\htile}{\mathcal{T}_{\mathcal{H}}}
\newcommand*{\eBST}{W^{\epsilon}}
\newcommand*{\ePT}{T^{\epsilon}}

\theoremstyle{definition}
\newtheorem{definition}{Definition}[section]
\newtheorem{theorem}{Theorem}[section]

\numberwithin{equation}{section}
\newenvironment{hproof}{%
  \proof}{\endproof}

\DeclareMathOperator{\Tr}{Tr}

\title{Approximate Bacon-Shor Code and Holography}

\author[a]{ChunJun Cao\note{Corresponding author.}}
\author[a,b,c,d]{and Brad Lackey}

\affiliation[a]{Joint Center for Quantum Information and Computer Science, University of Maryland, College Park, MD, 20742, USA}
\affiliation[b]{Institute for Advanced Computer Studies, University of Maryland, College Park, MD 20742, USA}
\affiliation[c]{Departments of Computer Science and Mathematics, University of Maryland, College Park, MD 20742, USA}
\affiliation[d]{Quantum Systems Group, Microsoft, Redmond, WA 98052, USA}

\emailAdd{ccj991@gmail.com}
\emailAdd{Brad.Lackey@Microsoft.com}

\abstract{We explicitly construct a class of holographic quantum error correction codes with non-trivial centers in the code subalgebra. Specifically, we use the Bacon-Shor codes and perfect tensors to construct a gauge code (or a stabilizer code with gauge-fixing), which we call the holographic hybrid code. This code admits a local log-depth encoding/decoding circuit, and can be represented as a holographic tensor network which satisfies an analog of the Ryu-Takayanagi formula and reproduces features of the sub-region duality. We then construct approximate versions of the holographic hybrid codes by ``skewing'' the code subspace, where the size of skewing is analogous to the size of the gravitational constant in holography. These approximate hybrid codes are not necessarily stabilizer codes, but they can be expressed as the superposition of holographic tensor networks that are stabilizer codes. For such constructions, different logical states, representing different bulk matter content, can ``back-react'' on the emergent geometry, resembling a key feature of gravity.  The locality of the bulk degrees of freedom becomes subspace-dependent and approximate. Such subspace-dependence is manifest from the point of view of the ``entanglement wedge'' and bulk operator reconstruction from the boundary. Exact complementary error correction breaks down for certain bipartition of the boundary degrees of freedom; however, a limited, state-dependent form is preserved for particular subspaces. We also construct an example where the connected two-point correlation functions can have a power-law decay. Coupled with known constraints from holography, a weakly back-reacting bulk also forces these skewed tensor network models to the ``large $N$ limit'' where they are built by concatenating a large $N$ number of copies. }

\begin{document} 
\maketitle




\section{Introduction}

The AdS/CFT correspondence is a concrete implementation of the holographic principle\cite{Susskind:1994vu,tHooft:1993dmi}, which connects a $d$-dimensional conformal field theory (CFT) on Minkowski spacetime with a theory of quantum gravity in $d+1$ dimensional asymptotically anti-de Sitter (AdS) space\cite{Maldacena:1997re,Witten:1998qj}. It provided a fruitful testing ground for various ideas in quantum gravity; in particular, there has been considerable progress in understanding how spacetime and gravity may emerge from quantum information quantities such as entanglement\cite{Ryu:2006bv,VanRaamsdonk:2010pw,Maldacena:2013xja,Faulkner:2013ica,Faulkner:2017tkh} and complexity\cite{Stanford:2014jda,Brown:2015bva}. 
One particular connection is made by \cite{Almheiri:2014lwa}, which shows that quantum error correction code (QECC), a crucial component for fault-tolerant quantum computation, is intimately connected with how operators on different sides of the duality are related. Since its initial discovery, it is shown that many other physical phenomena exhibit the same basic features of quantum error correction codes. These include emergent asymptotically flat space-time from entanglement~\cite{Cao:2017hrv}, thermalizing quantum systems, and black hole physics\cite{Kim:2020cds}. 

In the particular context of AdS/CFT, the boundary CFT degrees of freedom are identified with the physical degrees of freedom which one controls in a laboratory. The low energy subspace of the bulk AdS theory corresponds to the code subspace of the error correction code. Logical operations on that subspace are given by operators of the bulk effective field theory.
Along a somewhat different direction, it was realized \cite{Swingle:2009bg} that tensor networks, in particular, the ones like the multi-scale entanglement renormalization ansatz (MERA)\cite{Vidal2008}, capture both the entanglement and geometric aspects of holography in a discrete setting\footnote{Note that it is disputed whether the MERA truly corresponds to a spatial slice of the discretized AdS\cite{Beny:2011vh,Bao:2015uaa,Czech:2015kbp,Milsted:2018san}.}. Combining the observations from these lines of research, \cite{Pastawski:2015qua} constructed a class of explicit holographic tensor network toy models that is geometrically consistent with the duality and is an explicit construction of a quantum error correction code. Generalizations are then made in different directions by \cite{Yang:2015uoa, Hayden:2016cfa,Kim:2016wby, Kohler:2018kqk,Bao:2018pvs}. These tensor network models, other than making geometric connections with AdS/CFT, provide a graphical approach to understanding the construction and properties of quantum error correction codes.

 However, as one would reasonably expect, the quantum error correction codes that occur in nature generically should not be a pristine, exact code that we would ideally like to construct in a laboratory. Rather it is only approximate, where errors are correctable up to some tolerance. In fact, the deviation from the exact ``ideal'' codes is crucially related to the presence of gravitational interactions. It is known that AdS/CFT corresponds to an exact quantum error correction code only in the limit of $N\rightarrow \infty$, where the bulk gravitational interaction controlled by $G_N$ goes to zero. For finite $N$ or non-negligible gravitational interaction in the bulk, it has been argued that error correction must be approximate\cite{Cotler:2017erl, Hayden:2018khn,Faist:2019ahr}. 
  Despite these insights, the exact manner in which gravity arises in these tensor network quantum error correction codes remains unexplored. More generally, it is even less clear how, or if, gravity can be found in approximate quantum error correction codes. 
In this work, we will tackle these questions by connecting several key features of gravity to the approximate holographic quantum error correction codes we construct.

Specifically, we construct a holographic quantum error correction code toy model which we can explicitly decode using a local log-depth circuit. We call this the hybrid holographic code, which is built from the 4-qubit $[[4,1,2]]$ Bacon-Shor code and the perfect tensor. The latter is a $[[6,0,4]]$ stabilizer state obtainable from the perfect $[[5,1,3]]$ code\cite{Laflamme:1996iw,preskillLecture}. The hybrid holographic code is a subsystem code, or a gauge code, but reduces to a stabilizer code after gauge fixing. It reproduces key known features of holography, such as the Ryu-Takayanagi (RT) formula and entanglement wedge reconstruction, which are also present in \cite{Pastawski:2015qua,Hayden:2016cfa}. Furthermore, we show that it contains non-trivial centers in the code subalgebra, a feature needed to accommodate different semi-classical geometries in holographic error correction codes and the structure of gauge theories\cite{Harlow:2016vwg, Donnelly:2011hn, Donnelly:2016qqt}. 

Then we construct generalized versions of this code, where the code subspace is ``skewed'' from that of the hybrid holographic code, or the reference code. More concretely, the codewords in the skewed code are related to those of the reference code by some linear transformation. One such transformation is commonly  given by a unitary $U(\lambda)=\exp(-i \lambda\mathcal{K})$ that acts on the state in the physical Hilbert space, where $\lambda$ controls the amount of skewing and $\mathcal{K}$ is some Hermitian operator. These generalized codes can also be written as a superposition of different hybrid codes, each with a somewhat different choice of the code subspace and error subspaces. We find that the amount of skewing plays the role of $G_N$, which controls the strength of the ``gravitational interaction'' in this model. In the limit where the skewing is small, one can treat the generalized code as an approximate quantum error correction code (AQECC), whose logical operations and decoding can be performed by that of the reference code. However, these reference code operations will not act on or recover the original encoded information with perfect fidelity. In addition to preserving certain properties of the reference code, the generalized code displays some key features of gravity, consistent with our expectations in holography when one considers higher order corrections and quantum extremal surfaces. These include the breakdown of exact complementary recovery in the hybrid code, the subspace-dependence of entanglement wedge and bulk operator reconstruction, and back-reactions where different logical states can lead to different ``semi-classical geometries'' in the bulk. The last feature is similar to a so-called gravitational back-reaction, where matter, according to Einstein gravity, modifies the space-time geometry. The logical or bulk degrees of freedom also become weakly inter-dependent, forcing them to be delocalized; that is, a bulk degree of freedom that appears to be localized on some compact region actually weakly depends on the actions on other bulk degrees of freedom far away. We analyze this feature in different examples and conclude that the locality of these bulk degrees of freedom is approximate and state-dependent\cite{Papadodimas:2013jku}, consistent with our general expectations when local quantum field theories are coupled to gravity\cite{Donnelly:2016rvo}.

The generalized constructions with skewing can also support power-law decaying correlation functions on the boundary if we replace both the perfect tensor and the Bacon-Shor code by their skewed counterparts. The scaling dimension depends on the amount of skewing -- a small skewing, and hence $G_N$, leads to a large scaling dimension. This is consistent with our expectations in holographic CFTs that have a weakly coupled bulk dual\cite{Heemskerk:2009pn}. 

The paper is organized as follows. In section \ref{sec:2rev}, we clarify the notations and review the necessary backgrounds in holographic quantum error correction codes, on which we build this work. In section \ref{sec:3baconshor}, we review the fundamentals of the Bacon-Shor code and then discuss the construction, decoding, entanglement, and how non-trivial centers in the code subalgebra are identified in the 4-qubit code. We will then construct a concatenated version of this code that has properties better suited for our holographic model. In section \ref{sec:4aqec}, we discuss some basics of approximate quantum error correction codes and how such an approximate code can be constructed using skewing and the superposition of codes. We will then discuss their properties and contrast them with those of the exact codes. Some of these techniques, such as superpositions of codes, are more generally applicable to other codes as well. In section \ref{sec:5hybrid}, we construct the holographic tensor network using the Bacon-Shor code and the perfect tensor. As a QECC, we discuss its code properties, connection with holography such as the RT formula, and its decoding using the Greedy algorithm. In Section \ref{sec:6apphybrid}, we consider skewed versions of the hybrid code and identify its connections with gravity, such as state-dependence, back-reaction, dressed operators, and non-locality. We will also construct an example with power-law decaying connected two-point correlation functions. In section \ref{sec:7disc}, we conclude with some remarks of how this type of construction makes contact with concepts such as fixed-area states, super-tensor networks, the large $N$ limit, magic, and quantum error correction codes from tensor networks at large. In Appendix~\ref{app:codeprop}, we compute the encoding rate and bound the code distance of the central encoded qubit. We provide some details on the pushing properties in a double-copy Bacon-Shor code using gauge operators in Appendix~\ref{app:dcpushing}. In Appendix~\ref{app:tensorcontract}, we provide details for the tensor contraction properties used in Section~\ref{sec:6apphybrid}. Finally in Appendix~\ref{app:tracing}, we discuss how the tensor network tracing/gluing formalism applies to general stabilizer codes. We also provide detailed procedures on how the check matrices of the new codes are generated from the gluing operations.

\subsection{Summary and Readers' Guide}
Given the length of this paper and the variety of topics being covered, we provide a more detailed summary that also serves as a guide for navigating the relevant sections and results. We sort them by the potential interests of our targeted audiences in quantum information theory and in quantum gravity. 

\begin{figure}
    \centering
    \includegraphics[width=0.5\textwidth]{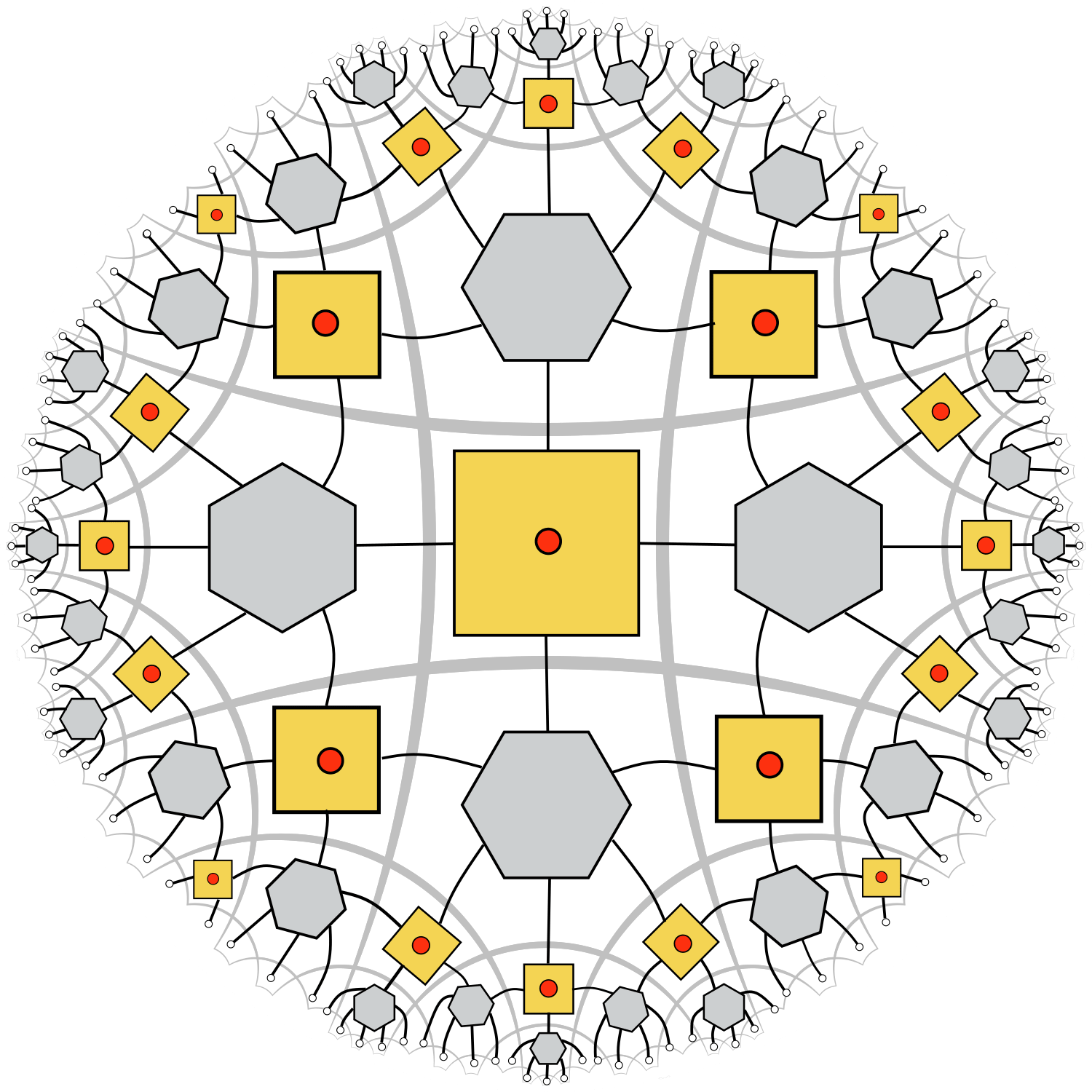}
    \caption{A tensor network representation of the hybrid holographic code.}
    \label{fig:46TNsum}
\end{figure}

We study the \textit{hybrid code} generated by tensor networks of the type shown in Figure~\ref{fig:46TNsum}. The 4-qubit Bacon-Shor codes, represented by yellow squares, are joined with perfect tensors, represented by grey hexagons. This is a graphical representation of a linear map, which encodes $k$ logical qubits in the bulk (red disks) into $n>k$ physical qubits on the boundary (edges that end on the boundary). We will analyze both the noiseless version of this code (Section~\ref{sec:5hybrid}), where the Bacon-Shor codes and perfect tensors are exact, and a noisy version of this code (Section~\ref{sec:6apphybrid}), where some of these tensors are contaminated and hence deviate from their standard forms. We call the contaminated forms ``skewed''.

To identify the logical and stabilizer operators of this code, we can apply the ``pushing rules'', which allow one to sequentially transform some operator in the interior of the network, \textit{e.g.} a logical operator inserted on one of the red dots, to one that is supported entirely on the boundary. The pushing rules for the Bacon-Shor code are given in Sections~\ref{subsec:bsoptpush} and \ref{subsubsec:multiBSCpush} while the rules for the skewed Bacon-Shor code are given in Sections~\ref{subsec:skewedBSCpush} and \ref{subsec:multics}. The pushing rules of the perfect tensor can be found in the Appendix of \cite{Pastawski:2015qua}. Section \ref{subsec:hybridpushing} discusses how they are used to obtain relevant operators for the hybrid code. 

\subsubsection{Quantum Information Perspective}
Despite a number of interesting work in quantum error correction code and tensor network, the field is still very much in its infancy. Here we construct a new subsystem code using a tensor network dual to a regular graph, which can be isometrically embedded in the hyperbolic space. We characterize some of its code property in~Section~\ref{sec:5hybrid} and its rate and distance in Appendix~\ref{app:codeprop}. The rate is a constant fraction, but depends on the degree of the dual regular graph as well as the cut off radius. In the tesellation we use, the encoding rate is $0.160\lesssim k/n \lesssim 0.445$. The distance of the most well-protected qubit grows at least logarithmically with system size $n$, but at most sublinearly $\sim n^{0.913}$. The bound is not tight.

Similar to the HaPPY code\cite{Pastawski:2015qua}, both the encoding and decoding maps of these codes can be explicitly constructed from the tensor networks using the Greedy algorithm. More specifically, for any tensor that can be locally contracted to an isometry, it is equivalent to adding a well-defined local unitary to the decoding quantum circuit. One can complete the decoding circuit by contracting these tensors in the network in sequence. This result is somewhat implicit in \cite{Pastawski:2015qua}, which we now make explicit in Section~\ref{subsec:distdecode}. In particular, the decoding can be achieved by a local log-depth circuit. 

The construction technique introduced in \cite{Pastawski:2015qua} generalizes code concatenation and applies to other quantum codes that are not the perfect code. Therefore, it can be used to generate new codes from existing codes with well-known properties. One can also easily characterize the new codes constructed from these tensor contractions both graphically and algebraically. On the graphically level, it simply amounts to operator pushing, or some form of stabilizer matching. On the algebraic level, it reduces to straightforward manipulations of the check matrices.  Regarding these tensor network reconstructions, we give detailed explanations in the beginning of Section~\ref{sec:5hybrid} and in Appendix~\ref{app:tracing}. 

In Section~\ref{subsec:skewedBSC} and Section \ref{subsec:skewedhybrid}, we also present a method for constructing models for approximate codes or other non-additive codes by taking the superposition of known stabilizer codes. Although we construct examples for the Bacon-Shor code, this formalism applies more generally to other codes as well. On a more speculative note, they can be useful tools for classifying approximate quantum error correction codes. These can also serve as useful models to (graphically) characterize how different types of (likely coherent) quantum noise impact the code properties and its associated encoding/decoding operations.

\subsubsection{Quantum Gravity Perspective}
For the quantum gravity audience who have been following the developments in AdS/CFT and quantum error correction codes, the hybrid code that we construct out of the Bacon-Shor code and the perfect tensor is in many ways similar to a version of the HaPPY code (Figure 17 of \cite{Pastawski:2015qua}) except we change the hyperbolic tessellation so that one can replace certain 5-qubit quantum codes with 4-qubit quantum codes. However, a key difference is that the hybrid code has non-trivial centers in the code subalgebra, which the HaPPY code lacks. We begin by analyzing the 4-qubit Bacon-Shor code in Section \ref{sec:3baconshor} as a simple toy model for holography with non-trivial code subalgebras, which is good for intuition-building. 
Section~\ref{sec:5hybrid} inherits these features in a more complex tensor network model. However, both of these codes are lacking in modelling gravitational features. To do so, we must move to skewed codes, which may be construed as noisy versions of the codes in Section~\ref{sec:3baconshor} and \ref{sec:5hybrid}.

The main results here are about a constructive approach that incorporates gravity in tensor networks that are also (approximate) quantum error correction codes. These discussions are found in Sections~\ref{sec:4aqec} where ``back-reactions'' can be seen in a 4-qubit model through the changes in the entanglement entropy associated with the area of the RT surface. This feature persists in Section~\ref{sec:6apphybrid} where we analyze the full tensor network. In particular, because the skewed code is also a superposition of different hybrid codes, this is an explicit construction of what one may consider a ``super tensor network''. One can identify the different fixed-area states using the decomposition of Hilbert spaces induced by the non-trivial centers in the code subalgebra. By replacing some of the 4-qubit Bacon-Shor codes in the tensor network with their approximate counterparts, we show in two examples in Section~\ref{subsec:subspacerec} where the boundary reconstruction of a bulk operator and the entanglement wedge depend on the bulk states. In particular, a zero bulk logical state is analogous to the vacuum state, while a logical state in an arbitrary superposition of 0 and 1 is analogous to having mass back-reacting on the background geometry. All of these observations are made explicit using operator pushing and the Greedy algorithm, which we review in Section~\ref{sec:5hybrid}. The logical $\tilde{X}$ operator of the skewed code now plays a role analogous to the mass creation operator. It can be decomposed into a bare operator, which is the $\tilde{X}_0$ of the reference code and a ``gravitational dressing'', which is tied to the noisiness of the code. If all of the Bacon-Shor codes are noisy, then the bulk degrees of freedom also do not live on tensor factors on the code subspace. They only factorize in the limit of zero noise, which heuristically maps to a local quantum field theory on curved background. 

Finally, the general utility of the model should not be confined to the specific examples we have analyzed. In principle, the construction and techniques used here also extend beyond a graph that resembles the hyperbolic space. Rather than interpreting it as a toy model which may or may not capture the desired features of AdS/CFT, one should regard such systems in their own right -- they are many-body quantum mechanical systems that exhibit hints of emergent geometry and gravity which can be implemented on a quantum computer.

\section{Basics}
\label{sec:2rev}
\subsection{Notations}
Before we review the basic concepts we will use in this paper, we clarify the convention of notations we use. A Hilbert space is denoted with script letters such as $\mathcal{H}, \mathcal{C}$. In the case where a Hilbert space admits a tensor product decomposition, e.g. $\mathcal{H}=\mathcal{H}_A\otimes \mathcal{H}_{A^c}$, we often use $A$ or $A^c$ as shorthand for the corresponding Hilbert space tensor factors. We use $A^c$ to indicate the complement of $A$. 

Logical operators $\bar{O}, \tilde{O}$ are denoted with a bar or tilde above. The same goes for logical states, such as $|\bar{i}\rangle,|\tilde{j}\rangle$. Barred operators such as $\bar{X}, \bar{Z}$ are used to denote logical operations on the original Bacon-Shor code or its generalizations. Tilde operators $\tilde{X},\tilde{Z}$ are used for all other types of logical operations, especially including those that act on concatenated codes. Similar notations are used for the logical states as well.

A few commonly used acronyms are the Bacon-Shor Tensor (BST), Skewed Bacon-Shor Tensor (SBST), perfect tensor (PT), imperfect tensor (IPT).

\subsection{Review of AdS/CFT and Holographic QECC}

In this section, we will review some basic aspects of AdS/CFT that we use in our work. This includes the subregion duality, Ryu-Takayanagi (RT)-Faulkner-Lewkowycz-Maldacena (FLM) formula and connections with holographic quantum error correction codes (QECC). Readers familiar with these concepts can proceed directly to Section~\ref{sec:3baconshor}.

The AdS/CFT correspondence is a non-perturbative consequence of string theory, which describes a duality between a conformal field theory of $d$-dimensional Minkowski space-time and a theory of quantum gravity of $d+1$ dimensions that lives in an asymptotically (locally) AdS spacetime. A CFT describes, for example, certain quantum systems near phase transitions. On the other hand, AdS is a space-time with a negative cosmological constant, which is somewhat opposite to the physical Universe we live in. Because the CFT can be understood to live on the asymptotic boundary of AdS space, it is also often known as the boundary theory, while the dual theory with gravitational interaction is referred to as the bulk.
This duality can be understood as a particular example of the holographic principle\cite{tHooft:1993dmi, Susskind:1994vu} where physics of a volume of space can be encoded on a lower-dimensional boundary of that region. In this paper, we will take holography to be synonymous with the one described by the AdS/CFT correspondence, as it is the only holographic implementation we discuss here. 

\begin{figure}[ht]
    \centering
    \includegraphics[width=0.5\textwidth]{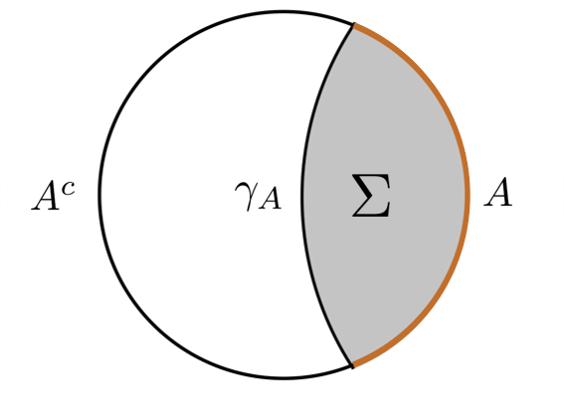}
    \caption{Here the disk represents a spatial slice of AdS. For a state associated with the boundary subregion $A$ (represented by the brown segment), its entropy is equal to the area of the minimal surface anchored to $\partial A$. Here it is given by the minimal geodesic $\gamma_A$ anchored to the endpoints of $A$. The shaded region $\Sigma$ that is bounded by $A$ and $\gamma_A$ is also defined as the entanglement wedge of $A$.}
    \label{fig:RTfig}
\end{figure}

Because both the bulk and the boundary descriptions are different manifestations of the same underlying theory, one should be able to ``translate'' bulk quantities into boundary ones and vice versa. Although our knowledge of the proposed duality is far from complete, some of those entries in the ``holographic dictionary'' have been identified \cite{Aharony:1999ti,Ryu:2006bv,Hubeny:2007xt}. For example, Ryu and Takayanagi~\cite{Ryu:2006bv} found that the von Neumann entropy of a quantum state on a subregion of the boundary is equal\footnote{When studying AdS/CFT, we frequently consider a perturbative expansion in a parameter $1/N$ where $N$ is taken to be large for a bulk that has weakly coupled gravitational dual, such that it can be analyzed using the known framework of semi-classical gravity. When there is a semi-classical bulk dual, this can be understood as an perturbative expansion in $G_N$, the gravitational constant. The RT formula holds to leading order ($N^2\sim 1/G_N$) in this perturbative expansion.} to the area of the minimal surface in the bulk that is homologous to the boundary region\footnote{We are assuming the state is dual to a well-defined bulk geometry in which the area of surfaces can be identified.}. An example of this is shown on a time-slice of AdS in Figure~\ref{fig:RTfig} for the subregion $A$, where
\begin{equation}
    S(A) = \frac{\min\mathrm{Area(\gamma_A)}}{4G_N}
\end{equation}
and $G_N$ is the gravitational constant. Here $\gamma_A$ labels the minimal surface (geodesic). The region $\Sigma$, which is bounded between the Ryu-Takayanagi (RT) surface and the boundary subregion $A$, is also referred to as the \textit{entanglement wedge} $W_{\cE}(A)$ of $A$\footnote{The entanglement wedge as defined is a region in space-time. However, for the purpose of understanding operator reconstruction from boundary subregions, one can time evolve operator at different times using the boundary CFT Hamiltonian, such that the relevant operators all lie on a single time slice. See \cite{Almheiri:2014lwa} for detailed definition and explanation.}.

One can improve the Ryu-Takayanagi (RT) formula to include other contributions such as bulk entropy in the enclosed bulk region, which was discussed in \cite{Faulkner:2013ana}. We will refer to this as the FLM (Faulkner-Lewkowycz-Maldacena) formula\footnote{ The FLM formula includes more corrections in the large $N$ expansion and is a more accurate version of the RT formula by including the sub-leading corrections to the order of $N^0$. },
\begin{equation}
    S(A) = \frac{\min \mathrm{Area}(\gamma_A)}{4G_N} + S_{\rm bulk}(\Sigma),
    \label{eqn:flm}
\end{equation}
where $\Sigma$ is the bulk region bounded by the RT surface and the boundary subregion that is homologous to the surface.

Intuitively speaking, this is the AdS/CFT analog of the generalized entropy\cite{Bekenstein:1972tm,Bekenstein73}, where the first term is tied to the geometry, similar to the black-hole entropy being proportional to its horizon area. The bulk term then captures the matter entropy of the effective field theory on a geometric background\footnote{One should take the regulated version of the bulk entropy; for example, the vacuum subtracted entropy. Intuitively, having bulk matter or entanglement that is different from the bulk vacuum state will contribute to this term.}.

In a semi-classical description of the bulk with weak gravitational interaction, one can approximate the bulk physics as having an effective field theory (EFT), which describes bulk matter, living on a geometric background. In this case, the geometric background is that of the AdS. Here the matter entropy from the FLM formula admits contributions from the bulk effective fields while the RT contribution can be computed from the geometric background. 

Being a duality between quantum theories, the dictionary between bulk and boundary also goes beyond scalar quantities like area and entropy\cite{Hamilton:2006az}. In particular, it was shown by \cite{Jafferis:2015del,Dong:2016eik} that any such bulk EFT operator in the entanglement wedge $W_{\cE}(A)$ of $A$ is also dual to a boundary operator on $A$ as long as certain assumptions are satisfied. In other words, having access only to the boundary region $A$, one can ``reconstruct'' the bulk operator in $W_{\cE}(A)$ and apply any such bulk operation by acting only on $A$.

\begin{figure}[ht]
    \centering
    \includegraphics[width=0.8\textwidth]{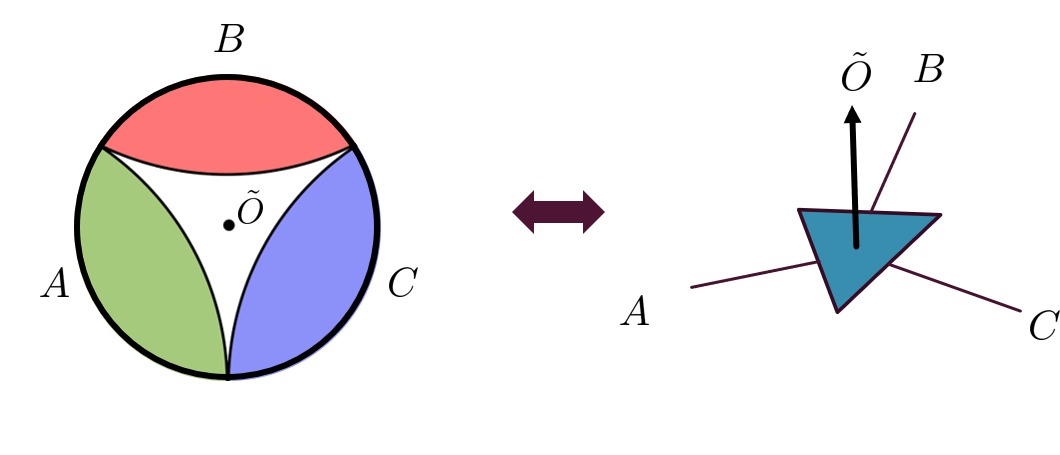}
    \caption{The left figure shows that the entanglement wedge of boundary subregions A, B, and C (coloured green, red and blue respectively) do not contain the central region where $\tilde{O}$ resides. However, their respective complementary wedges (that of BC, AC, and AB) do contain the central region. Hence we can reconstruct the operator $\tilde{O}$ on the union of any such two boundary subregions. This is similar to a qutrit code (right figure), where physical qutrits are represented as in-plane legs of the tensor network, and the logical qutrit is represented by a leg out of the plane. }
    \label{fig:qutritcode}
\end{figure}

In \cite{Almheiri:2014lwa}, the authors point out that this property in holography is similar to that of a quantum error correction code (QECC). To briefly review a motivating example, consider a spatial slice where the boundary has been divided into 3 equal subregions (Figure~\ref{fig:qutritcode}).

The entanglement wedges of $A, B, C$ are shaded. However, the bulk operator in the center can not be reconstructed on any of the subregions individually, because their entanglement wedges do not contain the central region. However, the union of any two of the subregions does. This is similar to a qutrit code\cite{Almheiri:2014lwa,Pastawski:2015qua}

\begin{equation}
    |i\rangle \xrightarrow{\rm encode} |\tilde{i} \rangle= \frac{1}{\sqrt{3}} \sum_{j=0}^2|j,j+i,j-i\rangle_{ABC},
\end{equation}

where each of $A, B$, and $C$ maps to a physical qutrit and $i,j\in GF(3)$ and $\tilde{O}$ is a logical operator. The code corrects one erasure error. In particular, there exists a representation of the logical operators on any two qutrits, while no operator supported on a single qutrit can access the code subalgebra. Therefore, from this simple example, we see that QECC has an elegant connection with holography, where the bulk operators or bulk EFT degrees of freedom correspond to logical operators or the encoded logical information. The boundary degrees of freedom correspond to the physical qubits (or qudits) one can prepare in a laboratory. More generally, the code subspace corresponds to a low energy subspace of bulk states, and the encoding map is a partial dictionary that translates bulk operators into boundary operators and vice versa\cite{Qi:2013caa}. A more comprehensive tensor network model of QECCs and holography were subsequently built using the $[[5,1,3]]$ (perfect) code, which corrects any two erasure errors\cite{Laflamme:1996iw}. Related constructions were also given in \cite{Hayden:2016cfa} and \cite{Kohler:2018kqk} where certain features are generalized. These constructions all reproduce some features of holography to different degrees. Most prominently, they produce an analog of the RT/FLM formula, entanglement wedge reconstruction, \textit{et cetera}.

In fact, given the possibility to reproduce the above features in somewhat general QECC toy models, there is no reason to think that such properties would only hold in the strictest context of the AdS/CFT correspondence\cite{Cao:2016mst, Cao:2017hrv,Giddings:2018koz,Singh2018}. For instance, \cite{Harlow:2016vwg} showed that properties such as the RT/FLM formula are, in fact, generic in a large class of QECCs. We review some of its key ideas below.

Consider a QECC defined in a finite-dimensional system $\mathcal{H}_{\rm code}\subset \mathcal{H}$ where we are given a von Neumann algebra $M$ over $\mathcal{H}_{\rm code}$. $M$ induces an essentially unique decomposition of the Hilbert space

\begin{equation}
    \mathcal{H}_{\rm code} = \bigoplus_{\alpha}\mathcal{H}_{a}^{\alpha}\otimes \mathcal{H}_{a^c}^{\alpha}
    \label{eqn:MCdecomp}
\end{equation}
such that $\forall \tilde{O}\in M$,

\begin{align}
    \tilde{O} =
    \begin{pmatrix}
    \tilde{O}_{a}^1\otimes \tilde{I}_{a^c}^1 & 0 &\dots\\
    0 & \tilde{O}_{a}^2\otimes \tilde{I}_{a^c}^2 & \dots\\
    \vdots & \vdots & \ddots
    \end{pmatrix},
    \label{eqn:MCblocdiag}
\end{align}
where $\tilde{O}_{a}^{\alpha}, \tilde{O}_{a^c}^{\alpha}$ are operators over $\mathcal{H}_a^{\alpha}$ and  $\mathcal{H}_{a^c}^{\alpha}$ respectively. 

In the same basis, for any $\tilde{Q}\in M'$, where $M'$ is the commutant of $M$,

\begin{align}
    \tilde{Q} =
    \begin{pmatrix}
    \tilde{I}_{a}^1\otimes \tilde{Q}_{{a}^c}^1 & 0 &\dots\\
    0 & \tilde{I}_{a}^2\otimes \tilde{Q}^2_{{a}^c} & \dots\\
    \vdots & \vdots & \ddots
    \end{pmatrix}.
\end{align}

 Now assume the physical degrees of freedom can be factorized into a subsystem $A$ and its complement $A^c$

\begin{equation}
    \mathcal{H}= \mathcal{H}_A\otimes \mathcal{H}_{A^c}.
\end{equation}

Suppose $\forall\tilde{O}\in M, |\tilde{\psi}\rangle \in \mathcal{H}_{\rm code}$, there exists $O_A$ that only has non-trivial support over $A$ where

\begin{align}
\begin{split}
    \tilde{O}|\tilde{\psi}\rangle &= O_A|\tilde{\psi}\rangle\\
    \tilde{O}^{\dagger}|\tilde{\psi}\rangle &= O_A^{\dagger}|\tilde{\psi}\rangle.
\end{split}
    \label{eqn:optrecov}
\end{align}

Then \cite{Harlow:2016vwg} showed that condition (\ref{eqn:optrecov}) is equivalent to the existence of a decoding unitary $U_A$ that only acts non-trivially on $A$, such that for any basis state $|\widetilde{\alpha, i,j}\rangle$ of the Hilbert space $\mathcal{H}_{a_{\alpha}}\otimes \mathcal{H}_{{a}^c_{\alpha}}$

\begin{equation}
    U_A |\widetilde{\alpha, i, j}\rangle = |\alpha, i\rangle_{A_{1^{\alpha}}} \otimes |\chi_{\alpha,j}\rangle_{A_{2^{\alpha}}A^c},
    \label{eqn:decodingunitary}
\end{equation}

where $\mathcal{H}_{A_{1^{\alpha}}}, \mathcal{H}_{A_{2^{\alpha}}}$ can be defined in some decomposition of $\mathcal{H}_A$

\begin{equation}
    \mathcal{H}_A = \bigoplus_{\alpha}(\mathcal{H}_{A_{1^{\alpha}}}\otimes \mathcal{H}_{A_{2^{\alpha}}})\oplus \mathcal{H}_{A_3}.
\end{equation}

Furthermore, the recovery is called \textit{complementary} if any $\tilde{O}'\in M'$ can also be represented as an operator that only acts non-trivially on $A^c$ such that,

\begin{align}
\begin{split}
    \tilde{O}'|\tilde{\psi}\rangle &= O_{A^c}|\tilde{\psi}\rangle\\
    \tilde{O}'^{\dagger}|\tilde{\psi}\rangle &= O_{A^c}^{\dagger}|\tilde{\psi}\rangle.
    \end{split}
    \label{eqn:optcomprecov}
\end{align}
Similarly, conditions (\ref{eqn:optcomprecov}) imply the existence of a decoding unitary $U_{A^c}$. Together with (\ref{eqn:decodingunitary}), one can write 

\begin{equation}
    U_AU_{A^c} |\widetilde{\alpha, ij}\rangle = |\alpha,i\rangle_{A_{1^{\alpha}}}|\alpha, j\rangle_{A^c_{1^{\alpha}}}|\chi_\alpha\rangle_{A_{2^{\alpha}}A^c_{2^{\alpha}}}.
\end{equation}

To see why such systems satisfy an FLM formula, consider any logical state $\tilde{\rho}\in L(\mathcal{H}_{\rm code})$ that may be mixed in general. The reduced state $\tilde{\rho}_A = \Tr_{A^c}[\tilde{\rho}]$ on $A$ admits a representation in $M$ such that
\begin{equation}
    \tilde{\rho}_A = \bigoplus_{\alpha} p_{\alpha}\tilde{\rho}_{a_{\alpha}}
    \label{eqn:ablockinfo1}
\end{equation}
is block diagonal in the decomposition (\ref{eqn:MCdecomp}). Above forms can be obtained by performing partial trace on each of the diagonal blocks, also referred to as the \textit{$\alpha$-blocks}, of $\tilde{\rho}$. 

Similarly, for the complementary subsystem $A^c$ with commutant $M'$,

\begin{equation}
    \tilde{\rho}_{A^c} = \bigoplus_{\alpha} p_{\alpha}\tilde{\rho}_{a_{\alpha}^c}.
    \label{eqn:ablockinfo2}
\end{equation}
Unit trace conditions on $\tilde{\rho}_A, \tilde{\rho}_{a_{\alpha}}$, and $\tilde{\rho}_{\bar{a}_{\alpha}}$ ensure that 
\begin{equation}
    \sum_{\alpha} p_{\alpha}=1.
\end{equation}

Here, $\{p_{\alpha}\}$ can be interpreted as a probability distribution over the $\alpha$-blocks\footnote{This is similar to the probability over different branches of the global wavefunction $\tilde{\rho}$ in a many-worlds picture of quantum mechanics.}.
Then using the decoding unitary $U_A, U_{A^c}$, we have

\begin{align}
    \tilde{\rho}_A &= \Tr_{A^c}\tilde{\rho} = U_A^{\dagger}(\oplus_{\alpha}p_{\alpha}\rho_{A_{1^{\alpha}}}\otimes \chi_{A_{2^{\alpha}}})U_A\\
    \tilde{\rho}_{A^c} &=\Tr_{A}\tilde{\rho} = U^{\dagger}_{A^c}(\oplus_{\alpha}p_{\alpha}\rho_{{A}^c_{1^{\alpha}}}\otimes \chi_{A^c_{2^{\alpha}}})U_{A^c},
\end{align}
where $p_{\alpha}\rho_{A_{1^{\alpha}}}, p_{\alpha}\rho_{{A}^c_{1^{\alpha}}}$ are precisely the encoded information in each block (\ref{eqn:ablockinfo1}, \ref{eqn:ablockinfo2}). The reduced state $\chi_{A_{2^{\alpha}}}$ is defined as $\chi_{A_{2^{\alpha}}}=\Tr_{A_{2^{\alpha}}^c}|\chi_{\alpha}\rangle\langle\chi_{\alpha}|$. The complement $\chi_{A^c_{2^{\alpha}}}$ is defined similarly by tracing out $A_{2^{\alpha}}$ instead.

Because the von Neumann entropy is preserved under unitary conjugation, this implies the following equalities

\begin{align}
    S(\tilde{\rho}_A) &= \Tr[\tilde{\rho}\mathcal{L}_A]+S(\tilde{\rho},M)\\
    S(\tilde{\rho}_{A^c}) &= \Tr[\tilde{\rho}\mathcal{L}_A]+S(\tilde{\rho},M').
    \label{eqn:HarlowFLM}
\end{align}

Here

\begin{equation}
    \mathcal{L}_A = \bigoplus_{\alpha} S(\chi_{A_{2^{\alpha}}})I_{a_{\alpha}{a}^c_{\alpha}}
\end{equation}
is an analog of the so-called area operator in quantum gravity. The first term on the right-hand side of the entropic relation plays the role of the RT $\mathrm{Area}/4G$ term, where the area term analogous to \ref{eqn:flm} is now is given by

\begin{equation}
\label{eqn:RTareaEnt}
    \mathrm{Area}\propto \Tr[\tilde{\rho}\mathcal{L}_A] = \sum_{\alpha}p_{\alpha} S(\chi_{A_{2^{\alpha}}}).
\end{equation}
This is simply a weighted sum over minimal areas over each branch/block of the wave function. 
The second term is more naturally linked to the bulk entropy contribution, 
\begin{equation}
    S(\tilde{\rho}, M) = \sum_{\alpha} p_{\alpha}S(\tilde{\rho}_{a_{\alpha}}) - \sum_{\alpha} p_{\alpha}\log p_{\alpha},
\end{equation}
where the first term is simply the weighted sum of the entropies of information  encoded on each $\alpha$-block, and the second term is a Shannon-like contribution called the entropy of mixing\cite{Almheiri:2016blp}. 

More generally, each $\alpha$-block can be interpreted as a different semi-classical geometry where the typical encoded information $\tilde{\rho}$ takes on a superposition of such geometries. In particular, they are related to the ``fixed-area states'' which the existing tensor network models describe\cite{Akers:2018fow,Dong:2018seb}. We will further comment on the relation between our construction in this work and the fixed area states in Section~\ref{sec:7disc}.

\section{Bacon-Shor Codes}
\label{sec:3baconshor}
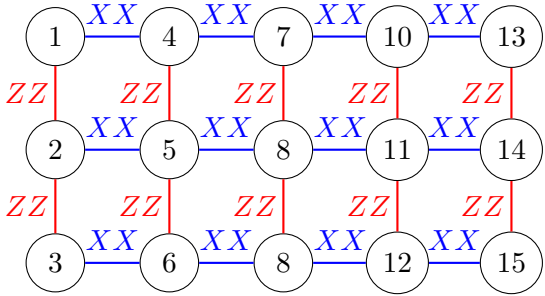
\begin{figure}[ht]
    \centering
    \begin{tikzpicture}[scale=1.5]
        \draw (0,0) node[circle,draw,minimum size=2em] (0) {$3$};
        \draw (0,1) node[circle,draw,minimum size=2em] (1) {$2$};
        \draw (0,2) node[circle,draw,minimum size=2em] (2) {$1$};
        \draw (1,0) node[circle,draw,minimum size=2em] (3) {$6$};
        \draw (1,1) node[circle,draw,minimum size=2em] (4) {$5$};
        \draw (1,2) node[circle,draw,minimum size=2em] (5) {$4$};
        \draw (2,0) node[circle,draw,minimum size=2em] (6) {$8$};
        \draw (2,1) node[circle,draw,minimum size=2em] (7) {$8$};
        \draw (2,2) node[circle,draw,minimum size=2em] (8) {$7$};
        \draw (3,0) node[circle,draw,minimum size=2em] (9) {$12$};
        \draw (3,1) node[circle,draw,minimum size=2em] (10) {$11$};
        \draw (3,2) node[circle,draw,minimum size=2em] (11) {$10$};
        \draw (4,0) node[circle,draw,minimum size=2em] (12) {$15$};
        \draw (4,1) node[circle,draw,minimum size=2em] (13) {$14$};
        \draw (4,2) node[circle,draw,minimum size=2em] (14) {$13$};
        \draw[thick,color=red] (0) -- (1) -- (2);
        \draw[thick,color=red] (3) -- (4) -- (5);
        \draw[thick,color=red] (6) -- (7) -- (8);
        \draw[thick,color=red] (9) -- (10) -- (11);
        \draw[thick,color=red] (12) -- (13) -- (14);
        \draw[thick,color=blue] (0) -- (3) -- (6) -- (9) -- (12);
        \draw[thick,color=blue] (1) -- (4) -- (7) -- (10) -- (13);
        \draw[thick,color=blue] (2) -- (5) -- (8) -- (11) -- (14);
        \draw (-0.25,0.5) node[color=red] {$ZZ$};
        \draw (-0.25,1.5) node[color=red] {$ZZ$};
        \draw (0.75,0.5) node[color=red] {$ZZ$};
        \draw (0.75,1.5) node[color=red] {$ZZ$};
        \draw (1.75,0.5) node[color=red] {$ZZ$};
        \draw (1.75,1.5) node[color=red] {$ZZ$};
        \draw (2.75,0.5) node[color=red] {$ZZ$};
        \draw (2.75,1.5) node[color=red] {$ZZ$};
        \draw (3.75,0.5) node[color=red] {$ZZ$};
        \draw (3.75,1.5) node[color=red] {$ZZ$};
        \draw (0.5,0.2) node[color=blue] {$XX$};
        \draw (0.5,1.2) node[color=blue] {$XX$};
        \draw (0.5,2.2) node[color=blue] {$XX$};
        \draw (1.5,0.2) node[color=blue] {$XX$};
        \draw (1.5,1.2) node[color=blue] {$XX$};
        \draw (1.5,2.2) node[color=blue] {$XX$};
        \draw (2.5,0.2) node[color=blue] {$XX$};
        \draw (2.5,1.2) node[color=blue] {$XX$};
        \draw (2.5,2.2) node[color=blue] {$XX$};
        \draw (3.5,0.2) node[color=blue] {$XX$};
        \draw (3.5,1.2) node[color=blue] {$XX$};
        \draw (3.5,2.2) node[color=blue] {$XX$};
    \end{tikzpicture}
    \caption{The $15$ qubit $3\times 5$ Bacon-Shor code and its gauge generators.}
    \label{fig:3x5Bacon-Shor}
\end{figure}

Generalized Bacon-Shor codes are constructed as subsystem codes represented on regular 2-dimensional grids (although higher-dimensional analogues have been proposed) \cite{Shor95,Bacon2006,Bravyi2011,Poulin2005,Jiang-Rieffel}. The general construction is to fix an $r\times s$ grid and place $n\leq rs$ qubits at some subset of vertices of the grid. For instance in Figure \ref{fig:3x5Bacon-Shor} we show the usual $3\times 5$ Bacon-Shor code where every vertex supports a qubit. The \emph{gauge operators} $\mathcal{G}$ are generated by weight two Pauli operators of the form $X\otimes X$ and $Z\otimes Z$ acting on pairs of adjacent qubits; if the qubits labeled $j$ and $k$ are horizontally adjacent then the gauge generator is $X_jX_k$, while for vertically adjacent qubits one instead would take $Z_jZ_k$. For example, in the $3\times 5$ Bacon-Shor code of Figure \ref{fig:3x5Bacon-Shor} its gauge operators are generated as
\begin{equation}
    \mathcal{G} = \langle X_1X_4, X_2X_5, \dots, X_{11}X_{14}, X_{12}X_{15}, Z_1Z_2, Z_2Z_3, \cdots, Z_{13}Z_{14}, Z_{14}Z_{15} \rangle.
\end{equation}
Notice that any two qubits in a row support a weight two $X\otimes X$ gauge operator obtained by multiplying together the generators for each adjacent pair of qubits between them. Similarly, there is a $Z \otimes Z$ gauge operator acting on any two qubits in a single column.

Clearly $\mathcal{G}$ is noncommutative, for instance $(X_1X_4)(Z_1Z_2) = - (Z_1Z_2)(X_1X_4)$ above, and so one cannot build a stabilizer code from it directly. The center $\mathcal{Z}(G)$ of $\mathcal{G}$ however is commutative, which we can take to be the stabilizers of the code. In the $3\times 5$ Bacon-Shor code above this is
\begin{equation}
    \mathcal{Z(G)} = \langle R_1R_2, R_2R_3, C_1C_2, C_2C_3, C_3C_4, C_4C_5\rangle
\end{equation}
where the ``row'' and ``column'' operators are defined as
\begin{equation}
    \begin{array}{rcl}
        R_1 &=& Z_1Z_4Z_7Z_{10}Z_{13}\\
        R_2 &=& Z_2Z_5Z_8Z_{11}Z_{14}\\
        R_3 &=& Z_3Z_6Z_9Z_{12}Z_{15}
    \end{array}\qquad\qquad
    \begin{array}{rcl}
        C_1 &=& X_1X_2X_3\\
        C_2 &=& X_4X_5X_6\\
        C_3 &=& X_7X_8X_9\\
        C_4 &=& X_{10}X_{11}X_{12}\\
        C_5 &=& X_{13}X_{14}X_{15}.
    \end{array}
\end{equation}
In fact, the stabilizer of generalized Bacon-Shor codes are always generated by products of row operators and products of column operators, and the method for determining these is straightforward. See section 3 of \cite{Bravyi2011}.

As a stabilizer code, a generalized Bacon-Shor code encodes a larger number of qubits. For example, the $3\times 5$ Bacon-Shor code above would encode $9$ logical qubits. However these codes are treated as \emph{subsystem} codes; there is a vast literature on this subject from a number of different perspectives, see for instance \cite{bacon1999robustness,Poulin2005}. For our purposes, we can take a constructive approach. The centralizer $\mathcal{C(G)}$ of $\mathcal{G}$ is the collection of Pauli operator that commute with all elements of $\mathcal{G}$, and in fact $\mathcal{Z(G)} = \mathcal{G} \cap \mathcal{C}(\mathcal{G})$. We can always write
\begin{equation}
    \mathcal{C(G)} = \langle \mathcal{Z(G)}, \overline{X}_1, \overline{Z}_1, \dots, \overline{X}_k, \overline{Z}_k\rangle
\end{equation}
where $\langle \overline{X}_j, \overline{Z}_j\rangle$ is the Pauli group for the $j^\text{th}$ bulk/logical qubit. As usual with the stabilizer codes, the logical Pauli elements above are only defined modulo the stabilizer. For the $3\times 5$ Bacon-Shor code there is only a single bulk qubit and its Pauli group is generated by $\overline{X} = C_1$, or indeed any product of an odd number of columns, and $\overline{Z} = R_1$ or any product of an odd number of rows. 

Yet, for subsystem codes, there is no canonical way to define a bulk subspace as there are no joint eigenspaces for all operators in $\mathcal{G}$. Nonetheless, we can ``promote'' $g_1,\dots,g_\ell \in \mathcal{G}$ that pairwise commute so that
\begin{equation}
    \mathcal{S}_{g_1,\dots,g_\ell} = \langle \mathcal{Z(G)}, g_1, \dots, g_\ell\rangle
\end{equation}
becomes a commutative group of rank $n-k$. Then the joint $+1$-eigenspace of these operators form a concrete subspace of our Hilbert space on which $\overline{X}_1, \overline{Z}_1, \dots, \overline{X}_k, \overline{Z}_k$ act as logical Pauli operators. The choice of $g_1,\dots,g_\ell$ is far from unique and might be interpreted as a particular choice of gauge for expressing our bulk subspace. For simplicity we will often write this subspace as the $g_1 = \cdots = g_\ell = +1$ gauge.

While much of our construction works for a variety of generalized Bacon-Shor codes, we can illustrate the salient points with just the $4$-qubit $2\times 2$ code. Our gauge group is
\begin{equation}
    \mathcal{G} = \langle Z_1Z_2, Z_3Z_4, X_1X_3, X_2X_4 \rangle
\end{equation}
leading to the stabilizer
\begin{equation}
    \mathcal{Z(G)} = \langle Z_1Z_2Z_3Z_4, X_1X_2X_3X_4 \rangle
\end{equation}
and logical operators
\begin{align}
    \overline{X} &= X_1X_2 = X_3X_4 \pmod{\mathcal{Z(G)}}\\
    \overline{Z} &= Z_1Z_3 = Z_2Z_4 \pmod{\mathcal{Z(G)}}.
\end{align}

\begin{figure}[ht]
    \centering
    \begin{tikzpicture}[scale=1.5]
        \draw (-1,1) node {a)};
        \draw (0,0) node[circle,draw,minimum size=2em] (0) {$2$};
        \draw (0,1) node[circle,draw,minimum size=2em] (1) {$1$};
        \draw (1,0) node[circle,draw,minimum size=2em] (3) {$4$};
        \draw (1,1) node[circle,draw,minimum size=2em] (4) {$3$};
        \draw[thick,color=red] (0) -- (1);
        \draw[thick,color=red] (3) -- (4);
        \draw[thick,color=blue] (0) -- (3);
        \draw[thick,color=blue] (1) -- (4);
        \draw (-0.25,0.5) node[color=red] {$ZZ$};
        \draw (1.25,0.5) node[color=red] {$ZZ$};
        \draw (0.5,-0.2) node[color=blue] {$XX$};
        \draw (0.5,1.2) node[color=blue] {$XX$};
        \draw (3,1) node {b)};
        \draw[thick] (4,0) -- (4,1) -- (5,1) -- (5,0) -- (4,0);
        \draw[fill=blue] (4,0) circle (3pt);
        \draw[fill=red] (4,1) circle (3pt);
        \draw[fill=red] (5,0) circle (3pt);
        \draw[fill=blue] (5,1) circle (3pt);
        \draw (4.5,0) node[circle,draw,fill=white] (a) {$4$};
        \draw (4,0.5) node[circle,draw,fill=white] (b) {$2$};
        \draw (5,0.5) node[circle,draw,fill=white] (c) {$3$};
        \draw (4.5,1) node[circle,draw,fill=white] (d) {$1$};
        \draw (3.75,-0.2) node[color=blue] {$XX$};
        \draw (3.75,1.2) node[color=red] {$ZZ$};
        \draw (5.25,-0.2) node[color=red] {$ZZ$};
        \draw (5.25,1.2) node[color=blue] {$XX$};
    \end{tikzpicture}
    \caption{Two views of the $2\times 2$ Bacon-Shor code (a) qubits at vertices with gauge generator on edges or (b) qubits as edges with gauge generators at vertices.}
    \label{fig:baconshor}
\end{figure}
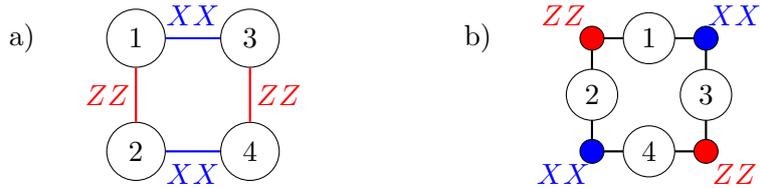

\begin{figure}[ht]
    \centering
    \includegraphics[width=0.6\textwidth]{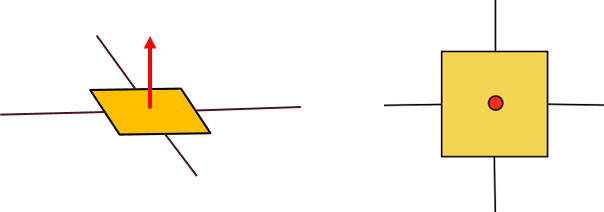}
    \caption{The tensor network representation of a Bacon-Shor code (or the multicopy stacked Bacon-Shor code which we will define later on). The (red) arrow/dot denote the single logical qubit that it encodes, and the four in-plane legs (black) denote the physical qubits.}
    \label{fig:BStensor}
\end{figure}

 In addition to the conventional presentation of the Bacon-Shor code, we will use its tensor network representation extensively in this work. In the tensor network representation (Figure~\ref{fig:BStensor}), we can rewrite the 4-qubit Bacon-Shor code as a tensor with four in-plane legs that represent the four physical qubits, as well as a ``bulk'' leg that sticks out of the plane, which represents the logical qubit. In this notation (Figure~\ref{fig:logicalPushBST}), the logical operator, which acts on the ``bulk'' degree of freedom can be represented as a logical operator that acts on the logical qubit (bulk leg), or its equivalent representation on the physical qubits (in-plane legs). 

\subsection{Entanglement Properties of the 2-by-2 Bacon-Shor Code}
\label{subsec:bs4}
Recently, the 5 qubit code (or perfect code) was used to construct a holographic quantum error correction code that reproduced interesting features such as subregion duality and entanglement wedge reconstruction. However, the code subalgebra has a trivial center, whereas it has been argued that more general holographic codes need to have a non-trivial center as well\cite{Harlow:2016vwg}. One such attempt is given by \cite{Donnelly:2016qqt}.
This is partly related to the fact that the perfect code corrects errors a little too well: on any given subregion there is either full access to the logical qubit or none whatsoever, but never in between. However, the 4 qubit Bacon-Shor code~(Figure~\ref{fig:baconshor}), having inherited features from repetition codes, has a non-trivial center for the code subalgebra in 2-qubit bipartitions. This is because for some subsystem one can access the logical $X$ or $Z$ subalgebra, but not the full Pauli algebra.

We first note that for any 3-1 bipartitioning of the 4 physical qubits, because the code corrects a located error on any one qubit, there must be maximal entanglement between the one qubit and its complement, regardless of the gauge choice. Without loss of generality, let $A$ contain 3 qubits and $A^c$ contain the remaining 1. This also implies that $A$ has full access to the logical Pauli algebra, while $A^c$ has none. In particular, the recovery is complementary in that the 3 qubit subsystem recovers the entire logical Pauli algebra, whereas the remaining 1 qubit subsystem recovers no logical information. This is a somewhat trivial example of the ``subsystem quantum erasure correction'' as defined in Section 4 of \cite{Harlow:2016vwg}, whose notation we will use below.  Explicitly, consider a tensor product bipartition of the Hilbert space, 
\begin{equation}
    \mathcal{H} = \mathcal{H}_A\otimes \mathcal{H}_{A^c}
\end{equation}
where as the code subspace 
\begin{equation}
    \mathcal{H}_{\rm code}= \mathcal{H}_a\otimes \mathcal{H}_{a^c}
\end{equation}
such that $A$ can access $\mathcal{H}_a$, which consists of one qubit, and $A^c$ can access $\mathcal{H}_{a^c}$, which is trivial in our example. 

Since both $\bar{X},\bar{Z}$ have representation over $A$, the encoded information can be decoded by acting only on $A$. We can decompose the Hilbert space $\mathcal{H}_A$ as
\begin{equation}
    \mathcal{H}_A = (\mathcal{H}_{A_1}\otimes\mathcal{H}_{A_2}).
\end{equation}
such that there exists decoding unitary $U_A$ which decodes the information on $\mathcal{H}_{\rm code}$ into $\mathcal{H}_{A_1}$. That is, for any basis $|\bar{i}\rangle$ of the code subspace 
\begin{equation}
U_A|\bar{i}\rangle_{AA^c} = |i\rangle_{A_1}\otimes |\chi\rangle_{A_2A^c}.
\end{equation}

\begin{figure}[th]
\centering
\begin{tikzpicture}[scale=1.000000,x=1pt,y=1pt]
\filldraw[color=white] (0.000000, -7.500000) rectangle (54.000000, 52.500000);
\draw[color=black] (0.000000,45.000000) -- (54.000000,45.000000);
\draw[color=black] (0.000000,30.000000) -- (54.000000,30.000000);
\draw[color=black] (0.000000,15.000000) -- (54.000000,15.000000);
\draw[color=black] (0.000000,0.000000) -- (54.000000,0.000000);
\filldraw[color=white,fill=white] (0.000000,-3.750000) rectangle (-4.000000,48.750000);
\draw[decorate,decoration={brace,amplitude = 4.000000pt},very thick] (0.000000,-3.750000) -- (0.000000,48.750000);
\draw[color=black] (-4.000000,22.500000) node[left] {$|\bar{i}\rangle$};
\draw (9.000000,30.000000) -- (9.000000,15.000000);
\begin{scope}
\draw[fill=white] (9.000000, 30.000000) circle(3.000000pt);
\clip (9.000000, 30.000000) circle(3.000000pt);
\draw (6.000000, 30.000000) -- (12.000000, 30.000000);
\draw (9.000000, 27.000000) -- (9.000000, 33.000000);
\end{scope}
\filldraw (9.000000, 15.000000) circle(1.500000pt);
\draw (27.000000,45.000000) -- (27.000000,15.000000);
\begin{scope}
\draw[fill=white] (27.000000, 45.000000) circle(3.000000pt);
\clip (27.000000, 45.000000) circle(3.000000pt);
\draw (24.000000, 45.000000) -- (30.000000, 45.000000);
\draw (27.000000, 42.000000) -- (27.000000, 48.000000);
\end{scope}
\filldraw (27.000000, 15.000000) circle(1.500000pt);
\draw (45.000000,45.000000) -- (45.000000,30.000000);
\begin{scope}
\draw[fill=white] (45.000000, 30.000000) circle(3.000000pt);
\clip (45.000000, 30.000000) circle(3.000000pt);
\draw (42.000000, 30.000000) -- (48.000000, 30.000000);
\draw (45.000000, 27.000000) -- (45.000000, 33.000000);
\end{scope}
\filldraw (45.000000, 45.000000) circle(1.500000pt);
\draw[color=black] (54.000000,45.000000) node[right] {$|i\rangle$};
\draw[color=black] (54.000000,30.000000) node[right] {$|0\rangle$};
\filldraw[color=white,fill=white] (54.000000,-3.750000) rectangle (58.000000,18.750000);
\draw[decorate,decoration={brace,mirror,amplitude = 4.000000pt},very thick] (54.000000,-3.750000) -- (54.000000,18.750000);
\draw[color=black] (58.000000,7.500000) node[right] {${\tfrac{1}{\sqrt{2}}(|00\rangle + |11\rangle)}$};
\draw[draw opacity=0.000000,fill opacity=0.100000,fill=blue] (3.000000,52.500000) rectangle (51.000000,7.500000);
\draw[color=blue] (8,60) node {$A$};
\draw (-15,60) node {a)};
\end{tikzpicture}\quad
\begin{tikzpicture}[scale=1.000000,x=1pt,y=1pt]
\filldraw[color=white] (0.000000, -7.500000) rectangle (18.000000, 52.500000);
\draw[color=black] (0.000000,45.000000) -- (18.000000,45.000000);
\draw[color=black] (0.000000,30.000000) -- (18.000000,30.000000);
\draw[color=black] (0.000000,15.000000) -- (18.000000,15.000000);
\draw[color=black] (0.000000,0.000000) -- (18.000000,0.000000);
\filldraw[color=white,fill=white] (0.000000,-3.750000) rectangle (-4.000000,48.750000);
\draw[decorate,decoration={brace,amplitude = 4.000000pt},very thick] (0.000000,-3.750000) -- (0.000000,48.750000);
\draw[color=black] (-4.000000,22.500000) node[left] {$|\bar{i}\rangle$};
\draw (9.000000,45.000000) -- (9.000000,15.000000);
\begin{scope}
\draw[fill=white] (9.000000, 45.000000) circle(3.000000pt);
\clip (9.000000, 45.000000) circle(3.000000pt);
\draw (6.000000, 45.000000) -- (12.000000, 45.000000);
\draw (9.000000, 42.000000) -- (9.000000, 48.000000);
\end{scope}
\filldraw (9.000000, 15.000000) circle(1.500000pt);
\draw[color=black] (18.000000,45.000000) node[right] {$|i\rangle$};
\filldraw[color=white,fill=white] (18.000000,-3.750000) rectangle (22.000000,33.750000);
\draw[decorate,decoration={brace,mirror,amplitude = 4.000000pt},very thick] (18.000000,-3.750000) -- (18.000000,33.750000);
\draw[color=black] (22.000000,15.000000) node[right] {${\left\{\begin{array}{rl}\tfrac{1}{\sqrt{2}}(|000\rangle + |111\rangle) & i=0\\\tfrac{1}{\sqrt{2}}(|011\rangle + |100\rangle) & i=1\end{array}\right.}$};
\draw[draw opacity=0.000000,fill opacity=0.100000,fill=blue] (3.000000,52.500000) rectangle (15.000000,37.500000);
\draw[draw opacity=0.000000,fill opacity=0.100000,fill=blue] (3.000000,22.500000) rectangle (15.000000,7.500000);
\draw[color=blue] (8,60) node {$A$};
\draw (-15,60) node {b)};
\end{tikzpicture}
\begin{tikzpicture}[scale=1.000000,x=1pt,y=1pt]
\filldraw[color=white] (0.000000, -7.500000) rectangle (18.000000, 52.500000);
\draw[color=black] (0.000000,45.000000) -- (18.000000,45.000000);
\draw[color=black] (0.000000,30.000000) -- (18.000000,30.000000);
\draw[color=black] (0.000000,15.000000) -- (18.000000,15.000000);
\draw[color=black] (0.000000,0.000000) -- (18.000000,0.000000);
\filldraw[color=white,fill=white] (0.000000,-3.750000) rectangle (-4.000000,48.750000);
\draw[decorate,decoration={brace,amplitude = 4.000000pt},very thick] (0.000000,-3.750000) -- (0.000000,48.750000);
\draw[color=black] (-4.000000,22.500000) node[left] {$|\pm\bar{x}\rangle$};
\draw (9.000000,45.000000) -- (9.000000,30.000000);
\begin{scope}
\draw[fill=white] (9.000000, 30.000000) circle(3.000000pt);
\clip (9.000000, 30.000000) circle(3.000000pt);
\draw (6.000000, 30.000000) -- (12.000000, 30.000000);
\draw (9.000000, 27.000000) -- (9.000000, 33.000000);
\end{scope}
\filldraw (9.000000, 45.000000) circle(1.500000pt);
\draw[color=black] (18.000000,45.000000) node[right] {$|\pm{x}\rangle$};
\draw[color=black] (18.000000,30.000000) node[right] {$|0\rangle$};
\filldraw[color=white,fill=white] (18.000000,-3.750000) rectangle (22.000000,18.750000);
\draw[decorate,decoration={brace,mirror,amplitude = 4.000000pt},very thick] (18.000000,-3.750000) -- (18.000000,18.750000);
\draw[color=black] (22.000000,7.500000) node[right] {${\tfrac{1}{\sqrt{2}}(|00\rangle \pm |11\rangle)}$};
\draw[draw opacity=0.000000,fill opacity=0.100000,fill=blue] (3.000000,52.500000) rectangle (15.000000,22.500000);
\draw[color=blue] (8,60) node {$A$};
\draw (-15,60) node {c)};
\end{tikzpicture}
\caption{The decoding operations in the $Z_1Z_2 = +1$ gauge for (a) a 3-1 bipartition (\ref{equation:decode3-1}), (b) one 2-2 bipartition (\ref{equation:decodeZZ13a},\ref{equation:decodeZZ13b}), and (c) the other 2-2 bipartition (\ref{equation:decodeZZ12}).}\label{figure:decodingZZ}
\end{figure}
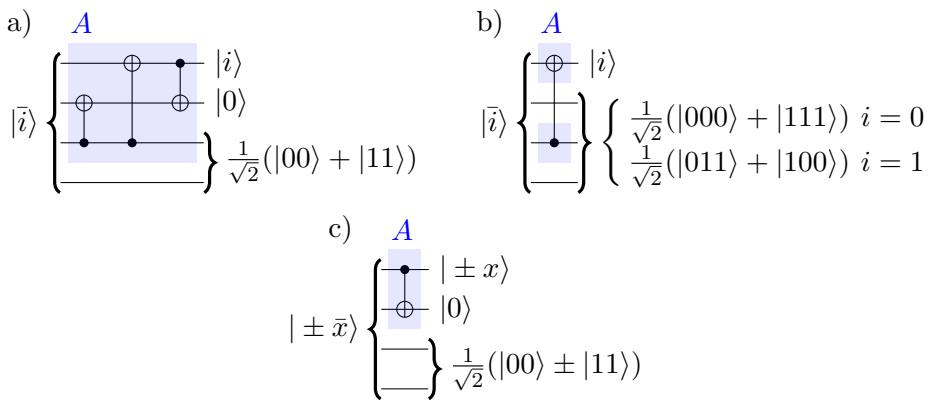

\begin{figure}[t]
    \centering
    \begin{tikzpicture}[scale=0.6]
\draw[thick] (-0.5,2.5) -- (3.5,2.5);
\draw[thick] (0.5,3.5) -- (2.5,1.5);
\draw[fill=yellow] (0,3) -- (2,3) -- (3,2) -- (1,2) -- (0,3);
\draw[color=red, thick] (1.5,2.5) -- (1.5,4);
\draw (1.5,4.4) node {$\bar{X}$};
\draw (3.5,1.5) node[rotate=-30] {\large $=$};
\draw[thick] (3.5,0.5) -- (7.5,0.5);
\draw[thick] (4.5,1.5) -- (6.5,-0.5);
\draw[fill=yellow] (4,1) -- (6,1) -- (7,0) -- (5,0) -- (4,1);
\draw[color=red, thick] (5.5,0.5) -- (5.5,2);
\draw (3.1,0.5) node {$X$};
\draw (6.7,-0.8) node {$X$};
\end{tikzpicture}
\begin{tikzpicture}[scale=0.6]
\draw[thick] (-0.5,2.5) -- (3.5,2.5);
\draw[thick] (0.5,3.5) -- (2.5,1.5);
\draw[fill=yellow] (0,3) -- (2,3) -- (3,2) -- (1,2) -- (0,3);
\draw[color=red, thick] (1.5,2.5) -- (1.5,4);
\draw (1.5,4.4) node {$\bar{Y}$};
\draw (3.5,1.5) node[rotate=-30] {\large $=$};
\draw[thick] (3.5,0.5) -- (7.5,0.5);
\draw[thick] (4.5,1.5) -- (6.5,-0.5);
\draw[fill=yellow] (4,1) -- (6,1) -- (7,0) -- (5,0) -- (4,1);
\draw[color=red, thick] (5.5,0.5) -- (5.5,2);
\draw (4.3,1.8) node {$Z$};
\draw (3.1,0.5) node {$Y$};
\draw (6.7,-0.8) node {$X$};
\end{tikzpicture}
\begin{tikzpicture}[scale=0.6]
\draw[thick] (-0.5,2.5) -- (3.5,2.5);
\draw[thick] (0.5,3.5) -- (2.5,1.5);
\draw[fill=yellow] (0,3) -- (2,3) -- (3,2) -- (1,2) -- (0,3);
\draw[color=red, thick] (1.5,2.5) -- (1.5,4);
\draw (1.5,4.4) node {$\bar{Z}$};
\draw (3.5,1.5) node[rotate=-30] {\large $=$};
\draw[thick] (3.5,0.5) -- (7.5,0.5);
\draw[thick] (4.5,1.5) -- (6.5,-0.5);
\draw[fill=yellow] (4,1) -- (6,1) -- (7,0) -- (5,0) -- (4,1);
\draw[color=red, thick] (5.5,0.5) -- (5.5,2);
\draw (3.1,0.5) node {$Z$};
\draw (4.3,1.8) node {$Z$};
\draw (6.7,-0.8) node {\phantom{$I$}};
\end{tikzpicture}
    \caption{The logical Pauli operators can be ``pushed'' to the in-plane legs in the tensor network representation. Here we have ``cleaned'' the rightmost leg -- all logical operations act trivially on this qubit.}
    \label{fig:logicalPushBST}
\end{figure}
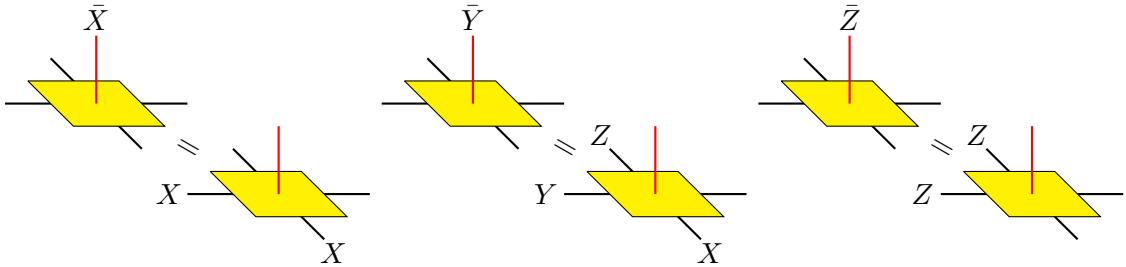

Without loss of generality, we can choose $\{|i\rangle\}$ to be the computational basis, and take $A=\{1,2,3\}$ and $A^c=\{4\}$. One can easily construct $U_A$, for example, by choosing the $ZZ=+1$ gauge. The computational basis vectors of the logical qubit is a GHZ state

\begin{align}
    |\bar{0}\rangle &= \tfrac{1}{\sqrt{2}} (|0000\rangle+|1111\rangle)\\
    |\bar{1}\rangle &= \tfrac{1}{\sqrt{2}} (|1100\rangle +|0011\rangle),
    \label{eqn:bs4ZZgauge}
\end{align}
where we label the position of the qubits according to either diagram in Figure \ref{fig:baconshor}. 
The decoding unitary $U_A$, as shown in Figure \ref{figure:decodingZZ}a, decodes the information as follows:
\begin{equation}\label{equation:decode3-1}
    U_A |\bar{\psi}\rangle = |\psi\rangle_{A_1}\otimes \left(|0\rangle\otimes \frac{1}{\sqrt{2}}(|00\rangle+|11\rangle)\right)_{A_2A^c}
\end{equation}
where $A_1=\{1\}$ is the first physical qubit, $A_2=\{2,3\}$, and $|\chi\rangle = \frac{1}{\sqrt{2}}(|0\rangle\otimes (|00\rangle+|11\rangle))$. As shown in equation \ref{equation:decode3-1}, this bipartition contributes one unit of entanglement (or one ebit) to the RT surface (c.f. (\ref{eqn:RTareaEnt})) because $S(\chi_{A_2})=S(\Tr_{A^c}[|\chi\rangle\langle\chi|])=1$, but none to the bulk entropy in the RT/FLM formula.

The previous 3-1 bipartition has similar properties as one would find in the 5-qubit code or other existing holographic error correction codes, where having access to a physical subsystem provides access to a subset of the bulk qubits through their corresponding Pauli algebra. However, it is not possible for a subregion, or its complement, to have access to only a subalgebra. In fact, such codes indicate that the code subspace must be factorizable, which is not the case in holography where the bulk can support a gauge theory (gravity). 
More generally, we expect bipartitions of QECC to have a decomposition where the physical Hilbert space is factorizable,
\begin{equation}
    \mathcal{H}=\mathcal{H}_A\otimes \mathcal{H}_{A^c}
\end{equation}
yet the code subspace is not necessarily so
\begin{equation}
    \mathcal{H}_{\rm code} = \bigoplus_{\alpha} \mathcal{H}_{a_{\alpha}}\otimes \mathcal{H}_{a^c_{\alpha}}.
\end{equation}
In particular, the code subalgebra $M_A$ that is supported on $A$ has a nontrivial center $\mathcal{Z}(M_A) \ne \{I\}$. It has been argued that different $\alpha$ can correspond to different semi-classical geometries in a holographic error correction code \cite{Akers:2018fow,Dong:2018seb}.

We see this property in 2-2 bipartitions of the 4-qubit code, where the entanglement entropies will now depend on the state of the gauge qubit. For our purposes, we primarily focus on two different choices that result in GHZ or Bell-like entanglement as they are easier for developing intuition. Nevertheless, the code itself is far more versatile and can accommodate a range of constructions, which will help produce the correct kind of correlation function. We will discuss this briefly later on. 

First, consider the $Z_1Z_2=+1$ gauge as we did above. 
In this gauge, $Z_1Z_2$ and $Z_3Z_4$ are promoted to stabilizer generators, which together with $X_1X_2X_3X_4$ generate the stabilizer group. While it is clear from the outset the total entanglement across any bipartition is of unit 1, it is worthwhile to understand what these terms correspond to in the RT-FLM formula. 

For a 2-qubit subsystem $A$ that includes just either the top or bottom row of Figure \ref{fig:baconshor}a, it is only possible to support the logical $\bar{Z}$ operator.\footnote{A similar argument applies when the 2-qubit subsystem contains the diagonal, where the entanglement computation is the same as that of partitioning the rows. As $Z_1Z_2$ and $Z_3Z_4$ are stabilizers in this gauge, $Z_1Z_4$ and $Z_2Z_3$ are also logical $\bar{Z}$ operators.} Without loss of generality, let $A = \{1,3\}$ be the top row with qubits 1 and 3.  
This subregion only has access to the commutative subalgebra generated by $\bar{Z}$. In general, whenever a subalgebra has a nontrivial center we can decompose our Hilbert space in such a way that the central elements act trivially (that is, as a multiple of the identity) on each summand. In this case, we can write 
\begin{equation}
    \mathcal{H}_{\rm code} = \mathcal{H}_{0} \oplus \mathcal{H}_1
\end{equation}
where $\mathcal{H}_{\alpha} = \text{span}\{|\bar{\alpha}\rangle\}$, and 
\begin{equation}
    \mathcal{H}_A = (\mathcal{H}_{A_1}^{0}\otimes \mathcal{H}_{A_2}^{0}) \oplus (\mathcal{H}_{A_1}^{1}\otimes \mathcal{H}_{A_2}^{1})
\end{equation}
Here $\mathcal{H}_{A_1}^{\alpha} = \text{span}\{|\alpha\rangle\}$ are one dimensional Hilbert spaces, and $\mathcal{H}_{A_2}^{\alpha} = \mathbb{C}^2$. Indeed, for basis $|\bar{0}\rangle, |\bar{1}\rangle$, we have decoding unitary $U_A=CNOT(3,1)$ with
\begin{align}
    U_A|\bar{0}\rangle &= |0\rangle \otimes \tfrac{1}{\sqrt{2}}(|000\rangle + |111\rangle)\label{equation:decodeZZ13a}\\
    U_A|\bar{1}\rangle &= |1\rangle \otimes \tfrac{1}{\sqrt{2}}(|011\rangle + |100\rangle)\label{equation:decodeZZ13b}
\end{align}
Tracing over the complement $A^c = \{2,4\}$ produces
\begin{align}
    \Tr_{A^c}(U_A|\bar{0}\rangle\langle\bar{0}| U_A^{\dagger}) &= |0\rangle\langle 0|\otimes \tfrac{1}{2}I_2\\ 
    \Tr_{A^c}(U_A|\bar{1}\rangle\langle\bar{1}| U_A^{\dagger}) &= |1\rangle\langle 1|\otimes \tfrac{1}{2}I_2.
    \label{eqn:decstatezz}
\end{align}
Therefore, the logical 0 and 1 states have a single unit of entanglement that can be purely attributed to the surface area term in the RT formula. 

More generally, any state $\bar{\rho}\in L(\mathcal{H}_{\rm code})$ can be reduced to the code subalgebra. For instance, consider a mixed state $\tilde{\rho} = p|\bar{0}\rangle\langle\bar{0}| + (1-p)|\bar{1}\rangle\langle\bar{1}|$, represented in the logical Z basis as
\begin{equation}
    \tilde{\rho} \mapsto 
    \begin{pmatrix}
    p & 0\\
    0 & 1-p
    \end{pmatrix}.
\end{equation}
Its total entanglement is 
\begin{align}
    S(\tilde{\rho}_A) &= \Tr[\mathcal{L_A}\tilde{\rho}] +S(M_A, \tilde{\rho}) \\
   &= S(\chi) +[- p\log p - (1-p)\log (1-p)],
\end{align}
where there is one unit of entanglement from the $\chi$-term that corresponds to the minimal area in the RT formula and an additional bulk entropy term that is Shannon-like. The maximal bulk entropy contribution is also unit 1. 

On the other hand, the entanglement across the columns, while having the same value, is very different in nature according to the RT formula. Indeed, when $A = \{1,2\}$ one has access to the subalgebra generated by $\bar{X}$. Now the decomposition for this subalgebra is along basis vectors 
\begin{align}
    |+\bar{x}\rangle &= \tfrac{1}{2}(|00\rangle + |11\rangle)\otimes(|00\rangle + |11\rangle)\\
    |-\bar{x}\rangle &= \tfrac{1}{2}(|00\rangle - |11\rangle)\otimes(|00\rangle - |11\rangle),
    \label{eqn:decstatexx}
\end{align}
which are both separable with respect to this bipartition. In particular, with decoding unitary $U_A=CNOT(1,2)$ we have
\begin{equation}\label{equation:decodeZZ12}
U_A|\pm\bar{x}\rangle = |\pm x\rangle\otimes |0\rangle \otimes \tfrac{1}{\sqrt{2}}(|00\rangle \pm |11\rangle),
\end{equation}
and tracing out the complement $A^c$ we find
\begin{align}
    \Tr_{A^c}(U_A|{+\bar{x}}\rangle\langle{+\bar{x}}| U_A^{\dagger}) &= |+\rangle\langle +|\otimes |0\rangle\langle 0|\\ 
    \Tr_{A^c}(U_A|{-\bar{x}}\rangle\langle{-\bar{x}}| U_A^{\dagger}) &= |-\rangle\langle -|\otimes |0\rangle\langle 0|.
\end{align}
Therefore, for $|\bar{0}\rangle, |\bar{1}\rangle$
\begin{equation}
    S(\tilde{\rho}_A) = S(\tilde{\rho}_A, M_A) = 1.
\end{equation}
So for $A = \{1,2\}$ the bulk entropy contribution is of unit 1, while the area contribution is zero. 
Indeed, the bulk contribution can also be computed explicitly. It is clear that when written in the $|\pm \bar{x}\rangle$ basis, the states $|\bar{0}\rangle,|\bar{1}\rangle$ reduces to the maximally mixed state in the $\{\bar{I},\bar{X}\}$ subalgebra. Hence the bulk term has one unit of entanglement from the Shannon term\footnote{Proof is left as an exercise for the reader.}.

Similarly, we can fix the $X_1X_3=+1$ gauge, with 
\begin{align}
    |\bar{0}\rangle &= \tfrac{1}{2}(|0000\rangle + |0101\rangle +|1010\rangle +|1111\rangle) = \tfrac{1}{2}(|00\rangle + |11\rangle)_{1,3}\otimes (|00\rangle + |11\rangle)_{2,4}\\
    |\bar{1}\rangle &= \tfrac{1}{2}(|0011\rangle + |0110\rangle +|1001\rangle +|1100\rangle) = \tfrac{1}{2}(|01\rangle + |10\rangle)_{1,3}\otimes (|01\rangle + |10\rangle)_{2,4}
\end{align}
In this case, we also have two Bell pairs, now with each Bell pair positioned along each row. Again, any single qubit is maximally entangled with the complement. For 2-2 split, the entanglement between the top and the bottom rows in the logical 0 or 1 state is zero, while the entanglement across columns is 2. 

Again, we wish to understand the nature of these entanglement contributions in terms of the FLM formula. For a subsystem of rows, it is clear that all entanglement is trivial. For a column subsystem, the decoding unitary is given by CNOT but conjugated by Hadamard. The code subalgebra is diagonal in the $|\pm \bar{x}\rangle$ basis, where these states are GHZ states. A quick decoding reveals that 

\begin{equation}
    U_A \Tr_{A^c}[|\pm \bar{x}\rangle\langle \pm \bar{x}|] U_A^{\dagger} = |\pm x\rangle \langle \pm x|\otimes\tfrac{1}{2} I_2
\end{equation}

Given that the logical 0 and 1 reduce to a maximally mixed state in the $X$ code subalgebra on the subsystem
\begin{equation}
    |\bar{\alpha}\rangle \langle \bar{\alpha}|\rightarrow \frac 1 2 |+\bar{x}\rangle\langle +\bar{x}|+\frac 1 2 |-\bar{x}\rangle\langle -\bar{x}|
\end{equation}
we again obtain the entropy contributions $S(\chi)=1, S(\tilde{\rho}_A, M_X)=1$, where the area term and the bulk term both contribute one unit of entanglement.


 \subsection{Operator Pushing in the Bacon-Shor code}
 \label{subsec:bsoptpush}

 Similar to the perfect tensor and the perfect code in \cite{Pastawski:2015qua}, we can also discuss operator pushing for the Bacon-Shor code, which has a convenient tensor network representation. Readers unfamiliar with this concept can find detailed explanations in Sec. 2 of~\cite{Pastawski:2015qua}. As the pushing rule amounts to simply finding an equivalent representation of a certain operator and how it acts on states within the code subspace, we can also consider pushing ``across'' a tensor. Here we find equivalent representations of physical operators that preserve the logical information or logical subspace. 
 In the former case, there are two methods for operator pushing, unlike the HaPPY code construction. The more restrictive of these we call \emph{state-invariant} pushing, which involves stabilizer multiplication exactly like the HaPPY construction. This type of operator pushing allows us to ``clean'' \cite{haahpreskill,BravyiTerhal,Flammiaetal} a particular subsystem on which any equivalent operator will act trivially. More precisely, this implies the following equality holds for any $|\psi\rangle$ in the code space:
 \begin{equation}
    O_{A,A^c}|\psi\rangle = I_A\otimes Q_{A^c}|\psi\rangle.
 \end{equation}
 
As the Bacon-Shor code detects one error on any qubit, one can clean any weight one operator by multiplying a suitable stabilizer so that its equivalent representations act trivially on that qubit. As such, any operator that has support over one site can be pushed to at most 3 other sites using stabilizers. Explicitly, we see that any operator $O$ on a physical qubit can be cleaned using the stabilizer elements. 
 
For concreteness, let $A$ be a single qubit subregion and $O_A\otimes I_{A^c}$ be any single qubit operator, which can be expanded in the Pauli basis as
 \begin{equation}\label{eqn:paulidecomp}
     O_A= \sum_{i=0}^{3} c_iP_i,
 \end{equation}
 where as usual $P_0=I$. Using stabilizers to give equivalent representations while preserving the state, we have that for any $ |\psi\rangle\in\mathcal{H}_{\rm code}$ and any single Pauli operator $P_A$ we can write
 \begin{equation}
     P_{A} |\psi\rangle  = P_{A^c}|\psi\rangle
 \end{equation}
 where $P_{A^c}$ is a suitable Pauli operator supported on $A^c$.\footnote{Namely, we simply choose a $P_{A^c}$ so that $P_A\otimes P_{A^c}$ is a stabilizer, which always exists as the Bacon-Shor code detects one error on any qubit.}
 Applying this term by term in the decomposition (\ref{eqn:paulidecomp}), we obtain 
 
 \begin{equation}
     (O_A\otimes I_{A^c})|\psi\rangle = [I_{A}\otimes(c_0 I_{A^c}+c_1  (XXX)_{A^c} +c_2 (YYY)_{A^c}+c_3  (ZZZ)_{A^c})]|\psi\rangle,
 \end{equation}

\begin{figure}[ht]
    \centering
    \includegraphics[width=0.7\textwidth]{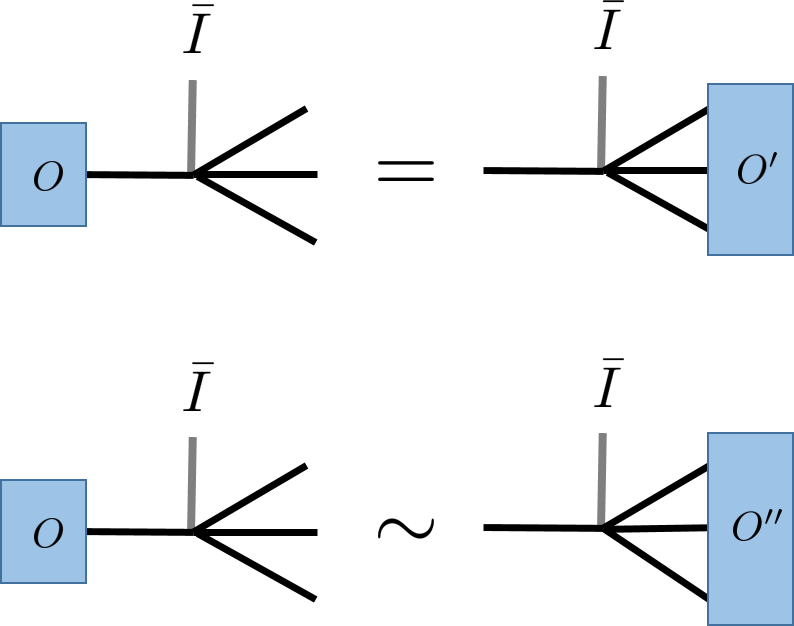}
    \caption{Top: pushing using stabilizers. Bottom: pushing using gauge operator. Note that the state in the top are precisely equal when we use stabilizers. The bottom will change the state from one member of the equivalence class to another while preserving the entanglement structure, as the gauge operators are transversal. Both pushing preserves the entanglement structure of the state. We do not leave the orbit where entanglement can change. This is important to note.}
    \label{fig:BSMT}
\end{figure}
 
 The second, more general method, for operator pushing need not leave the state $|\psi\rangle$ invariant, as long as the logical information in the subsystem is preserved. Such kind of \textit{state-covariant} pushing is done by multiplying gauge operators instead of stabilizer elements.
 
\begin{figure}[ht]
     \centering
     \includegraphics[width=0.7\textwidth]{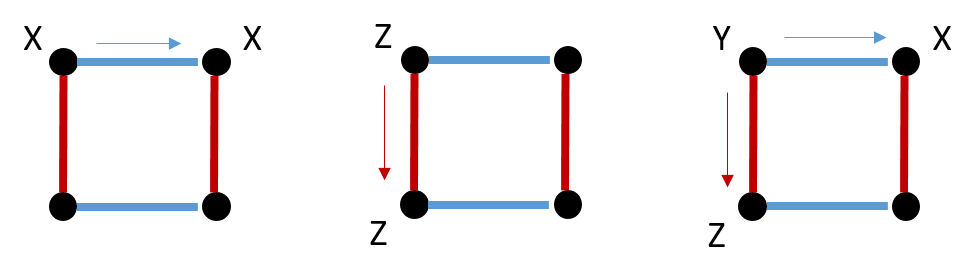}
     \caption{Pushing using gauge operators. Pushing of a weight one Pauli operator by multiplying gauge operators. This can ``push'' the operator onto at most two sites, but it can not be pushed across the diagonal.}
     \label{fig:gaugepushing}
 \end{figure}

Equivalently, we require that 
\begin{equation}
    O|\psi\rangle = I_A\otimes Q'_{A^c}|\psi'\rangle
\end{equation}
where $|\psi'\rangle\sim|\psi\rangle$ are gauge equivalent. For a single Bacon-Shor code, we first consider how single site Pauli operators can be cleaned using gauge elements. Weight one $X$ and $Z$ type operators can be pushed to another weight one $X$ or $Z$ operator, at the expense of possibly changing the gauge of the logical qubit by multiplying by the appropriate gauge operator. A single $Y$ operator can be pushed to two sites by multiplying both the gauge $X$ and $Z$ operators (Figure \ref{fig:gaugepushing}). However, this pushing is not always possible for arbitrary input and output locations because the Bacon-Shor code is not symmetric across all physical qubits, unlike the perfect code.

For example, suppose we wish to push $X_1$ into $X_3$ using the gauge operator $X_1X_3$. If our code is in the $Z_1Z_2 = +1$ gauge then
\begin{align}
    X_1(\alpha|\bar{0}\rangle + \beta|\bar{1}\rangle)_{Z_1Z_2=+1} &= \tfrac{\alpha}{\sqrt{2}}(|1000\rangle + |0111\rangle) + \tfrac{\beta}{\sqrt{2}}(|1011\rangle + |0100\rangle)\\
    &= X_3\left(\tfrac{\alpha}{\sqrt{2}}(|1010\rangle + |0101\rangle) + \tfrac{\beta}{\sqrt{2}}(|1001\rangle + |0110\rangle)\right).
\end{align}
But this last state is precisely $(\alpha|\bar{0}\rangle + \beta|\bar{1}\rangle)_{Z_1Z_2=-1}$, and so pushing using the gauge operator $X_1X_3$ preserves our logical information while changing the gauge in which it is represented. This is not surprising as the action of $X_1X_3$ on the eigenstates of $Z_1Z_2$ is to interchange the $+1$-eigenstates and $-1$-eigenstates. If the code is in the $X_1X_3 = \pm 1$ gauge, then the gauge is also preserved. Interestingly, pushing using $X_1X_3$ in the $X_1X_3 = -1$ gauge introduces a global phase (of $-1$) on the logical subsystem.

\begin{figure}[ht]
    \centering
    \includegraphics[width=0.7\textwidth]{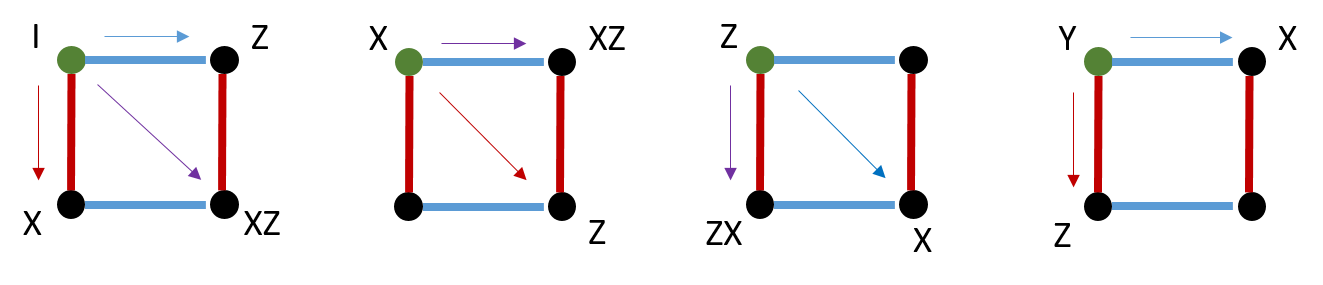}
    \caption{Any Pauli operator (including the identity) can be pushed across the tensor by switching the original state $|\psi\rangle$ to $|\psi'\rangle$.}
    \label{fig:general_gauge_pushing}
\end{figure}

Any single site operator $O$ can be pushed to the remaining 3 sites while moving the state $|\psi\rangle$ to a gauge equivalent representation $|\psi'\rangle$ using the $Y$ type gauge operator $Y_G$. However, as we have just seen, some gauge operators involve a change of gauge and others do not, with potentially with a global phase change. To be concrete let us illustrate this using a change of gauge $Z_1Z_2 = +1$ to $Z_1Z_2 = -1$ with the addition of a global $-1$-phase change. That is
\begin{equation}
    |\bar{\psi}\rangle_{Z_1Z_2 = +1} = |\psi\rangle \mapsto \bar{Y}_G|\psi\rangle = |\psi'\rangle = -|\bar{\psi}\rangle_{Z_1Z_2 = -1}.
\end{equation}
As we have seen above we can implement the change of gauge with $X_1X_3$, and we can add the global phase change with an additional $Z_1Z_2$. Thus we can implement this change of gauge with the operator $\bar{Y}_G = (Z_1Z_2)(X_1X_3) = -iY_1Z_2X_3$. We can derive the rules for push Pauli operators from, say, $A = \{1\}$ by correcting this representation using stabilizers: 
\begin{align}
    I_1 &\mapsto -i X_2Z_3Y_4\\
    X_1 &\mapsto -i Y_3Z_4\\
    Y_1 &\mapsto -i Z_2X_3\\
    Z_1 &\mapsto -i Y_2X_4
\end{align}
as illustrated in Figure \ref{fig:general_gauge_pushing}. 

Therefore, the overall operator $O_A$ can be pushed across term by term as before. We identify a gauge transformation $G$ that implements the change of change $|\psi\rangle \stackrel{G}{\mapsto} |\psi'\rangle$. Then for each Pauli term $P_A$ in the expression for $O_A$ we use the pushing rules to rewrite $G = P_A\otimes Q_{A^c}$, giving
\begin{equation}
    O_A |\psi\rangle = \sum_{i=0}^3 c_i Q_i|\psi'\rangle
\end{equation}
where each $Q_i$ is now supported on $A^c$.

Both the state-invariant and the state-covariant pushing operations preserve the entanglement structure of the state because the gauge operators used for pushing are transversal.

Lastly, we consider the subspace-preserving pushing, which cleans a subsystem while allowing the state to change, as long as it remains in the code subspace. That is,

\begin{equation}
        O|\psi\rangle = I_A\otimes Q''_{\bar{A}}|\psi''\rangle
\end{equation}
where $|\psi\rangle, |\psi''\rangle\in \mathcal{H}_{\rm code}$. In this scenario, operator pushing can be achieved using stabilizer, gauge group elements, as well as logical operators. 

If we only require to preserve the logical subspace, but not necessarily the encoded logical information, then we may allow ourselves logical operators in addition to the gauge operators for pushing. This will be useful in Appendix~\ref{app:codeprop} when we discuss the code distance of the central bulk qubit. It allows us to push any weight one Pauli to any two sites. For example, we apply an additional $\bar{X}$ or $\bar{Z}$ operator to move the support across the diagonal in Figure \ref{fig:gaugepushing} to qubit 4. Because of the additional logical operators one can use for operator pushing, the push rule for this type inherits all of the previous pushing rules. However, only weight 1 Pauli operators can be pushed to  two arbitrary sites, whereas generic single qubit operator $O$ cannot as when it is decomposed as Pauli components, different logical operators may need to be applied to push different components, thereby leading to a different state in each term of the superposition.

\subsection{Multi-copies and Code Concatenation}
\label{subsec:multic}

Equipped with the knowledge of the previous section, we know that there are aspects of the Bacon-Shor code that is lacking as a toy model for holography. One of which is the interpretation of entanglement entropies across different cuts. For example, although the $ZZ=+1$ gauge gives the same entanglement for all bipartitions, the interpretation of that entropy can be different for the same logical state. In particular, bipartition along columns and along rows lead to entanglement entropies that are attributed to the area term or the bulk entropy. One way of mitigating this problem is to concatenate multiple copies of the code where each copy has its qubits shuffled from another.
A simple choice is to concatenate two rotated copies of the Bacon-Shor code. 

\begin{figure}[h]
    \centering
    \includegraphics[width=0.7\textwidth]{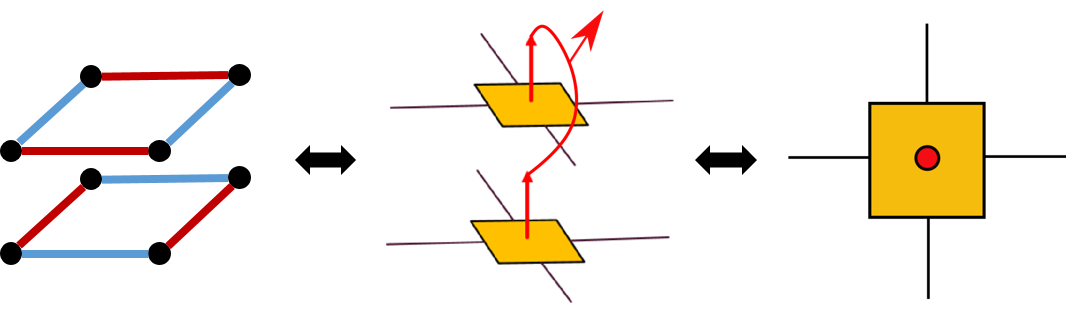}
    \caption{One can construct a double-copy Bacon-Shor Tensor (BST) by stacking two 4-qubit Bacon-Shor codes. The top is rotated by 90 degrees relative to the bottom(left). They are then concatenated by a repetition code. In the tensor network, this is equivalent to contracting their bulk outputs with another 2-legged tensor representing the repetition code (center). Finally, one can still represent the concatenated tensor as a single 4-legged tensor (right) with one bulk output, just like the single-copy BST.}
    \label{fig:doublecopy}
\end{figure}

For example, let us concatenate the two logical qubits of the Bacon-Shor codes with a repetition code, such that
\begin{align}
    &|\bar{0}\bar{0}\rangle \rightarrow |\tilde{0}\rangle\\
    &|\bar{1}\bar{1}\rangle \rightarrow |\tilde{1}\rangle,
\end{align}
with logical operators $\tilde{Z}=\bar{I}\bar{Z},\bar{Z}\bar{I}$ and $\tilde{X} = \bar{X}\bar{X}$. Note that for any element in the original stabilizer group, $S_i, S_j\in \mathcal{S}$, the tensor product $S_i\otimes S_j$ of the Bacon-Shor stabilizers remain a stabilizer for this code. There is an additional $\bar{Z}\bar{Z}$ stabilizer given by the concatenation. We can similarly define the gauge operators as they preserve the logical information of each copy. 
For the double copy code, the gauge group is $\mathcal{G}^{(2)}=\{g_i\otimes g_j, g_i\bar{Z}\otimes g_j\bar{Z}, \forall g_i, g_j \in \mathcal{G}\}$. It includes the tensor product of the original gauge group elements and the gauge equivalent forms of the $\bar{Z}\bar{Z}$ operator.

In this construction, suppose we represent each Bacon-Shor code as a 4-legged Bacon-Shor tensor (BST),  then the concatenated code is by joining the bulk legs with another tensor with two input legs and one output (Figure \ref{fig:doublecopy}). This second tensor represents the repetition code. Overall, we will suppress such details in the double or multi-copy tensor construction, and represent both the single and double copy (or multicopy) tensor as a four-legged tensor, but with different bond dimensions on the legs~(Figure \ref{fig:BStensor}). 

The logical $\tilde{Z}$ is now accessed by having control over the logical $\bar{Z}$ of either copy, which means that having two legs of the tensor accesses the $\tilde{Z}$ code subalgebra. Logical $\tilde{X}$ on the other hand requires access to the logical $\bar{X}$ of both copies, which means one needs at least 3-legs of the tensor to access the $\tilde{X}$ code subalgebra. Consequently, having any 3 legs will also access the full logical Pauli algebra. Therefore, for any 2-2 bipartition by rows and columns, one can only access the Z subalgebra. If we choose the $ZZ=+1$ gauge, then any two legs will access the Z code subalgebra because of the additional $Z$ type stabilizers. The logical computational basis states $|\tilde{0}\rangle, |\tilde{1}\rangle$ now have zero bulk entropy. The entanglement entropy can be entirely attributed to the area term in the FLM formula. In the $ZZ=+1$ gauge, it clearly has twice the amount of the single-copy entanglement. 

While the double copy code is simpler to analyze, an obvious drawback is the differences in entanglement entropies across different 2-2 bipartitions when we choose other gauges. Although this problem is absent in the $ZZ=+1$ gauge, it is not in the $XX=+1$ gauge. In the latter case, because each copy consists of two Bell pairs. The relative rotation ensures that cuts along rows or columns grant 2 units of entanglement while cutting along the diagonal grants 4. This issue can be addressed by stacking more copies of the code, with the physical qubits in different permutations. The model we present here stacks six copies of the Bacon-Shor code, similar to the double-copy construction, where each copy represents a distinct arrangement of the qubits. 

\begin{figure}[h]
    \centering
    \includegraphics[width=0.8\textwidth]{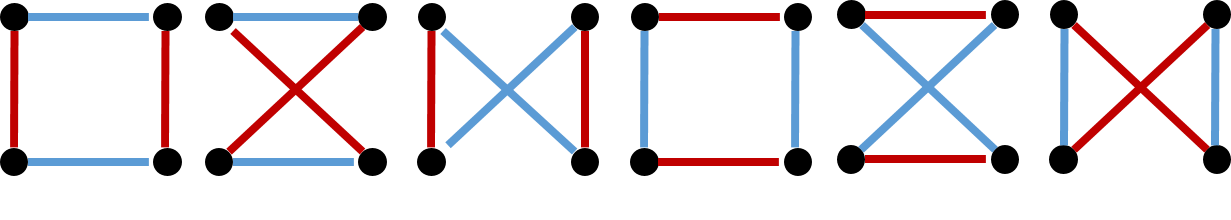}
    \caption{Red lines indicate Z type gauge operators and blue lines are X type gauge operators. The locations of the qubits are shuffled and there are a total of 6 distinct configurations depending the locations of the gauge operators.}
    \label{fig:my_label2}
\end{figure}

With the gauges we consider, the entanglement entropy across any bipartitioning cut is uniform. Then we concatenate these codes using a repetition code 

\begin{equation}
|\tilde{0}\rangle =  |\bar{0}\rangle^{\otimes 6} ,~~ |\tilde{1}\rangle =  |\bar{1}\rangle^{\otimes 6}.
\label{eqn:concatenation}
\end{equation}

With this concatenation, we again form another four-legged tensor with bond dimension $\chi=2^6$. Suppose $A$ is a 3-leg subsystem, then the von Neumann entropy $S(A)=6$ because we are tracing out (6 qubits) for each leg. However, now for any bipartition with a 2-2 split, $S(A)=8$ in the $XX=+1$ gauge, but $S(A)=6$ in the $ZZ=+1$ gauge. In this repetition code, the logical operators are 
\begin{equation}
    \tilde{X} = \bar{X}^{\otimes 6},~~ \tilde{Z} = \bar{Z}_i
\end{equation}
where $\bar{Z}_i$ denotes applying a single $\bar{Z}$ operator on the $i$th copy and $\bar{I}$ for the rest. The index $i$ runs from 1 to 6. The stabilizers are again defined by taking the six-fold tensor product of the Bacon-Shor stabilizer elements as well as the stabilizers from the repetition code, which consists of $\bar{Z}$ operators of even weight. Elements of the gauge group of this code are similarly generated.

Again, in the tensor network representation, the behaviour of this code is not too different from the double-copy code. While having one leg does not access any non-trivial code subalgebra, having any two legs of the tensor as a subsystem accesses the logical $\tilde{Z}$ operator. However,  to perform a logical $\tilde{X}$ operation, one needs to apply $\bar{X}$ on each copy. This again implies that a minimum of 3 legs is necessary to access the $X$ subalgebra as well. Therefore, in this concatenated code, accessing any two legs accesses the Z subalgebra, and accessing 3 legs or above have full control of the code subalgebra.

Regarding connections with the RT-FLM formula, the interpretation of the entropy terms is also similar to the double copy construction. For any 3-1 split, the entirety of the entanglement is coming from the area term, because they are simply 6 tensor copies of the original Bacon-Shor code. 
For 2-2 split, having access to \textit{any} two legs accesses the Z subalgebra for any gauge. This is somewhat different from the double-copy construction, where a similar level of access would require the $ZZ=+1$ gauge.  As such, the logical computational basis $|\tilde{\alpha}\rangle, \alpha = 0,1$ have zero bulk entropy contribution regardless of the bipartition. Therefore, all units of the entanglement entropy are attributed to the area-like RT term.

One can also generalize this construction to having $N$ stacks of the Bacon-Shor code that may be related to one another through spatial or gauge transformations. Then, a multi-copy Bacon-Shor tensor can be produced through the same kind of repetition code concatenation.
They can be thought of as stacked versions of the single-copy Bacon-Shor code, where a number of tensor copies are taken, then a two-dimensional subspace of the tensor copied code subspace is chosen by the repetition code as the final  code space. The tensor network representation of the multi-copy code is precisely the same as that of the single copy~(Figure~\ref{fig:BStensor}), where the in-plane legs are again physical qubits and the bulk leg denotes logical qubit. The difference is that each in-plane leg can contain a number of physical qubits. For instance, in the double-copy code, each in-plane leg represents two qubits stacked on top of each other. In the tensor network language, the bulk leg has bond dimension $\chi_b=2$ whereas each in-plane leg has bond dimension $\chi_p=2^N$, where $N$ is the number of tensor copies.  It is easy to check that the logical operators are supported in a similar way as the six-copy code, where any two legs access the code $\tilde{Z}$ subalgebra while having access to any 3 legs accesses the full logical Pauli algebra. 

In such codes, the bulk entropy contribution is zero when the logical states are initialized in the 0 or 1 state while maintaining a non-trivial center in the code Z-subalgebra. Note that although in a 3-1 split, there is no bulk entropy contribution as long as the logical information is pure,  it is possible to pick up a bulk term in a 2-2 split when we consider a superposition $a |\tilde{0}\rangle+b|\tilde{1}\rangle$. This is because the pure logical state reduces to a density matrix on the $Z$-code subalgebra.

One can derive these entanglement properties explicitly by decoding. For these multi-copy codes created through concatenation, the decoding unitary can be constructed in a simple way. In the 2-2 split, we first identify the copy that has access to the logical $\bar{Z}$ operator, then apply the decoding unitary to that copy and identity to all other copies. We apply the convention of (\ref{eqn:HarlowFLM}) to identify the RT and bulk contributions in the FLM formula. 
When we have access to 3 legs or above, we apply the decoding unitary of the concatenated code, which again allows us to apply the results (\ref{eqn:HarlowFLM}) for different contributions in the FLM formula. The concatenated code can be decoded by first applying the Bacon-Shor decoding unitary $U(\textrm{BSC})$, given in (\ref{equation:decode3-1}), to each copy. Then one applies the repetition code decoding unitary $U(\textrm{R})$ which acts on the decoded information of the Bacon-Shor code.

More concretely, consider $|\tilde{\psi}\rangle = a|\tilde{0}\rangle+b|\tilde{1}\rangle$, which is the encoded state on 4 legs. One applies 
\begin{equation}
    U^{\otimes N}(\textrm{BSC}) (a|\tilde{0}\rangle+b|\tilde{1}\rangle)\rightarrow (a|\bar{0}\rangle^{\otimes N}+b|\bar{1}\rangle^{\otimes N} )\otimes |\chi\rangle,
\end{equation}
followed by
\begin{equation}
    U(\mathrm{R})(a|\bar{0}\rangle^{\otimes N}+b|\bar{1}\rangle^{\otimes N} ) \rightarrow (a|0\rangle +b|1\rangle) |0\rangle^{\otimes (N-1)},
\end{equation}
where $U(R)$, without loss of generality, consists of a sequence of CNOT gates that act on the first and the $j$-th qubit with the first qubit acting as the control. We omit the subsystem on which $|\chi\rangle$ is supported because $U(R)$ acts trivially on it.

\subsubsection{Multi-copy Code Operator Pushing}
\label{subsubsec:multiBSCpush}
\begin{figure}[h]
    \centering
    \includegraphics[width=0.4\textwidth]{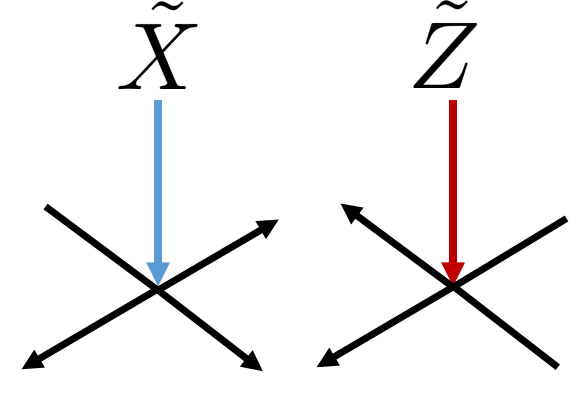}
    \caption{ Logical $\tilde{X}$ pushes to any 3 legs while logical $\tilde{Z}$ pushes to any 2 legs. This does not imply the operator is weight $3N$ or $2N$, because some are identity operators even though they act non-trivially on a tensor leg. For example, in the double-copy, logical $\tilde{X}$ has weight 4 while logical $\tilde{Z}$ is weight 2.}
    \label{fig:multicopy_pushing}
\end{figure}

The pushing rules for multi-copy codes are similar as those for the single-copy.
The logical operator pushing of a multi-copy code is shown in Figure~\ref{fig:multicopy_pushing}. 

The logical operator $\tilde{Z}$ is supported on two legs because such operators take on the form where $\bar{Z}$ acts on one of the copies while $\bar{I}$ act on the rest. It can be pushed to any two legs if a row, column, or diagonal supports $\bar{Z}$ in at least one of the copies. Logical $\tilde{X}$, on the other hand, is given by the string of all $\bar{X}$. Because having access to 3 qubits on each copy must access a row, a column, and a diagonal, it guarantees access to the $\bar{X}$ operator of that copy. Therefore, in the stacked multi-copy code, the logical $\tilde{X}=\bar{X}^{\otimes N}$ operator can be supported on any 3 legs. Trivially, because $\tilde{Z}$ is also supported on any 3 legs, any logical operator can be pushed to 3 legs.

Because we also inherit all the gauge operators from the individual Bacon-Shor codes by taking their tensor products, the pushing rules of the single-copy tensor similarly apply here: any operator $O$ acting on any single leg can be pushed to an operator that acts on the other 3 legs using stabilizer or gauge elements while preserving the encoded information. If $O$ is a Pauli operator, then the pushed operator is also a Pauli operator that may be supported on fewer than 3 legs of the tensor if we use gauge operators for pushing. However, the specifics will depend on the operator being pushed, as well as the relative spatial arrangements of each copy.

Lastly, but non-trivially, we should understand whether a single site operator can be pushed to (at most) any two sites in the multi-copy code while preserving only the logical subspace but allowing the logical information to change. Again, this implies one can use both gauge and logical operators to ``clean'' the physical  qubits on which operators are applied. This does not follow trivially from the single copy pushing, because whatever logical operators applied to each copy do not necessarily constitute a logical operator of the concatenated code. For example, the logical operator of the repetition code $\tilde{X}$ is a tensor product of $\bar{X}$ on each copy, whereas $\bar{X}\bar{I}\dots$ is an error operator. Therefore, logical $\bar{X}$ has to be applied all other copies if even one copy uses $\bar{X}$ for pushing. We show in Appendix~\ref{app:dcpushing} that for arbitrary inputs in a double-copy code the single site Pauli operator can be pushed to any two sites while the state remains in the logical subspace. However, this property does not hold for more general multi-copy constructions. 

All of the above representations are independent of the gauge choice. More specialized pushing rules may hold for a specific gauge, but not in general. For example, in the $ZZ=+1$ gauge, which we will use extensively in this work, $Z_1Z_3, Z_2Z_4$ which act on the rows are also stabilizer operators in addition to the original weight 4 stabilizer elements. As a result, logical $\bar{Z}$ operators in this gauge are also manifestly supported on the diagonal. In the tensor network language, this means for a double-copy code, $\tilde{Z}$ can be pushed to any two in-plane legs while leaving the state invariant. Note that it is possible to push a logical $\tilde{Z}$ operator to the diagonal without fixing the $ZZ=+1$ gauge. However, there the state is typically covariant under the pushing, which preserves the logical information, but not the state itself.

\section{Approximate Bacon-Shor Code}
\label{sec:4aqec}

Quantum error correction codes can be defined as an encoding channel $\mathcal{N}: L(\mathcal{H}_A)\rightarrow L(\mathcal{H}_B)$ that admits a decoding map $\mathcal{D}:L(\mathcal{H}_B)\rightarrow L(\mathcal{H}_A)$ where

\begin{equation}
    \mathcal{D}\circ \mathcal{N} = Id.
\end{equation}

The channel $\mathcal{N}$ is related to the encoding isometry $W:\mathcal{H}_A\rightarrow \mathcal{H}_B\otimes \mathcal{H}_E$ via Stinespring dilation, where

\begin{equation}
    \mathcal{N}(\rho) = \Tr_E W\rho W^{\dagger}.
\end{equation}

In the Bra-ket notation, we can write 
\begin{equation}
    W= \sum_i |\psi_i\rangle\langle \bar{i}|,
\end{equation}
where $\langle\psi_i|\psi_j\rangle=\delta_{ij}$ and $\{|\bar{i}\rangle\}$ is an orthonormal basis that spans the logical subspace. In our case, we can think of the combined $BE$ system as the physical qubits on which information is unitarily encoded, and $E$ as the erasures. Some erasure errors can be corrected because the $E$ contains no access to the encoded information. $\mathcal{D}$ then decodes that information from the remainder of unaffected qubits.

Here we consider codes where reconstruction from subsystem $B$ is only approximate. Namely

\begin{equation}
    ||\mathcal{D}\circ\mathcal{N} - Id||< \epsilon
    \label{eqn:aqecchannel}
\end{equation}
and some amount of information leaked into $E$.

In this section, we construct approximate Bacon-Shor codes by perturbing the exact 4-qubit codes we discussed in Sec~\ref{subsec:bs4}. For these codes, the decoding is imperfect given some proper subsystem $B$, unlike the scenario we studied in Section~\ref{sec:3baconshor}. However, we can construct a simple decoding map that recovers the information up to small errors. We will discuss how this approximate code, which we call the ``skewed code'' can be built from a superposition of similar stabilizer codes. We will then discuss the entanglement and operator pushing properties, followed by decoding. Finally, we comment on the construction and properties of its multi-copy generalizations.

\subsection{The Skewed Code and the Superposition of Codes}
\label{subsec:skewedBSC}

Our ansatz is the perturbation from the original code by skewing the vectors that define the code words. In general, an error correcting code is just a subspace $\mathcal{C}\subset\mathcal{H}$ in which the logical information is encoded. As such, it need not have good error correction properties or efficient encoding/decoding. We consider one particular construction which we call the skewed Bacon-Shor Code, but generalizations to other codes are obvious. For the sake of simplicity, we do not define these skewed code as a gauge code. In this case, logical states are mapped to physical states, instead of equivalent classes of physical states.

Suppose $\mathcal{C}_0$ is the code subspace of a (gauge fixed) Bacon-Shor code as defined previously. Then the skewed code has subspace $\mathcal{C}$ which, as a Hilbert space, may overlap with $\mathcal{C}_0$ but does not coincide with it. We treat the code space as being defined by an encoding map $V:\mathcal{H}_{\rm logical}\rightarrow\mathcal{H}$, where $\mathcal{H}_{\rm logical}\simeq \mathcal{C}$. Concretely,
\begin{equation}\label{eqn:SBSCmap}
    V=|\psi_0\rangle\langle\bar{0}|+|\psi_1\rangle\langle\bar{1}|,      
\end{equation}
where $\mathcal{C}=\mathrm{span}\{|\psi_i\rangle,~ i= 0,1\}$. We do not require that the states $|\psi_i\rangle$ be mutually orthogonal, therefore this mapping need not be an isometry. However, we do require that they are linearly independent, thus forming a basis for the two-dimensional vector space.
  
The skewed code itself may not have nice error detection properties or decoding. However, when the skewing is small and the two code spaces largely coincide, then we may treat the original code with subspace $\mathcal{C}_0$, which we will call the \textit{reference code}, as an error correction code that approximates the skewed code $\mathcal{C}$. Conversely, the skewed code is an approximate Bacon-Shor code (with possible gauge fixing). In particular, while the skewed code may not allow exact error detection/correction or decoding from a subsystem, it is clear how to perform such operations approximately. Indeed, it inherits all properties of the reference code, such as decoding and logical operations, but at the cost of a small error. 

In the language of quantum channels we used above, the encoding channel $\mathcal{N}$ is an operation
\begin{equation}
    \rho \rightarrow V \rho V^{\dagger}
\end{equation}
followed by an erasure channel 
\begin{equation}
    \mathcal{E}(\rho) = \Tr_E V\rho V^{\dagger},
\end{equation}
where $V$ is a mapping from $\mathcal{H}_{\rm logical} =\mathcal{H}_A$ to $\mathcal{H}=\mathcal{H}_B\otimes\mathcal{H}_E$.

We are then to decode the logical information by acting only on $B$. For example, one can think of $B$ as any three of the physical qubits while $E$ as the remaining one physical qubit of the Bacon-Shor code. However, the one qubit erasure error can only be corrected approximately.
  
A convenient representation of the skewed code is to expand it as a superposition of stabilizer codes. 
Indeed, for generic codes, there is no requirement for the encoding map to be Bacon-Shor or even isometric. 
One way to construct (\ref{eqn:SBSCmap}) is to deform a reference code for which we understand the code properties well. When the deformations are small, we can say that the reference code approximates the new code $V$ we construct.
 
Consider the class of skewed codes that can be constructed as a superposition of stabilizer codes, 
 \begin{equation}
     V= \sum_i a_i W_i
 \end{equation}
 where $W_i$ is an isometric stabilizer encoding map and $a_{i} \in \mathbb{C}$. 
 Note that the superposition of isometric encoding maps need not be isometric. In the case where we do not restrict the values of $a_i$, we expect such a decomposition to be possible for any encoding map, because the stabilizer states form an overcomplete basis of the Hilbert space. For our work, when $i \not= 0$ we treat $W_i$ as noise or contamination of the reference code $W_0$ regardless of the amplitudes $a_i$. 
 
 However, for codes where the noise terms are sufficiently small, we are guaranteed a decoding channel given by the decoding unitary of the reference code. Suppose the reference code can be decoded by $U_D$ which has support subsystem $B$, then for any $\sigma\in L(\mathcal{H}_{\rm code})$, a decoding channel for the skewed code is nothing but
 
 \begin{equation}
     \mathcal{D}(\sigma) = U_D (\Tr_{A^c}\sigma) U_D^{\dagger}.
 \end{equation}

Then as long as $a_i$ (for $i \not= 0$) sufficiently small, this reference code decoding channel satisfies (\ref{eqn:aqecchannel}) for any fixed $\epsilon$.

Let us construct some concrete examples for the 4-qubit Bacon-Shor code. Consider an encoding map is given by
\begin{equation}
    V= \frac{1}{\sqrt{2}}(|0000\rangle+|1111\rangle)\langle \bar{0}| +(\alpha|1100\rangle+\beta|0011\rangle)\langle\bar{1}|
        \label{eqn:ghzcode}
\end{equation}
and $\alpha\neq \beta$. This is not the Bacon-Shor code in any gauge: while $|\phi_0\rangle$ and $|\phi_1\rangle$ are stabilized by $Z_1Z_2Z_3Z_4$, the later is not stabilized by $X_1X_2X_3X_4$. Indeed, the code subspace $\mathcal{C}$ is ``skewed'' and does not coincide with the normalizer of the gauge group $\mathcal{G}$. While the decomposition into a superposition of codes is not unique\footnote{While the non-uniqueness would trivially follow from the over-completeness of stabilizer states, here it is possible to construct explicit examples. For instance, it admits superposition into trivial repetition codes where each codeword is a product state of 0 and 1s. }, this encoding map can be written as a superposition of 3 different Bacon-Shor-like stabilizer codes 

\begin{equation}
    V=a_0 W_0+a_1 W_1 +a_2 W_2, ~~a_0=\frac{\alpha+\beta}{\sqrt{2}}, a_1=\frac{1-\sqrt{2}\beta}{2}, a_2= \frac{1-\sqrt{2}\alpha}{2}
\end{equation}
such that for  $|\alpha-\beta|<\epsilon, |a_1|,|a_2|<\sqrt{\epsilon}$.

$W_0$ is our standard encoding unitary for the Bacon-Shor code in the $ZZ=+1$ gauge
\begin{equation}
    W_0 = \frac{1}{\sqrt{2}}(|0000\rangle+|1111\rangle)\langle \bar{0}|+\frac{1}{\sqrt{2}}(|1100\rangle+|0011\rangle)\langle \bar{1}|,
\end{equation}
which has stabilizer generators $Z_1Z_2, Z_3Z_4, X_1X_2X_3X_4$. The logical operators are generated by $\bar{Z}_L=Z_1Z_3, Z_2Z_4$ and $\bar{X}_L=X_1X_2, X_3X_4$.

Similarly, $W_1$ is also a stabilizer code with $ZZ$ generators on the columns of the square, i.e. $Z_1Z_2, Z_3Z_4$, and $-X_1Y_2Y_3X_4$. 

\begin{equation}
        W_1 = \frac{1}{\sqrt{2}}(|0000\rangle+|1111\rangle)\langle \bar{0}|+\frac{1}{\sqrt{2}}(|1100\rangle-|0011\rangle)\langle \bar{1}|
\end{equation}

The logicals are $\bar{Z}_L=Z_1Z_3$ and $\bar{X}_L=X_1X_2Z_3$ or equivalent representations by multiplying stabilizers. In fact, we can unfix the gauge and turn it into a gauge code.

Let the following be basis states of the logical tensor the gauge qubit. 
\begin{align}
    |0000\rangle+|1111\rangle \rightarrow |00\rangle_{LG}\\
    |1100\rangle-|0011\rangle\rightarrow|10\rangle_{LG}\\
    |0101\rangle-|1010\rangle \rightarrow |01\rangle_{LG}\\
    |1001\rangle+|0110\rangle\rightarrow |11\rangle_{LG}
\end{align}
We have gauge generators $\bar{Z}_G = Z_1Z_2, Z_3Z_4$, $\bar{X}_G=X_1Z_2X_3I_4, I_1X_2Z_3X_4$ and logical operators $\bar{Z}_L=Z_1I_2Z_3I_4, I_1Z_2I_3Z_4$, $\bar{X}_L=X_1X_2Z_3I_4, -I_1Z_2X_3X_4$.
The stabilizer code corresponds to fixing the $Z_1Z_2, Z_3Z_4=1$ gauge.

$W_2$ is a slightly modified version of $W_1$, except with an extra minus sign for $\bar{X}_L$ and in the relevant codewords. 

\begin{equation}
        W_2 = \frac{1}{\sqrt{2}}(|0000\rangle+|1111\rangle)\langle \bar{0}|+\frac{1}{\sqrt{2}}(-|1100\rangle+|0011\rangle)\langle \bar{1}|
\end{equation}

All stabilizers and $\bar{Z}$ are the same, because its logical 0 is a +1 eigenstate of these operators. $\bar{X}=-XXZI$ or equivalent representations.
We may consider $W_1,W_2$ code as ``noise'' terms, especially if $|\alpha-\beta|\ll 1$ such that $V$ is an approximate version of $W_0$. 

One can check that this skewed code can not correct one erasure error perfectly. Let $B$ be a 3-qubit subsystem and $E$ be the remaining qubit that is erased. It is found by \cite{SchuNiel96} that perfect error correction is possible on $B$ if and only if 

\begin{equation}
    I(R:E) = S(R)+S(E)-S(B) =0
\end{equation}
in the state 

\begin{equation}
    |\phi \rangle = |0\rangle|\bar{0}\rangle+|1\rangle|\bar{1}\rangle_{RBE}.
\end{equation}
$R$ is a reference qubit that is maximally entangled with the $BE$ system. For generic $\alpha, \beta$, one can show that $I(R:E)>0$; hence a part of the encoded information is lost to $E$, impeding perfect recovery on $B$.

For another example, consider perturbations when the code is in the $XX=+1$ gauge. Let
 \begin{align}
      |\Phi^{\pm}_x\rangle &= |++\rangle\pm|--\rangle\\
      |\Psi^{\pm}_x\rangle &= |+-\rangle \pm |-+\rangle.
 \end{align}
 
 We can define the following perturbed codewords 
 
 \begin{align}
     |\Phi^+_x\rangle^{\otimes 2}&\rightarrow |\bar{0}\rangle\\
     (\alpha|\Phi^-_x\rangle+\beta|\Psi^-_x\rangle)^{\otimes 2}&\rightarrow |\bar{1}\rangle
 \end{align}
 where again the two states have different entanglement spectra. The overall encoding unitary can be decomposed into the sum of isometric encoders,
 \begin{equation}
     V=\sum_{i=0}^4 a_i W_i
 \end{equation}
 where each branch covers a $W_{i\ne 0}$ single error subspace.

The isometries can be defined as 

\begin{align}
    W_0 &= |\Phi^{+}_x\rangle^{\otimes 2}\langle \bar{0}|+|\Phi^-_x\rangle^{\otimes 2}\langle\bar{1}|\\
    W_1 &= |\Phi^{+}_x\rangle^{\otimes 2}\langle \bar{0}|+|\Phi^-_x\rangle|\Psi^-_x\rangle^\langle\bar{1}|\\
    W_2 &= |\Phi^{+}_x\rangle^{\otimes 2}\langle \bar{0}|+|\Psi^-_x\rangle|\Phi^-_x\rangle\langle\bar{1}|\\
    W_3 &= |\Phi^{+}_x\rangle^{\otimes 2}\langle \bar{0}|+|\Psi^-_x\rangle^{\otimes 2}\langle\bar{1}|\\
    W_4 &= |\Phi^{+}_x\rangle^{\otimes 2}\langle \bar{0}|-|\Psi^-_x\rangle^{\otimes 2}\langle\bar{1}|,   
\end{align}
where
\begin{equation}
    a_0=\alpha^2,  a_1= a_2=\alpha\beta, a_3=\frac{1-\alpha^2-2\alpha\beta+\beta^2}{2}, a_4=\frac{1-\alpha^2-\beta^2-2\alpha\beta}{2}
\end{equation}

Although in the above examples, we have chosen the perturbations such that they only deform one of the basis states from that of the reference code, this need not be the case. In fact, the encoding does not even need to be isometric.

 Consider a superposition of encoding isometries 
 
 \begin{align}
     W_0 = |\Phi^+_x\rangle^{\otimes 2}\langle \bar{0}|+|\Phi^-_x\rangle^{\otimes 2}\langle\bar{1}|\\
      W_1 = |\Phi^-_x\rangle^{\otimes 2}\langle \bar{0}|+|\Phi^+_x\rangle^{\otimes 2}\langle\bar{1}|,
 \end{align}
where we can take $W_0$ to be the reference code, and $W_1$, which has an extra minus sign on its logical $\bar{Z}$ operator, to be the perturbation. 

The encoding map $V=a_0W_0+a_1W_1$ yields

\begin{align}
    |\bar{0}\rangle &= \frac{a_0+a_1}{2}(|++++\rangle+|----\rangle) +\frac{a_0-a_1}{2}(|--++\rangle+|++--\rangle)\\
    |\bar{1}\rangle &= 
    \frac{a_0+a_1}{2}(|++++\rangle+|----\rangle) +\frac{a_1-a_0}{2}(|--++\rangle+|++--\rangle),
    \label{eqn:bellaqec}
\end{align}
 
 which is non-isometric because the inner product of the states
 \begin{equation}
    \langle\bar{0}|\bar{1}\rangle= |a_0+a_1|^2-|a_0-a_1|^2\neq 0.
 \end{equation}

 There are many more interesting examples one can construct. We will consider a few other examples of approximate code in Section~\ref{subsec:powerlaw} where these perturbations are also responsible for the power-law correlations. As we can choose our code space and codeword arbitrarily, the kind of perturbation one could consider is far more general than the examples we listed here. Much work is needed to characterize the type of perturbations different codes can have and their corresponding properties. Although we perform no such comprehensive characterization here, we wish to highlight the fact that there is a large class of approximate quantum codes whose specific properties may differ even though the general reference code decoding and syndrome measurement can work up to $\epsilon$ error. Understanding the specificity of different classes of approximate code may enable us to find better hardware-specific error correction codes and decoding procedures where certain errors are more likely to occur than others.

\subsection{Entanglement Properties}

For the following analysis, again consider the code defined in (\ref{eqn:ghzcode}). Let $A$ be the top row, which still has access to the $M_A=\{\tilde{I}, \tilde{Z}\}$ subalgebra over the code subspace. 
Recall that the codewords
\begin{align}
\begin{split}
    |\bar{0}\rangle &= \frac{1}{\sqrt{2}}(|0000\rangle+|1111\rangle)\\
    |\bar{1}\rangle &= \alpha|1100\rangle+\beta|0011\rangle
    \end{split}
\end{align}
 are $\pm 1$ eigenstates of the logical operator $\bar{Z} = Z_1Z_3, Z_2Z_4$.  The encoding is still isometric, as $\langle \phi_0|0\rangle_C=\langle\phi_0|1\rangle_C=0$, $\langle\bar{1}|\bar{0}\rangle=0$. However, it is clear that their entanglement entropies across the cut are different for these states. As a result, we conclude that $\bar{X}$ the actual logical operator for the above code must act globally (on all 4 qubits), otherwise the entanglement cannot be changed across all bipartitions.

The reduced density matrix again admits a block diagonal structure, where 
\begin{equation}
    \Tr_{{A}^c}[|\bar{0}\rangle\langle \bar{0}|] = 
    \begin{pmatrix}
    \frac 1 2 & 0 \\
    0 & \frac 1 2
    \end{pmatrix}
    \oplus \mathbf{0}
\end{equation}

and $\mathbf{0}$ is a 2 by 2 zero matrix. 
Similarly, 

\begin{equation}
    \Tr_{{A}^c}[|\bar{1}\rangle\langle \bar{1}|] = \mathbf{0} \oplus 
    \begin{pmatrix}
    |\alpha|^2 & 0 \\
    0 & |\beta|^2
    \end{pmatrix}
\end{equation}

It is simple to check that the original decoding unitary for 2-2 decomposition (Figure~\ref{figure:decodingZZ}) is still valid such that 
\begin{equation}
    U_A\Tr_{{A^c}}[|\bar{\alpha}\rangle\langle\bar{\alpha}|] U^{\dagger}_A \rightarrow |\alpha\rangle\langle\alpha|\otimes \chi_{\alpha},
\end{equation}
where $\alpha = 0,1$.

Again, the bulk entropy $S(\bar{\rho},M_A)=0$ for $|\bar{0}\rangle, |\bar{1}\rangle$. The only pieces that contribute to the total entropy are RT terms $S(\chi_{\alpha})$ where

\begin{equation}
\chi_0=
\begin{pmatrix}
\frac 1 2 & 0 \\
0 &\frac 1 2
\end{pmatrix}
\label{eqn:chi0}
\end{equation}
and 
\begin{equation}
\chi_1=
\begin{pmatrix}
|\alpha|^2 & 0 \\
0 & |\beta|^2
\end{pmatrix}
\label{eqn:chi1}
\end{equation}

in the computational basis. 

The RT contribution to entropy for a general state is now state-dependent. For a state

\begin{equation}
    |\bar{\psi}\rangle = a|\bar{0}\rangle + b|\bar{1}\rangle,
\end{equation}
the total entanglement entropy is now is then given by 

\begin{equation}
    S(\bar{\rho}_A)= H(p_{\alpha})+\sum_{\alpha}p_{\alpha} S(\chi_{\alpha})
    \label{eqn:entropyRT}
\end{equation}
where $p_0=|a|^2, p_1=|b|^2, S(\chi_{0})=-|\alpha|^2\log_2 |\alpha|^2-|\beta|^2\log_2|\beta|^2$, and $S(\chi_1)=1$. The Shannon-like $H(p_{\alpha})= -p_0\log p_0 -p_1\log p_1$ is the entropy of mixing, which is no longer vanishing.  In this model, changing the bulk encoded information will therefore also change the geometric piece ($\mathrm{Area}/4G$)=$\sum p_{\alpha}S(\chi_{\alpha})$, akin to back-reaction. In this gauge, the diagonal entanglement computation is the same as rows, because $Z_1Z_2,Z_3Z_4$ remain stabilizer elements, just like the reference code. 

However, bipartition by columns is much more complicated, because of each column only accesses the X subalgebra approximately. Here we will not discuss the column bipartition because have yet to construct a simple, explicit decoding unitary which extracts the partial logical information. It is also unclear if such a decoding unitary exists. Without ensuring the existence of such a unitary, we cannot apply the results of \cite{Harlow:2016vwg}. However, this problem is absent in the multi-copy code, where for such a stacked code, each column still only accesses the $\tilde{Z}$-subalgebra, therefore the breakdown of the entanglement entropy can be unambiguously computed. We will comment on this in \ref{subsec:multics}.

For 3-1 bipartitions, no subsystem accesses the entire code subspace exactly and complementary recovery breaks down as a result. We cannot simply use \ref{eqn:HarlowFLM} to compute the bulk and RT contribution because we do not have the exact decoding of the entire logical information in either subsystem. Even though approximate recovery is valid, we do not know how the results of \cite{Harlow:2016vwg} can generalize to these scenarios. However, it is possible to apply such a method to certain states where exact decoding \textit{is} possible. This yields an FLM formula that holds for these states or subspaces. We will discuss this form of state/subspace-dependent decoding and the RT formula in more detail in section~\ref{subsec:SEERT}.

We note that the 3 qubit subsystem does inherit the access to the code Z subalgebra. Therefore, for $|\bar{\alpha}\rangle, \alpha=0,1$, and their mixture, there is a decoding unitary that extracts the logical information exactly and leaves behind an entangled state across the bipartition. For these specific states, we can still define the RT and bulk entropy contributions. In this case, the entropy contribution  and interpretation for the 3 qubit subsystem are exactly the same as the row bipartition we have discussed. For pure states, the entropy is equal to the state-dependent RT contributions $S(\chi_{\alpha})$, where $\chi_{\alpha}$ are defined in (\ref{eqn:chi0},\ref{eqn:chi1}). It can also pick up a bulk entropy term if the encoded state is mixed 
\begin{equation}
    \tilde{\rho}= p |\bar{0}\rangle\langle\bar{0}|+(1-p)|\bar{1}\rangle\langle\bar{1}|.
\end{equation}
Then the overall von Neumann entropy is identical to (\ref{eqn:entropyRT}) for $p_0=p, p_1=1-p$. 
For the one qubit subsystem, it also retains a state-dependent RT-like contribution $S(\chi_{\alpha})$, where $\chi_{\alpha}$ are given by (\ref{eqn:chi0},\ref{eqn:chi1}). However, there is zero bulk entropy contribution for both pure and mixed states.

Because we rely on \cite{Harlow:2016vwg} to properly assign meaning to the entropy terms, when we discuss the connection to the RT formula and bulk entropy in the skewed code, we will focus primarily on analyzing the states or subspaces where exact recovery of information from a fixed subsystem is possible in later parts of this work. In the code construction (\ref{eqn:ghzcode}), these are the states $|\bar{\alpha}\rangle$ and their mixtures. However, we acknowledge the importance to explore generalizations of \cite{Harlow:2016vwg} to approximate codes and how the break down of terms can be implemented in an AQECC where it also properly identify the $1/N$ corrections. We will leave the entropic interpretations of more general states for future work. 

In summary, we find that in designing a Bacon-Shor code with perturbations, thus rendering it an AQEC, it is possible to construct examples where the RT entropy contribution depends on the logical state, which is analogous to back-reaction.

\subsection{Operator Pushing}
\label{subsec:skewedBSCpush}
Recall that a skewed code with encoding map $V$ can be decomposed into a superposition of stabilizer code encoding isometries $W_i$

\begin{equation}
    V=\sum_i a_i W_i.
\end{equation}

Then, any operator pushing can be performed on the stabilizer branches separately. Formally, we can write
 \begin{equation}
     V\bar{O} = \sum_i a_i O_i W_i = \sum_j O'_j \Pi_j V,
     \label{eqn:genoptpushskew}
 \end{equation}
 where $\Pi_j$ are a set of mutually orthogonal projectors onto the physical Hilbert space, $O'_j \Pi_j= \sum_i O_i \Pi_j W_iW_i^{\dagger}\Pi_j^{\dagger}.$
Because of the projectors, the physical operators can be global operators, instead of having support only over a subregion. Note that the operator pushing for each term in the sum (\ref{eqn:genoptpushskew}) can be non-unique, due to equivalent representations related by stabilizers. Furthermore, there is added non-uniqueness in how $V$ is decomposed. This is because different codes would have different stabilizers, and hence different representations of logical operators/pushing rules. Hence the overall representation for a single bulk/logical operator is also non-unique for the skewed code.

The operator can also be pushed through ``approximately'' at the expense of altering the state slightly. That is, we can factor out the physical operator on the right-hand side but at the same time altering the encoding map by a small amount. In this case, the pushed operator is that of the reference code.

\begin{equation}
    V\bar{O} = a_0 O_0W_0+\sum_{i\ne 0} a_i O_0 W'_i = O_0 V'
\end{equation}

We will refer to this as the reference code pushing, where the resulting operator on the boundary is exactly the same as that of the reference code. However, the logical information of the actual encoding map is not preserved throughout. In other words, given an approximate code defined by $V$, we can choose to act on it as if it were encoded by $W_0$, the reference encoding map, such that the reference code logical operators will not preserve the code subspace $\mathcal{C}$ exactly, but will ``leak'' outside the subspace with some error $\epsilon\sim \sum_{i\ne 0} |a_i|$. In doing so, reference logical operators can take the code words as defined in $V$ to those in $V'$, which spans $\mathcal{C}'$. While $\mathcal{C}$ and $\mathcal{C}'$ have substantial overlaps, they are not identical.

\begin{figure}
    \centering
    \includegraphics[width=0.4\textwidth]{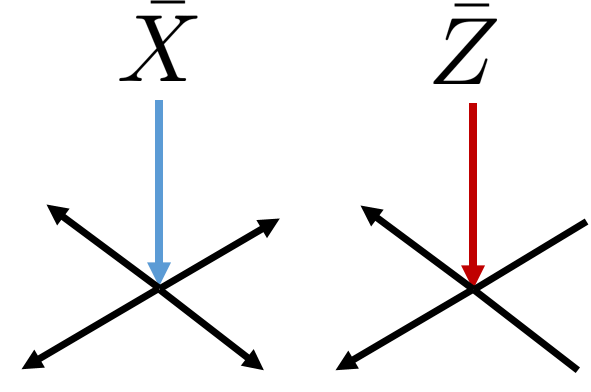}
    \caption{A logical operator in the construction (\ref{eqn:ghzcode}) admits pushing rules given by the above tensor network representation. The rule also applies for multi-copy skewed codes, where a number of copies were replaced by this type of skewed code. In the multi-copy case, a logical $Z$ operator can be pushed to \textit{any} 2 legs. }
    \label{fig:AQECBStensor}
\end{figure}

For the sake of concreteness, let us again focus on the (skewed) 4-qubit Bacon-Shor code examples we have discussed and see how operator pushing is modified for logical operators. Recall that for the skewed code we introduced in (\ref{eqn:ghzcode}), a single bulk logical $\bar{X}$ operator is pushed to all 4 legs exactly, although it can be pushed to 2 legs approximately, but up to error controlled by the noise terms. We can see this quite simply by looking at the entanglement of the codewords. Because the logical 0 and 1 states have different entanglement across all bipartitions for $\alpha\ne\beta$, the $\bar{X}$ operator for this code must have support over all 4 physical qubits. On the other hand, the logical $\bar{Z}=Z_1Z_3,Z_2Z_4$ operator remains unchanged in this construction because the codewords remain $\pm 1$ eigenstates of the logical $\bar{Z}$ operators. The pushing rule is thus modified. In the tensor network picture, this is given by Figure \ref{fig:AQECBStensor}.

We can derive the above observations quite explicitly using the stabilizer isometry decomposition for the $\bar{X}$ operator. 
By breaking up the non-stabilizer encoding map into a superposition of stabilizer encoding isometries, we can apply the standard pushing rules from section~\ref{subsec:bsoptpush} to each term in the superposition,

\begin{align}
\begin{split}
    V\bar{X} &= a_0 W_0 \bar{X}+a_1 W_1 \bar{X}+a_2 W_2\bar{X} \\
    &= a_0 X_1X_2 W_0 + a_1 X_1X_2Z_3 W_1-a_2 X_1X_2Z_3 W_2\\
    &= X_1X_2 (a_0 W_0+a_1 Z_3W_1-a_2 Z_3W_2)\\
    &= X_1X_2 V',
    \label{eqn:pushingchangestate}
\end{split}
\end{align}

The bulk-to-boundary pushing is performed independently on each branch, according to the code/stabilizer structure of that branch. If we insist on pulling out a factor $X_{\rm phys}=X_1X_2$, which is the operator for the reference code $W_0$, then we find that it is at a cost of altering the encoding map such that $V^{\dagger}V'\ne I$.
Conversely, it confirms that applying the reference code logical operator performs the skewed code logical operation approximately, provided that $a_{i\ne 0}$ are small. It also incurs a small error by taking the states out of the defined code subspace.

To construct an exact logical $\bar{X}$ operator, we note that it must be non-local.
Again, the logical $\bar{X}$ has physical representation
\begin{align}
    V\bar{X} &= \Big ((a_0 X_1X_2+(a_1-a_2)X_1X_2Z_3 )\Pi_0 +X_1X_2\Pi_1+ \frac{(a_1+a_2)}{a_1-a_2}X_1X_2Z_1\Pi_2\Big) V\\
    &=X_{\rm phys}V
    \label{eqn:exactSBSTpushing}
\end{align}

The projectors $\Pi_i =|P_i\rangle\langle P_i|$ satisfy $\Pi_i\Pi_j=\delta_{ij}$. $i=0,1,2$ labels projection onto the following mutually orthogonal states respectively.

\begin{align}
    |P_0\rangle =\frac{1}{\sqrt{2}}(|0000\rangle+|1111\rangle)\\
    |P_1\rangle=\frac{1}{\sqrt{2}}(|1100\rangle+|0011\rangle)\\
    |P_2\rangle=\frac{1}{\sqrt{2}}(|1100\rangle-|0011\rangle)
\end{align}

 The projection makes the physical operator $X_{\rm phys}$ global. Similarly, by restricting to local operators on a subregion, it incurs a small error penalty by dropping the terms that have coefficients which depend on $|a_{i\ne 0}|\ll 1$. Recall that for $|\alpha-\beta|<\epsilon, |a_1|,|a_2|<\sqrt{\epsilon}$. Indeed, these additional terms that depend on $a_i$ vanish in the limit $\alpha,\beta\rightarrow 1/\sqrt{2}$.

Having looked at the pushing of logical operators, now we discuss single site Pauli operators that are pushed through such tensors while preserving the logical information, similar to the exact case. Because of the $Z$ type stabilizers, any single site $Z$ type operators can be pushed through the approximate Bacon-Shor tensor to one site in the single-copy, and at most three sites in the multi-copy. However, the same does not apply for $X$ type operators, because not all codewords are stabilized by $XXXX$. For such kind of operators, it may be possible to push it across the tensor, but whether it leaves the state invariant or covariant is entirely state-dependent. For instance, the state $|\bar{0}\rangle$ is stabilized by all stabilizers of the reference code, while no other states are. Therefore, a general single site operator can be cleaned or pushed through the tensor\footnote{An important caveat is we only consider equivalent operators that have equal norm as the operator being pushed. There may be operators that are pushed across such tensors whose norms are not preserved.} only if the logical state is in $|\bar{0}\rangle$. The sole exception is, of course, a single Pauli $Z$ operator, which can be pushed in the same way as the reference code regardless of the logical state. This is related to the fact that such kind of skewed code does not detect a single qubit error exactly.

Indeed, with the exception of $|\bar{0}\rangle$, no single qubit is maximally entangled with the rest of the system for other logical states. These states do not function as an isometry when we treat it as a map from 1 qubit to 3 qubits. As before, it is still possible to push a single site operator through this tensor using the reference code stabilizers at the expense of incurring some error. This generically alters the encoded state and picks up a component that is out of the code subspace. For approximate codes with small noise, this component is of order $\epsilon$. 

Thus far, we have analyzed the behaviour for (\ref{eqn:ghzcode}), whose pushing rules are summarized in Figure~\ref{fig:AQECBStensor}. However, if we use different skewed codes, then the pushing rule in the tensor network language will also change.  For instance, in the skewed code \ref{eqn:bellaqec}, the logical $\bar{Z}$ operator can not be supported on a single row, which means that the operator cannot be pushed unitarily to two legs.

\subsection{Decoding}
\label{subsec:SBSdecode}

Unlike the exact Bacon-Shor code, having 3 physical qubits here does not allow us to decode any logical information exactly. However, we can decode the information on this subsystem if the logical state is guaranteed to be in a specific subset. We will refer to this as state-dependent decoding. For example, for codes like (\ref{eqn:ghzcode}), we can explicitly extract the logical information using the decoding unitary (\ref{figure:decodingZZ}) if the logical state is $|\bar{0}\rangle$ or $|\bar{1}\rangle$. Consequently, we can also decode any thermal state that is diagonal in this basis.  

It is clear that the decoding unitary remains valid for $|\bar{0}\rangle$ because it is unaltered. Let $A=\{1,2,3\}$ be a subsystem consists of the first 3 qubits and $U_A$ is defined in Figure \ref{figure:decodingZZ}a. Then

\begin{equation}
    U_A|\bar{0}\rangle \rightarrow |0\rangle\otimes|0\rangle|\Phi^+\rangle,
    \label{eqn:approxdecode0}
\end{equation}
where $|\chi_0\rangle=|\Phi^+\rangle$ is a Bell pair. 

For the logical 1 state, the same unitary yields
\begin{equation}
    U_A(\alpha|1100\rangle +\beta|0011\rangle)=\alpha|1100\rangle+\beta|1111\rangle \rightarrow |1\rangle\otimes|0\rangle \otimes(\alpha|00\rangle+\beta|11\rangle),
    \label{eqn:approxdecode1}
\end{equation}
where $|\chi_1\rangle=\alpha|00\rangle+\beta|11\rangle$ is no longer maximally entangled. 

As for the thermal state,
\begin{equation}
    U_A(p_0 |\bar{0}\rangle\langle \bar{0}|+p_1 |\bar{1}\rangle\langle\bar{1}|)U_A^{\dagger} =p_0 |0\rangle\langle 0|\otimes \rho_0\otimes \chi_0 +p_1|1\rangle\langle 1|\otimes \rho_0 \otimes\chi_1,
\end{equation}
the reduced state onto the first qubit recovers the original thermal state. However, now we have a separable state where there is classical correlation between the subsystems. In contrast, when decoding the exact Bacon-Shor code, we recover product states between the thermal state $\tilde{\rho}$ on qubit 1 and the $\chi\otimes \rho_0$ on the rest of the system. However, we must be careful in applying this result, as we can not decode arbitrary states that are superpositions of $|\bar{0}\rangle,|\bar{1}\rangle$. Therefore, we also cannot decode the thermal state if it is not diagonal in the computational basis.

\subsection{Multi-copy Skewed Code}
\label{subsec:multics}
A multi-copy skewed code is constructed in the same way as the multi-copy code, except some copies are replaced by their skewed counterparts. Then we use the same repetition code concatenation and select a subspace of the collective multi-copy logical subspace. For example, we can construct a skewed double-copy code by taking the exact double-copy code (Figure~\ref{fig:doublecopy}) and replacing one of the Bacon-Shor codes by the skewed code in the $ZZ=+1$ gauge~(\ref{eqn:ghzcode}). For the sake of simplicity, we use the skewed code (\ref{eqn:ghzcode}) and the same repetition code for concatenation. Both of these choices can be generalized.
 Overall, the pushing and entanglement properties in the multi-copy skewed code are very similar to that of a single-copy skewed code. A key difference is that the multi-copy is more ``isotropic'' from the point of view of entanglement and operator pushing. In particular, it allows us to define the RT-entanglement for any 2-2 bipartition of the physical legs without ambiguity, which is not possible in the single-copy skewed code.

 We still represent the multi-copy skewed code with the same 4-leg tensor, like its reference code counterpart. We will call it the Skewed Bacon-Shor Tensor (SBST). It is easy to check that the pushing rules for this tensor are essentially identical to that of the single-copy skewed code. The logical $\tilde{X}$ operator requires access to $\bar{X}$ operator from each copy, the bulk $\tilde{X}$ pushes to all 4 legs of the tensor. The logical $\bar{Z}$ and $\tilde{Z}$ remain unchanged, preserving the same pushing rule for logical Z of the reference code. Therefore, bulk-to-boundary pushing is again given by Figure~\ref{fig:AQECBStensor} for the skewed single-copy code. However, like the reference multi-copy code, the $\tilde{Z}$ operator can be supported on any two legs, whereas for the skewed single copy, $\bar{Z}$ cannot be supported the two legs that correspond to rows. Similar to the single copy, a generic single site operator may be pushed unitarily across the tensor in a state-dependent manner. However, there may be more $Z$-type Pauli strings that can be cleaned off a single leg.

The entanglement entropy of the multi-copy code can be easily computed  in the GHZ-like construction, but to properly define the RT-like entanglement contribution, we need to identify the decoding process. As we have identified the decoding process for certain states in section~\ref{subsec:SBSdecode}, a similar process is applicable for the multi-copy code. Again, we focus on $|\tilde{\alpha}\rangle$ for which there is subsystem decoding. The extension is simple: we only need to apply the same decoding unitary shown in Sec.~\ref{subsec:multic}.

First consider a subsystem consisting of 3 legs of the SBST. For any logical states in the computational basis $|\tilde{\alpha}\rangle$,  there exists a decoding unitary $U_A$ supported on the 3-leg subsystem $A$(or the corresponding $3N$ qubits),  

\begin{equation}
    U_A |\tilde{\alpha}\rangle \rightarrow |\alpha\rangle_{A_1}|\chi\rangle_{A_2A^c}= |\alpha\rangle_{A_1} \bigotimes_i^N|\chi^{i}_{\alpha}\rangle_{A_2A^c}
    \label{eqn:decodedNstate}
\end{equation}
where $i$ labels the $N$ tensor copies. $|\chi_{\alpha}\rangle$ is the state left over after extracting the logical information. For instance, we can construct a decoding unitary that decodes all copies of the information, then $|\chi_{\alpha}\rangle$ is simply the tensor product of the Bell-like states from each copy in \ref{eqn:approxdecode0}, \ref{eqn:approxdecode1} tensoring some $|0\rangle$ ancillary states.

If we have access to only 2 legs, then we do not have to decode all copies, because we need only decode a single copy whose $\bar{Z}$-subalgebra is accessible and leave the rest unchanged.  Without loss of generality, assume the $\tilde{Z}$ has non-trivial support over the $i$-th copy. Namely, $\tilde{Z} =  I^0\otimes \dots\otimes \bar{Z}^i\otimes I^{i+1}\otimes\dots$. Hence, let $U_A= I_A^0\otimes\dots U_A^i\otimes \dots$, and act on the state $|\tilde{\alpha}\rangle$. We then obtain a state of the form (\ref{eqn:decodedNstate}) except $|\chi_{\alpha}\rangle$ is the tensor product of a Bell-like state (times the ancilla) at the $i$-th copy and a number of GHZ-like states on the remaining copies.

To compute the RT contribution to the entanglement entropy, let $\chi_{\alpha}^i = \Tr_{A^c}[|\chi_{\alpha}^i]\rangle\langle\chi_{\alpha}^i|]$. We see that, in this case, the von Neumann entropy of the subsystem $A$ is entirely coming from the RT contribution.

\begin{equation}
    S(A) = S(\chi_{\alpha}) 
    \label{eqn:RTNcopyent}
\end{equation}
and 
\begin{equation}
    S(\chi_{\alpha}) \equiv \sum_i^N S(\chi^i_{\alpha}).
    \label{eqn:NRTdef}
\end{equation}
 Here, $S(\chi_{\alpha}^i)$ only depends on $\alpha$ if the $i$-th copy is a skewed code. Thus, the interpretation of the entanglement is identical to that of a single copy skewed code, except we have to sum over all $i$ to obtain the final entropy contribution.

For all of the above bipartitions, it is also easy to generate bulk entropy by going to a thermal state
 
 \begin{equation}
     \tilde{\rho} = \sum_{\alpha} p_{\alpha}|\tilde{\alpha}\rangle\langle\tilde{\alpha}|,
     \label{eqn:thermalNstate}
 \end{equation}
 
 which is diagonal in the computational basis. Then again we obtain the overall von Neuman entropy is given by (\ref{eqn:entropyRT}), where the RT term $S(\chi_{\alpha})$ is defined in \ref{eqn:NRTdef}. 
 
 Note that, in the 2-2 bipartition, the recovery is also complementary for \textit{arbitrary} bulk/logical states because of the complementary access to the code $\tilde{Z}$-subalgebra. The FLM formula and entropy definitions apply even for a pure state $|\tilde{\psi}\rangle = a|\tilde{0}\rangle+b|\tilde{1}\rangle$. In this case, the bulk entropy contribution can be computed by reducing the pure logical state onto the Z subalgebra, then computing the entropy of this reduced state. This again yields a state of the form (\ref{eqn:thermalNstate}) for any entropy computations.

\section{Hybrid Holographic Tensor Network}
\label{sec:5hybrid}
\subsection{Tensor Network Construction}

\begin{figure}[ht]
    \centering
    \includegraphics[width=0.3\textwidth]{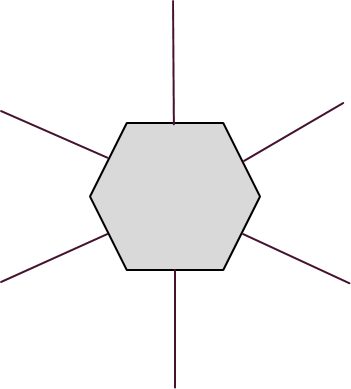}
    \caption{A perfect tensor, which is a stabilizer state maximally entangled across any bipartitions. Each leg represents a physical qubit.}
    \label{fig:PTensor}
\end{figure}

We now construct a hybrid holographic tensor network, or a holographic hybrid code, with exact Bacon-Shor codes and perfect tensors $|T\rangle$. The so-called perfect tensors \cite{preskillLecture, Pastawski:2015qua} are 6-qubit stabilizer states that are maximally entangled across all bipartitions. It is also known as a $[[6,0,4]]$ code \cite{preskillLecture} where exactly one state is encoded. The state can be understood as an isometry if we divide its degrees of freedoms as inputs and outputs. The specific correspondence is given by the channel-state duality. See Appendix of \cite{Pastawski:2015qua} for a brief review of its construction and relevant properties in the tensor network.
The tensor network representation of this object is a six-legged tensor (Figure~\ref{fig:PTensor}). Similar operator-pushing rules also apply. In this case, it can clean up to 3-site operators using stabilizer multiplication. In the tensor network language, any operator supported on 3 or fewer legs can be pushed to the remaining legs. In particular, Pauli strings are pushed to Pauli strings.

\begin{figure}[ht]
    \centering
    \includegraphics[width=0.6\textwidth]{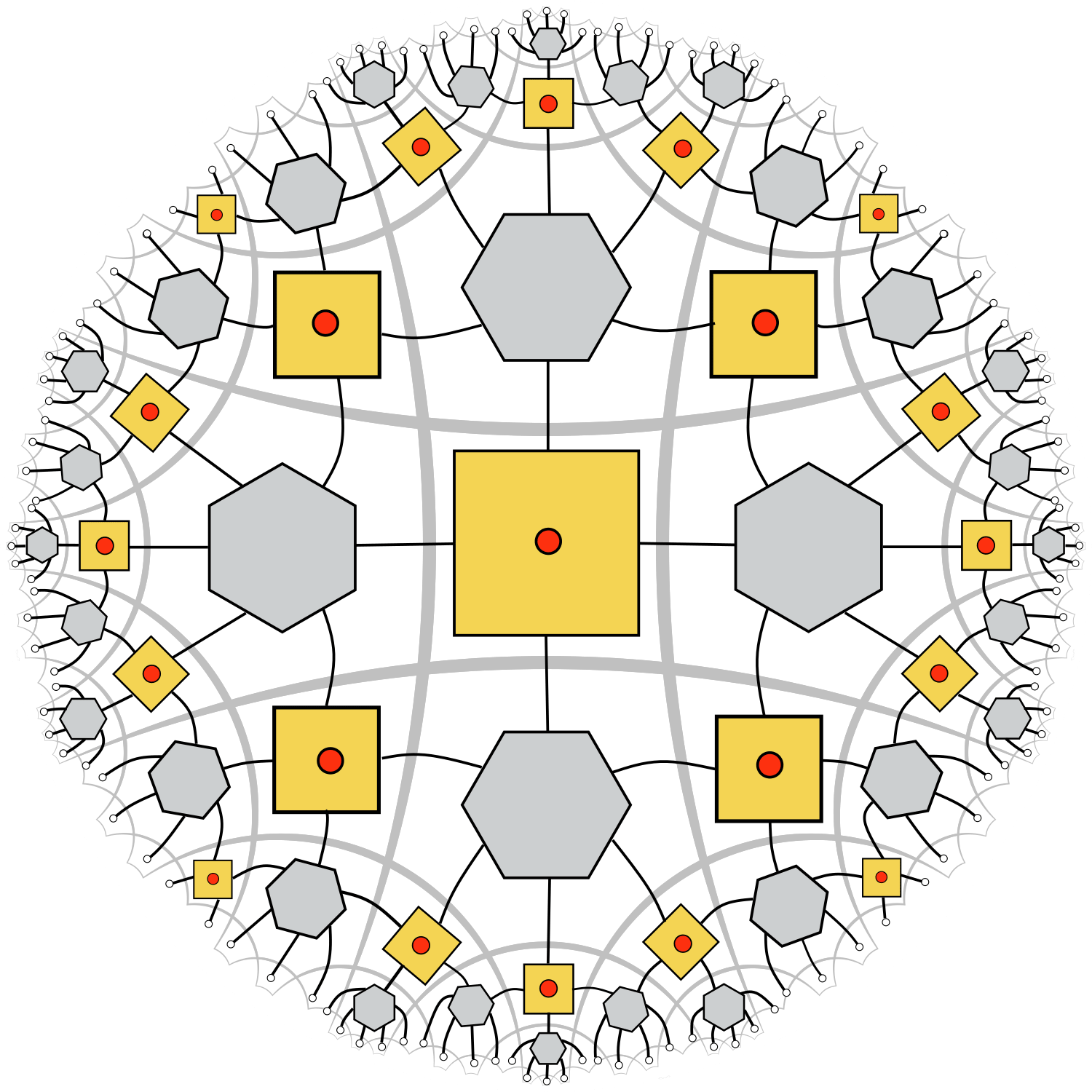}
    \caption{Each square tile is associated with a 4-leg tensor. In the case of a single copy construction, each 4-leg tensor is precisely a Bacon-Shor tensor and a hexagon is a perfect tensor. In the double or multi-copy construction, each 4-leg tensor is replaced by the multi-copy Bacon-Shor code whereas the hexagons are multi-copy perfect tensors. The planar graph $G$ is marked in light grey.}
    \label{fig:46TN}
\end{figure}

To construct the holographic tensor network, first consider a semi-regular tiling of the hyperbolic disk with squares and hexagons, where the boundaries of the tiles are demarcated by a regular planar graph $G$ of degree $d$. For example, see Figure (\ref{fig:46TN}) where $d=4$ and Figure (\ref{fig:hybrid_contraction}) where $d=6$. 
For each square tile $\stile$, we place a 4-qubit Bacon-Shor code with encoding isometry $W_{i_{\stile}}$, where the physical qubits are placed on the edges of the square. Similarly, for each hexagonal tile $\htile$, we place a perfect tensor $|T_{j_{\htile}}\rangle$, again with physical qubits placed on the edges of the hexagon, which are shared with adjacent squares.

Each edge now contains two qubits, one from the Bacon-Shor tensor and the other from the perfect tensor. We can glue together these tensors, or the individual blocks of squares and hexagons, by projecting the two qubits on each edge to a Bell state

\begin{equation}
    |\Phi^+\rangle = \frac{1}{\sqrt{2}}(|00\rangle+|11\rangle)
\end{equation}
and normalize. 
More precisely, the encoding isometry of the holographic tensor network is 
\begin{equation}
    V=\mathcal{N} (\bigotimes_E \langle \Phi^+_E|) \bigotimes_{i,j}(W_{i_{\stile}}\otimes |T_{j_{\htile}}\rangle )
\end{equation}
where the projection $\langle \Phi^+_E|$ is performed for each edge $E$ that is shared by two polygons. $W_i$ are encoding isometries associated with the 4-qubit Bacon-Shor code, or more precisely, a gauge fixed version of the Bacon-Shor code\footnote{In principle, one should construct a gauge code by considering the class of states or encoding isometries tied to each Bacon-Shor tensor. Therefore $W_i$ in fact represents an equivalence class of encoding isometries. Here we abuse the notation and sacrifice some mathematical precision for a simpler presentation.}. This procedure eliminates all the physical qubits in the interior of the tessellation. The ones remaining are edges that lie on the boundary of the tessellation, which are not projected out. They are the physical degrees of freedom of this error correction code, similar to the HaPPY construction \cite{Pastawski:2015qua}. If the Bacon-Shor code/tensors are in a particular gauge, say $ZZ=+1$, then this is nothing but an $[[n,k]]$ stabilizer code where there are $n$ boundary qubits and $k$ bulk qubits\footnote{If instead of fixing a gauge, we promote the gauge qubits to logical qubits, then it becomes an $[[n,2k]]$ stabilizer code where each bulk leg represents two logical qubits.}. The new stabilizers/gauge operators of this code can be generated by taking the individual stabilizers of each individual tensors, and matching the Pauli operators on where the gluing occurred. We will discuss the stabilizers, gauge operators and logical operators of this code in more detail in section~\ref{subsec:hybridpushing}. We also present some results in how their check matrices transform under these gluing operations in Appendix~\ref{app:tracing}, which describes the procedure of stabilizer matching in a more algebraic form. More recently, the construction and properties of such kind of codes are also summarized in \cite{Farrelly:2020mxf}.

The specific $n$ and $k$ can be determined by simply counting boundary edges and tiles in the tensor network. There is a bulk logical degree of freedom associated with each square, but the hexagons contribute none. More precisely, it is an isometric mapping between the logical degrees of freedom in the bulk to the physical degrees of freedom on the boundary~(Figure~\ref{fig:46TN}).

The in-plane graph geometry of the tensor network is given by the dual graph $G^*$ of the planar graph that marks the semi-regular tiling. Explicitly, we condense grey hexagons and yellow squares into vertices and leave the edges unchanged. At each boundary degree of freedom, we attach an additional vertex. These are the physical degrees of freedom we construct in a laboratory. Each Bacon-Shor tensor (BST)  \footnote{ We might also treat the bulk degree of freedom as another vertex attached to the BST, which would make it degree 5. However here we focus on the in-plane geometry so we will ignore the bulk degrees of freedoms.} is represented by a vertex of degree 4, while a perfect tensor is a vertex of degree 6. 
Although this tensor network has a fixed graph structure, it is not necessarily representative of the entanglement pattern of the state. In particular, as we may fix different gauge representatives in the construction, the entanglement pattern can also be drastically different.

\subsection{Double and Multi-copied Code}
As mentioned before, a single Bacon-Shor block breaks rotational invariance. Depending on the bipartition, it can also produce bulk entropy of order one despite having a logical pure state. A potential solution is to replace the BST with its multi-copy cousin that we have introduced in the previous sections. 

For the sake of simplicity, we will use the double-copy in the tensor network construction to build a double copy tensor network. Similar to the single copy construction above, here we take double copied Bacon-Shor tensor and a double tensor copy of the perfect tensor for our construction. We can represent the double-copy tensors as one stacked on another, as in Figure \ref{fig:doublecopy}. However, for clarity of presentation, we still represent the double copy tensors with the same degrees/number of edges, except each edge now has a larger bond dimension that captures the doubling of the physical qubits. Working as above, each square is assigned a double-copy BST, a hexagon is assigned a double copy perfect tensor $|T\rangle\otimes|T\rangle$, and common edges are projected onto a doubled Bell state 
\begin{equation}
|\Phi^{+}_2\rangle    = |\Phi^+\rangle^{\otimes 2}.
\end{equation}

That is, this double copy construction is formed by stacking two copies of the hybrid holographic code, except the BSTs in one copy are rotated relative to those in the other copy before contracting the tensors (gluing their legs together). The code subspace before concatenation is given by the tensor product of all bulk qubits. The bulk space is realized as
\begin{equation}
    \mathcal{H}_{\rm bulk} = \mathrm{span}\{\bigotimes_{x\in\mathcal{B}} |\bar{i}_x\bar{j}_x\rangle \::\: i,j = 0,1\}
\end{equation}
where we write $\mathcal{B}$ for the set of squares in the tessellation.

Concatenation by the repetition code then further selects a subspace $\mathcal{C}\subset \mathcal{H}_{\rm bulk}$ to be the final code subspace we work with, where $i_x=j_x$. Namely:
\begin{equation}
    \mathcal{C} = \mathrm{span}\{\bigotimes_{x\in\mathcal{B}} |\bar{i}_x\bar{i}_x\rangle: i_x=0,1\}.
\end{equation}

The tensor network representation of such a double-copy holographic code is identical to Figure \ref{fig:46TN}, although the specific tensors that are represented as squares and hexagons are double copy tensors, instead of single copy ones as above.

More general multi-copy holographic hybrid codes can also be similarly generated.

\subsection{Operators from Pushing}
\label{subsec:hybridpushing}
\subsubsection{Bulk to Boundary Pushing}
Despite a somewhat different choice of quantum codes, the hybrid code has many commonalities with the HaPPY code. The encoding map defines a general $[[n,k,r,d]]$ gauge code, or a $[[n, k, d]]$ stabilizer code after gauge fixing. The tensor network allows us to derive the logical operator, stabilizers, and gauge operators of this code in a geometric fashion through operator pushing, where one can treat the individual tensors as isometries once suitable input and outputs are chosen. Each such operator can be pushed from the bulk to the boundary, where we obtain their physical representations. This provides an elegant geometric picture for identifying the logical operators of the various encoded qubits.
For each logical, or bulk, operator, there also exists a class of equivalent boundary operators related by multiplication of stabilizers, or more generally, gauge operators. 

Given encoding isometry $V$, we can find the physical representation of a logical operator $\bar{O}$ by identifying and $O$ such that 
\begin{equation}
    V\bar{O} = OV.
\end{equation}
A convenient representation of $\bar{O}$ can be generated through tensor pushing \cite{Pastawski:2015qua}, which is intimately related to the cleaning property of a code\cite{haahpreskill,BravyiTerhal,Flammiaetal}.  
Recall that for perfect tensor, because it is maximally entangled across any bipartitions, any number of inputs that have support on at most $d-1=3$ legs can be pushed to the remaining output legs while leaving the tensor invariant. The perfect tensor can be treated as an isometry when its legs are split into input and output sets where the inputs have 3 legs or fewer. In other words, for any operator that has support over $A$ with 3 legs or fewer, one can find their equivalent forms that only act non-trivially on $A^c$.

We derived similar pushing rules for the Bacon-Shor code in section \ref{sec:3baconshor}. Because the code detects a single error, we know that any operator that has support over a single qubit has equivalent forms over only the complementary region. For example, we can multiply any weight one Pauli operator by an appropriate stabilizer, resulting in an equivalent operator supported on the 3-qubit complementary region. This tensor is also an isometry if we partition the 4 legs into 1 input and 3 outputs.

We are now ready to generate a physical representation of an arbitrary logical operator using the tensor network. Given a bulk logical operator, we first identify its representation in the bulk of the tensor network. Its support is then pushed through the tensor network by applying the pushing rules locally on each tensor. We then progressively clean the operators that are supported on the interior legs by pushing them across the tensor until all non-trivial operators are entirely supported on boundary (physical) degrees of freedom. The resulting operator then implements the desired logical operator on the particular bulk qubit. As with HaPPY code, and in holography, we find that a bulk degree of freedom can be supported over a subregion of the boundary. This subregion may or may not be connected, which is related to the cleaning property of the hybrid code.

\begin{figure}[ht]
    \centering
    \includegraphics[width=0.7\textwidth]{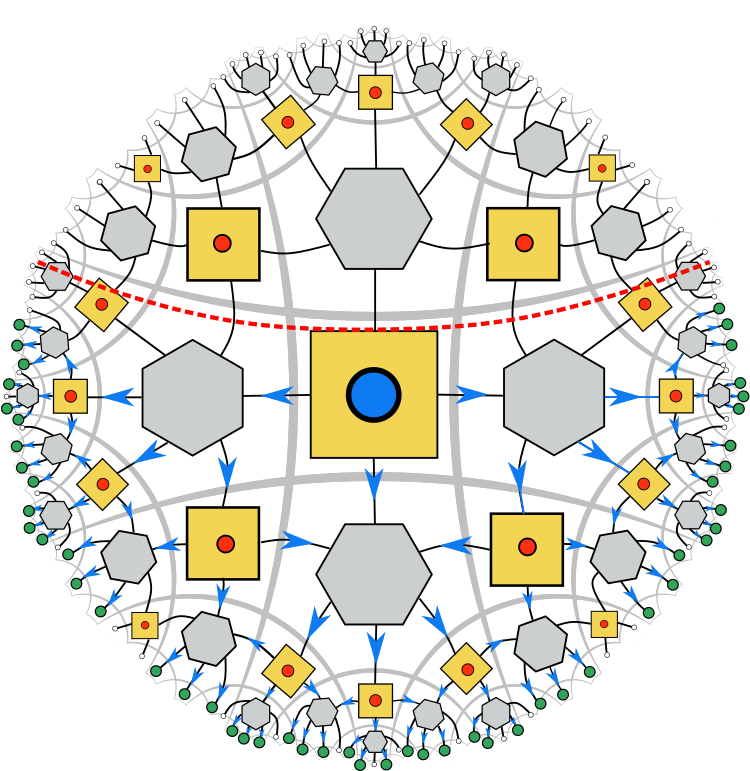}
    \caption{The bulk operator on the central tensor can be supported on a single connected interval on the boundary}
    \label{fig:singlepush}
\end{figure}
For instance, the central bulk qubit can be accessed from a boundary subregion that is a single continuous interval (Figure~\ref{fig:singlepush}). For a logical/bulk operator (coloured in blue in this figure), we know that it is supported on at most 3 legs of the tensor in either the single or multi-copy constructions. Then, by treating the nearby tensors as isometries with inputs, we can use the isometric properties of the nearby tensors and sequentially ``push'' the operator to their respective outputs. This continues down the layers following the (blue) arrows and eventually reaching the boundary. The resulting operator over the boundary qubits (green) is one way to  represent the blue logical operator on the physical Hilbert space. In holography, it simply implies that the central operator (blue) can be supported on the boundary subregion below the dashed red curve. 

\begin{figure}[ht]
    \centering
    \includegraphics[width=0.7\textwidth]{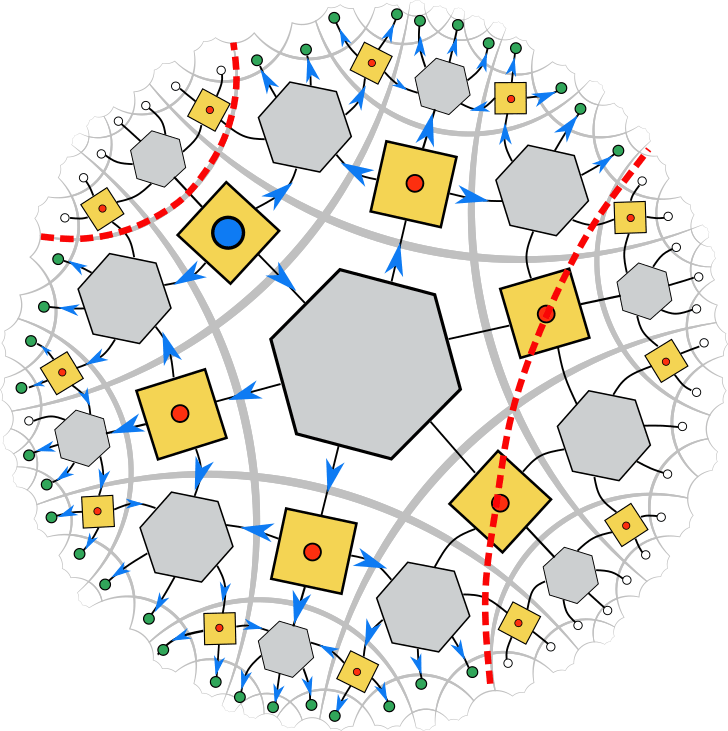}
    \caption{Representation of the same logical operator, but support over two disjoint boundary intervals.}
    \label{fig:2intervals}
\end{figure}

\begin{figure}[ht]
    \centering
    \includegraphics[width=0.7\textwidth]{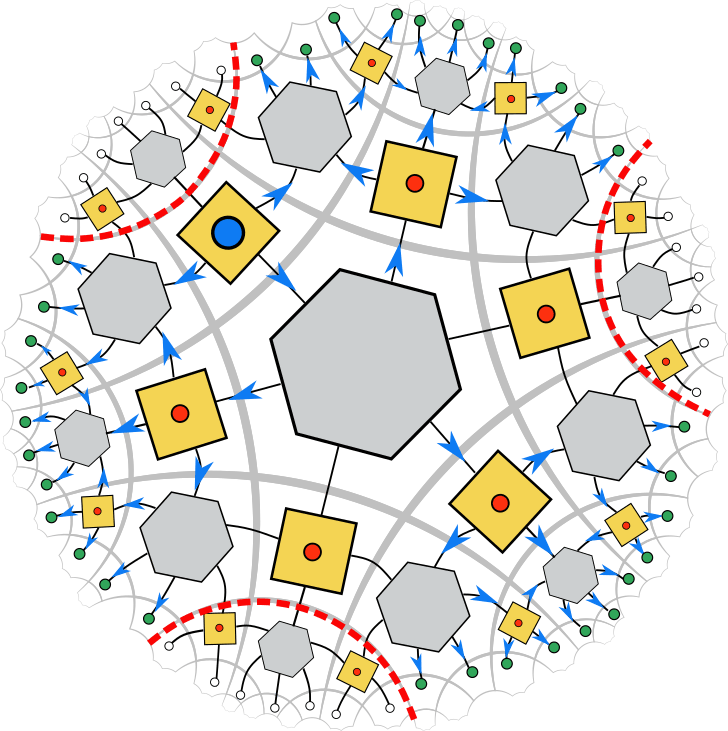}
    \caption{Representation of the same bulk operator over three boundary intervals.}
    \label{fig:3intervals}
\end{figure}

This representation is, of course, non-unique, as we can choose operator-pushing in different ways. Geometrically, the non-uniqueness leads to boundary representations on different, and potentially disjoint, intervals. This is consistent with observations in holography, where an operator can have different representations on different subregions of the boundary. Some examples are shown in Figures~\ref{fig:2intervals} and~\ref{fig:3intervals}, where we shifted the central tile so we can see the differences more clearly.

\begin{figure}[ht]
    \centering
    \includegraphics[width=0.7\textwidth]{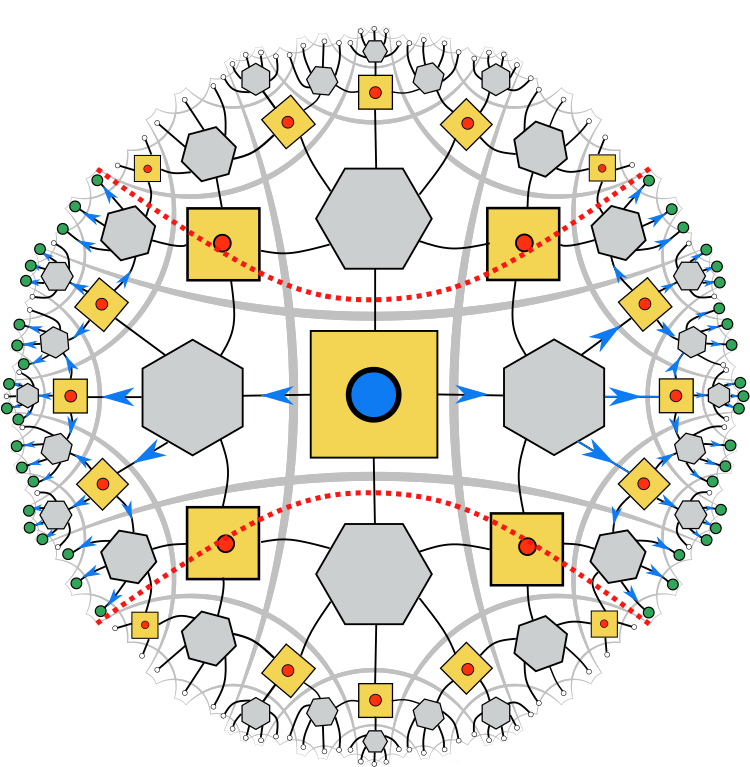}
    \caption{Central bulk operator has support on 2 legs. The operator can be supported on two disjoint intervals on the boundary, separated from its complement by the red dashed lines.}
    \label{fig:weight2push}
\end{figure}
If we further restrict ourselves to only $X$ or $Z$ logical operators in the single-copy or $Z$ logical operators in the multi-copy, then the central subalgebra can also be supported as in Figure \ref{fig:weight2push} where each logical operator can be pushed to 2 legs.

\begin{figure}[ht]
    \centering
    \includegraphics[width=0.6\textwidth]{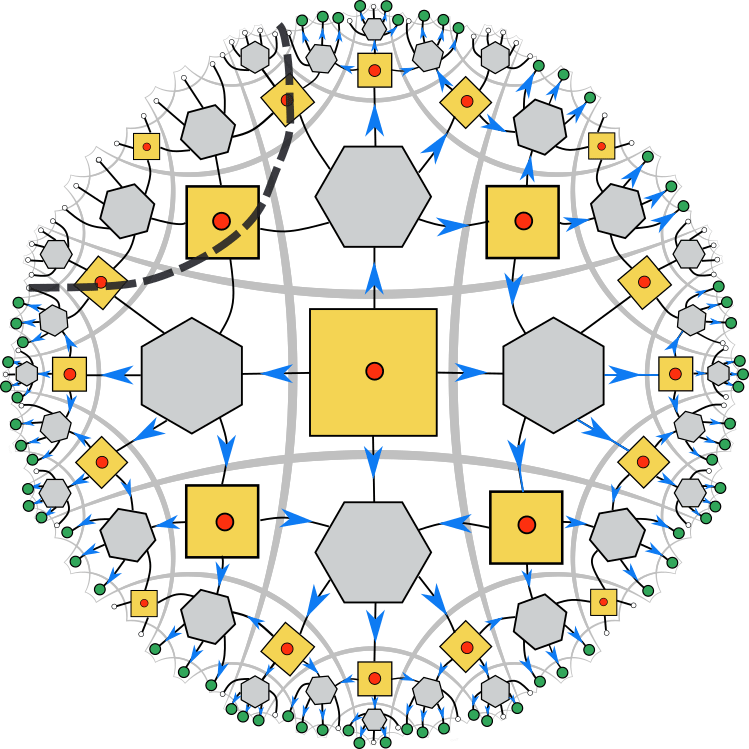}
    \caption{A weight-4 operator can be pushed to the boundary. It need not act non-trivially on the logical subspace.}
    \label{fig:weight4pushing}
\end{figure}

One can also generate stabilizer and gauge operators of the holographic hybrid code similarly. For instance, instead of pushing a logical operator through the central tile, we can apply a stabilizer to the central tile, which now is supported over all 4 legs. For example, one such pushing is shown in Figure \ref{fig:weight4pushing}. It is then subsequently pushed to the boundary using stabilizers on other tensors like before. Because this boundary operator by construction leaves any state $|\tilde{\psi}\rangle\in \mathcal{H}_{\rm code}$ invariant, it is a stabilizer operator of the hybrid code. 

Gauge operators of this code can be generated similarly. Recall that a gauge operator preserves the logical information, but can take a state to its equivalent form. We can construct them by simply taking gauge operators in the bulk (for example the weight two gauge generator $Z_1Z_2$ on a Bacon-Shor code) then pushing them to the boundary using stabilizer elements. For example, hypothetically suppose\footnote{...because no gauge operator of the Bacon-Shor code is supported on opposite legs of the square. However, we can rearrange the locations of the tensor legs such that the operators push out at the right locations.} the weight 2 operator acting on the central tile in Figure \ref{fig:weight2push} is instead a gauge element, then one can construct an operator that modifies the state of the gauge qubit on the central tile but leaving the other ones invariant. Such an operator is a gauge operator because it takes us to an equivalent state, which can be constructed by replacing the state of the central tile by its gauge equivalent form through the gauge operator, then gluing it with the rest of the network. 

As shown in Section \ref{subsec:bsoptpush}, it is also possible to push or clean using gauge operator multiplication instead of stabilizers. This does not leave the physical state invariant, but rather covariant under the pushing. For each Bacon-Shor tensor that uses a gauge operator for pushing, it takes that tile to a gauge equivalent form. If the operator pushed through was a logical operator, then the boundary operator is indeed a logical operator but up to a gauge from the stabilizer-pushing construction. If the operator pushed through is a gauge or stabilizer element, then the overall operator is a gauge operator, which preserves the logical information but not the state itself.

The minimum weight of the logical operator supported on the boundary is called the distance $d$ of the code. Here because we encode multiple qubits, and each qubit may have different distances, we focus on characterizing the code distance of the central qubit, which is the farthest from the boundary and thus the best protected against errors. Although a comprehensive understanding of the code distances for different logical qubits is still lacking, we can easily bound the distance $d_o$ of the central bulk qubit. This bound is far from tight, however. Schematically, the code distance grows at least logarithmically with the system size $N$ and at most sublinearly with the system size,

\begin{equation}
    O(\log(N))\leq d_o \leq O(N^{\gamma}),
\end{equation}
where $\gamma = (\log_5 \eta)^{-1}\approx 0.913$.

More specifically, here we give the bounds in the number of layers $n$, which is roughly related to the system size via $N\approx C \eta^n$, where $C$ is a constant of order one and $\eta \approx 5.83$ which depends on the current semi-regular tiling.

The distance of the middle qubit is bounded
\begin{equation}
   2(2n+3) \leq d_o\leq 3(5^n)
\end{equation}

The calculations on the encoding rate and distance are examined in detail in Appendix~\ref{app:codeprop}. Moreover, from the upper bound of the code distance, we know that in the infinite layer limit where the cut off of the tensor network approaches the asymptotic boundary of hyperbolic space, 
\begin{equation}
    \frac{d_o}{N} \xrightarrow {n\rightarrow \infty}  0,
\end{equation}
thus the logical operator for the central bulk qubit has support over a measure zero set on the boundary qubits, consistent with the observation of uberholography in AdS/CFT \cite{PastawskiPreskill}.

\subsubsection{Boundary to Bulk Pushing}
\begin{figure}[ht]
    \centering
    \includegraphics[width=0.7\textwidth]{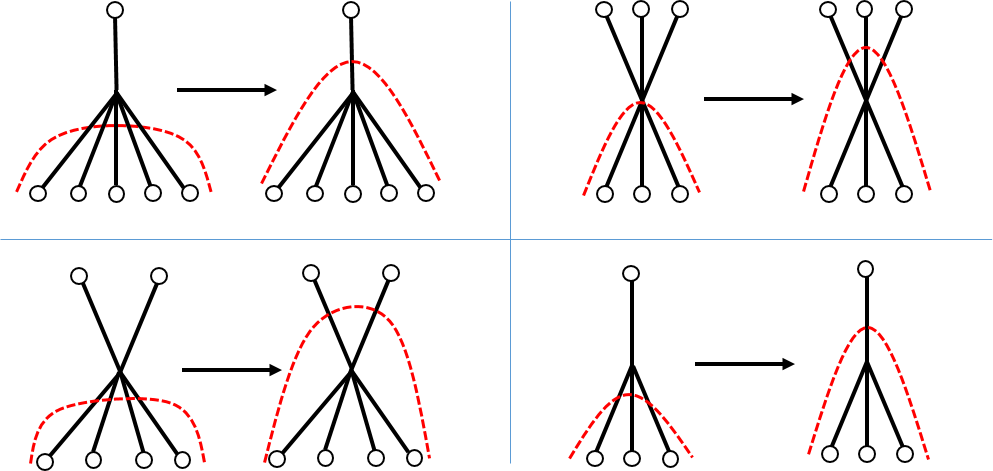}
    \caption{We start with an initial surface and update the curve $\gamma_A$ (red dashed curve) via these moves until the algorithm terminates at a local minimum. }
    \label{fig:local_moves}
\end{figure}

Now we would like to identify the bulk/logical degrees of freedom one has access to, given a particular subsystem $A$ on the boundary. To do so, we use the ``Greedy algorithm'' introduced in \cite{Pastawski:2015qua} to identify the accessible bulk region, which is called the ``Greedy wedge'' in the tensor network. The Greedy wedge is analogous to the entanglement wedge in AdS/CFT\footnote{It may differ from the entanglement wedge when the boundary subregions are disjoint. This is because the Greedy algorithm runs locally in the tensor network, and certain bulk regions, though recoverable, are not included in the Greedy local moves.}. This consists of applying the local moves in Figure \ref{fig:local_moves} progressively to update the cut $\gamma_A$ (red curve in Figure \ref{fig:local_moves}) until no such moves are available. The cut then demarcates the Greedy wedge, whose bulk degrees of freedom are accessible from $A$. 

We can visualize the local moves in the tensor network notation below.

Geometrically, the cut $\gamma_A$ (red dashed curves) marked in the figure can be updated according to these moves locally. Physically, it is simply reverse operator pushing, where having access to a subsystem $A$ (and therefore able to apply any operator on it) one can equivalently apply a desired operator on the complement $A^c$ where we do not have access. The curve $\gamma_A$ separates the degrees of freedom we can locally access and the ones that we cannot. 

More precisely, let us label the degrees of freedom below the curve $\gamma_A$ in Figure \ref{fig:local_moves} as $A$ and the rest of $A^c$. The statement of these local moves is that given any $O_{A^c}$ we wish to implement on $A^c$, we may find operator $O_A$ which only acts on $A$ such that for each state $|\tilde{\psi}\rangle$ which is either a state in the code subspace of the Bacon-Shor code, or a perfect tensor,

\begin{equation}
   O_{A^c} |\tilde{\psi}\rangle = O_{A}|\tilde{\psi}\rangle.
   \label{eqn:localmove_opt}
\end{equation}

\begin{figure}[ht]
    \centering
    \includegraphics[width=0.7\textwidth]{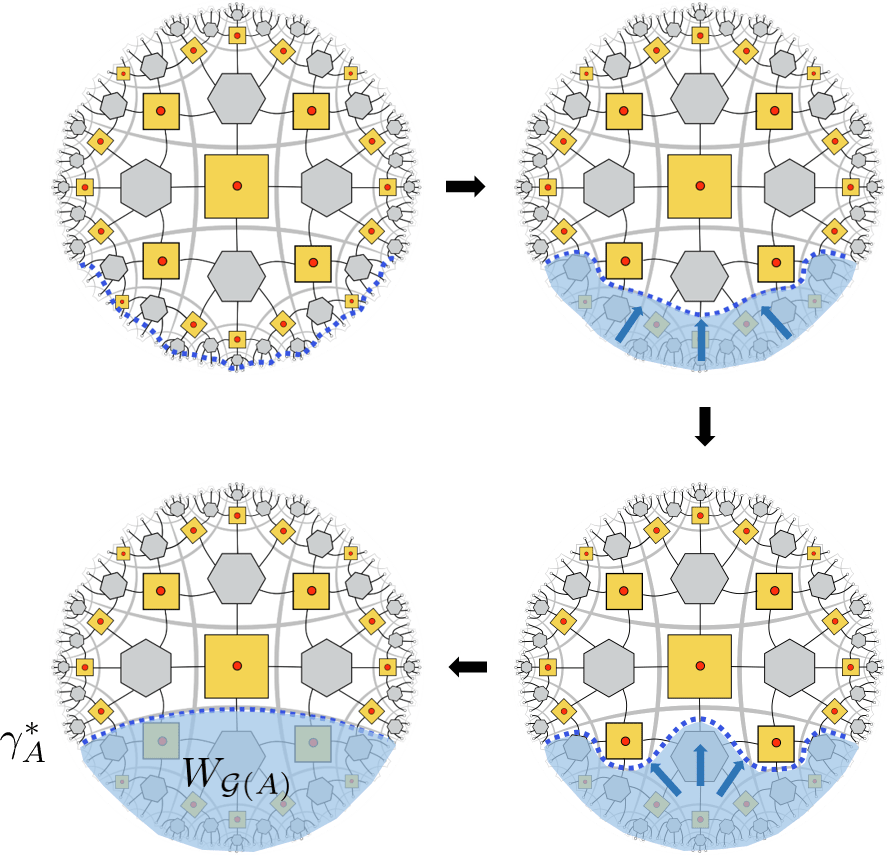}
    \caption{Starting from a boundary region $A$, we can use the Greedy algorithm to obtain a bulk region that is recoverable from $A$. Using local moves, the curve $\gamma_A$ bounding the wedge can be gradually updated until it reaches the local minimum $\gamma_A^*$. The final wedge boundary between $\gamma^*_A$ and the boundary subregion $A$ is the Greedy wedge $W_{\mathcal{G}}(A)$.}
    \label{fig:greedy_update}
\end{figure}

For the holographic tensor network, one can iteratively apply these local moves to update $\gamma_A$ once a boundary region $A$ is fixed (Figure~\ref{fig:greedy_update}). Because the local moves minimize the number of cuts through the dual graph $G^*$ associated with the tensor network, the algorithm terminates at a locally minimum cut. Let the curve $\gamma_A$ at termination be defined as the Greedy geodesic $\gamma_A^*$ bounding the region $A$. We define the region bounded by the Greedy geodesic $\gamma^*_A$ and the boundary subregion $A$ as the Greedy wedge $W_{\mathcal{G}(A)}$. By construction, all bulk degrees of freedom that lie in $W_{\mathcal{G}}(A)$ can be recovered from the boundary region $A$. We will discuss the details of such decoding processes in Sec~\ref{subsec:distdecode}.

One such example is shown in Figure~\ref{fig:greedy_update}, where blue regions demarcate the bulk degrees of freedom that can be recovered from the boundary subregion. In this case, it is easy to check that the bulk Greedy wedges that are separately recoverable from boundary regions $A$ and its complement $A^c$ cover the entire hyperbolic disk, i.e., $W_{\mathcal{G}}(A)\cup W_{\mathcal{G}}(A^c)= \mathbb{H}^2$. 

The recovery is called complementary for such bipartitions of the boundary into $A, A^c$ where any bulk degree of freedom is recoverable in either $A$ or $A^c$. For such cases, both $W_{\mathcal{G}}(A)$ and $W_{\mathcal{G}}(A^c)$ search for local cut minimizers within their wedges. The fact that they are complementary and meet at a common Greedy geodesic $\gamma^*$ implies that the cut is the global minimum cut for the graph $G^*$ with the chosen boundary bipartition. Therefore, the locally minimal Greedy geodesic also coincides with the global minimal geodesic of the tessellation. This is because, in the hyperbolic plane, there is a unique geodesic anchored to a fixed boundary interval.

\begin{figure}[ht]
    \centering
    \includegraphics[width=0.6\textwidth]{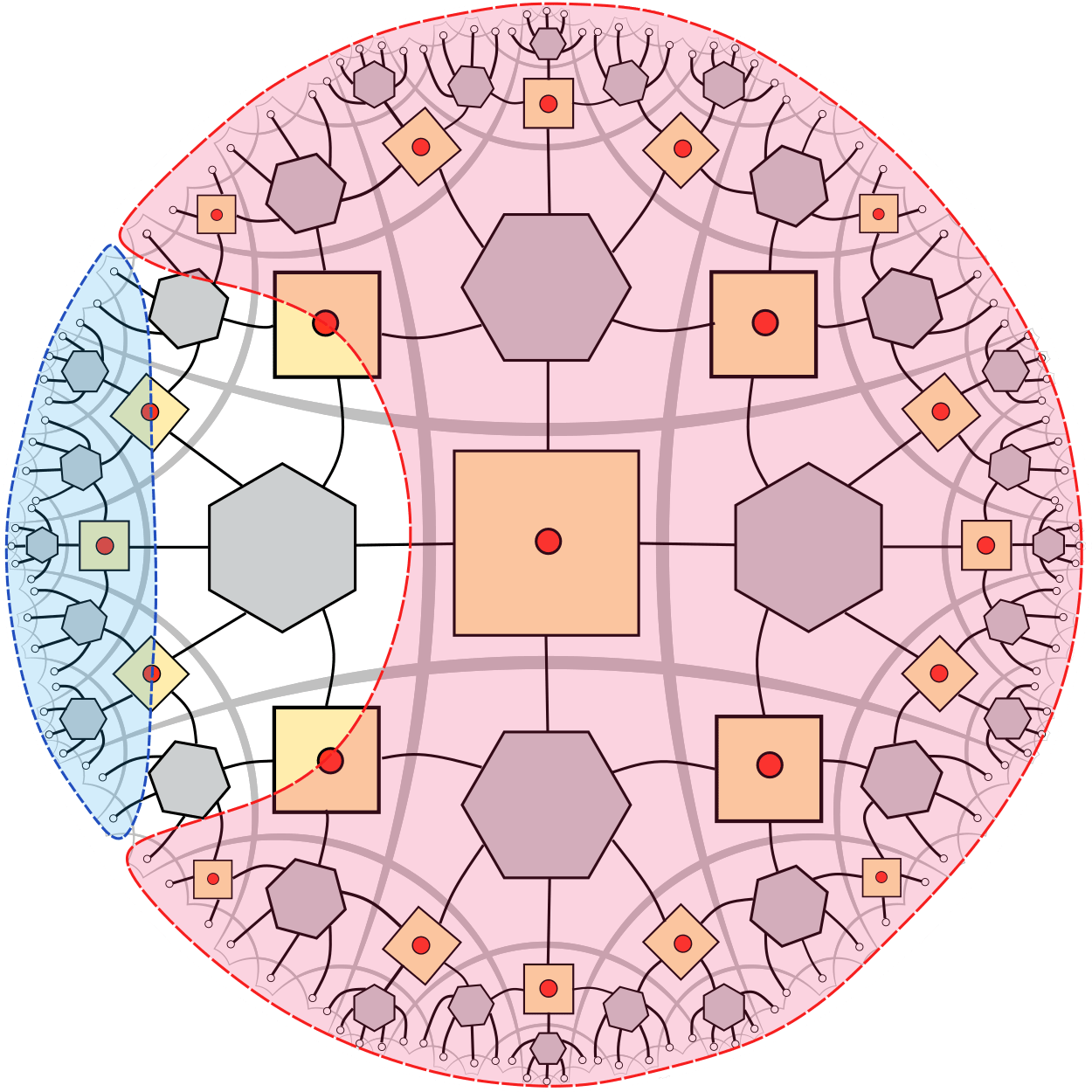}
    \caption{Complementary recovery based on local moves can fail even with boundary bipartitions that are not disjoint.}
    \label{fig:noncompdecodestep}
\end{figure}

Although the hybrid code is clearly complementary for some biparitions of the boundary, it is not obviously true for all bipartitions, even when the boundary subregion is connected. This can be seen in Figure \ref{fig:noncompdecodestep}, where we neither boundary region accesses the central region from the local moves using the Greedy algorithm.

\subsection{Distillation and Decoding}
\label{subsec:distdecode}

Equivalently, (\ref{eqn:localmove_opt}) and Theorem 5.2 of \cite{Harlow:2016vwg} imply the existence of unitary $U_A$ which (i) ``distills'' an entangled state across the bipartition (as marked by the cut $\gamma_A$) and (ii) extracts the encoded information if there is any. For some $\gamma_A$ that separates $A$ and $A^c$, a local move is possible if and only if
\begin{equation}
    U_A |\tilde{\psi}\rangle_{AA^c} = |\psi\rangle|a_0\rangle_{A_1} |\chi\rangle_{A_2A^c}.
\end{equation}
For explicit examples in this notation, see (\ref{eqn:decstatezz},\ref{eqn:decstatexx}) in section \ref{subsec:bs4}. 
We will denote the entanglement we ``distill'' across the cut by $|\chi\rangle$. The specifics of $|\chi\rangle$ will depend on the cut in (\ref{fig:greedy_update}). $|a_0\rangle$ is some ancillary state. If the tensor is a code, then it is also the decoding unitary which recovers the encoded information $|\psi\rangle$.

\begin{figure}[ht]
    \centering
    \includegraphics[width=\textwidth]{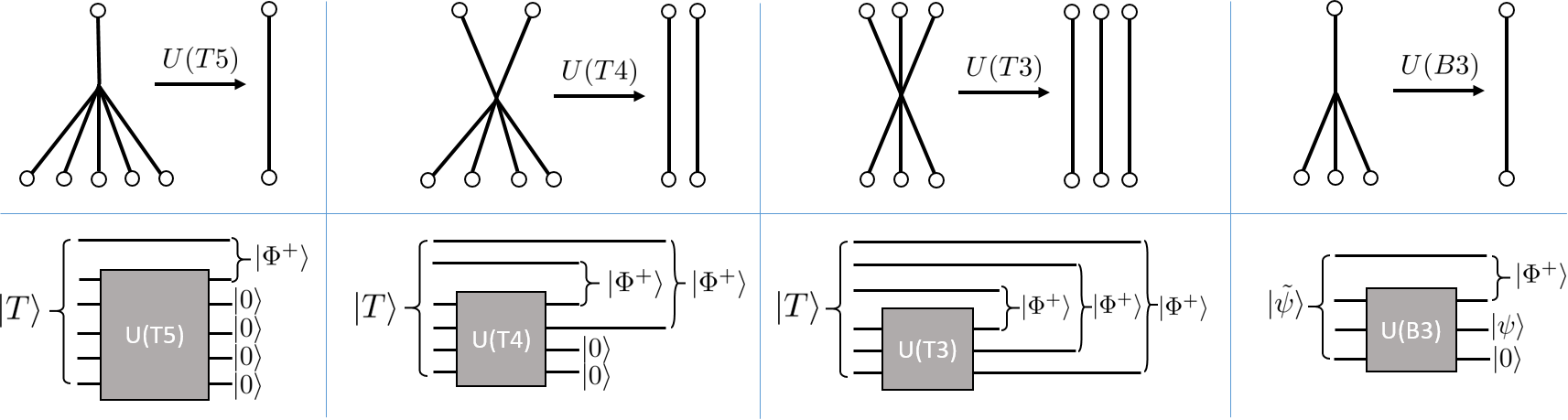}
    \caption{Local decoding/distillation procedure. We called the bottom degrees of freedoms as $A$ in the main text.
    Top diagrams are transformations in the tensor network, where the local unitary acts on bottom degrees of freedom. The bottom diagrams are the corresponding circuit representations.}
    \label{fig:local_decode}
\end{figure}

The local decoding unitaries of perfect tensors and the Bacon-Shor tensor are shown in Figure \ref{fig:local_decode}.
In particular, these local unitaries act on the bottom degrees of freedom, denoted $A$ as in Figure \ref{fig:local_decode}, in such a way that we can decompose a space containing the perfect tensor $|T\rangle\in\mathcal{H}_{PT}$ as
\begin{equation}
    \mathcal{H}_{PT} = \mathcal{H}_{A_1}\otimes \mathcal{H}_{A_2}\otimes \mathcal{H}_{A^c}
\end{equation}
so that
\begin{equation}
U_B |T\rangle = |a_0\rangle_{A_1} \otimes |\Phi\rangle^k_{A_2A^c}.
\end{equation}
Here $|\Phi\rangle$ are Bell states entangled across $A_2$ and $A^c$, and $k\leq 3$ depends on the number of degrees of freedom in $A$. $|a_0\rangle$ are some ancillary degrees of freedom that are decoupled from the rest of the system, which we can simply discard. These properties were discussed in \cite{preskillLecture,Pastawski:2015qua}.

The local move for the BST also implies the existence of such unitary that ``distills'' a Bell pair entangled between $A$ and $A^c$. We have shown in section \ref{subsec:bs4} that for any state  $|\tilde{\psi}\rangle \in \mathcal{H}_{\rm BSC}$ in the Bacon-Shor code subspace, $U_A=U(B3)$ is simply the decoding unitary which acts on 3 qubits. In a similar way,

\begin{equation}
    U_A|\tilde{\psi}\rangle =   |\psi\rangle \otimes |\Phi\rangle
\end{equation}
except now we extract the encoded information $|\psi\rangle$ from such a local unitary action. 
Therefore for each local move in Figure \ref{fig:local_moves}, there is also a corresponding local unitary transformation that distills the Bell-like entanglement in Figure \ref{fig:local_decode}. 

\begin{figure}[ht]
    \centering
    \includegraphics[width=0.8\textwidth]{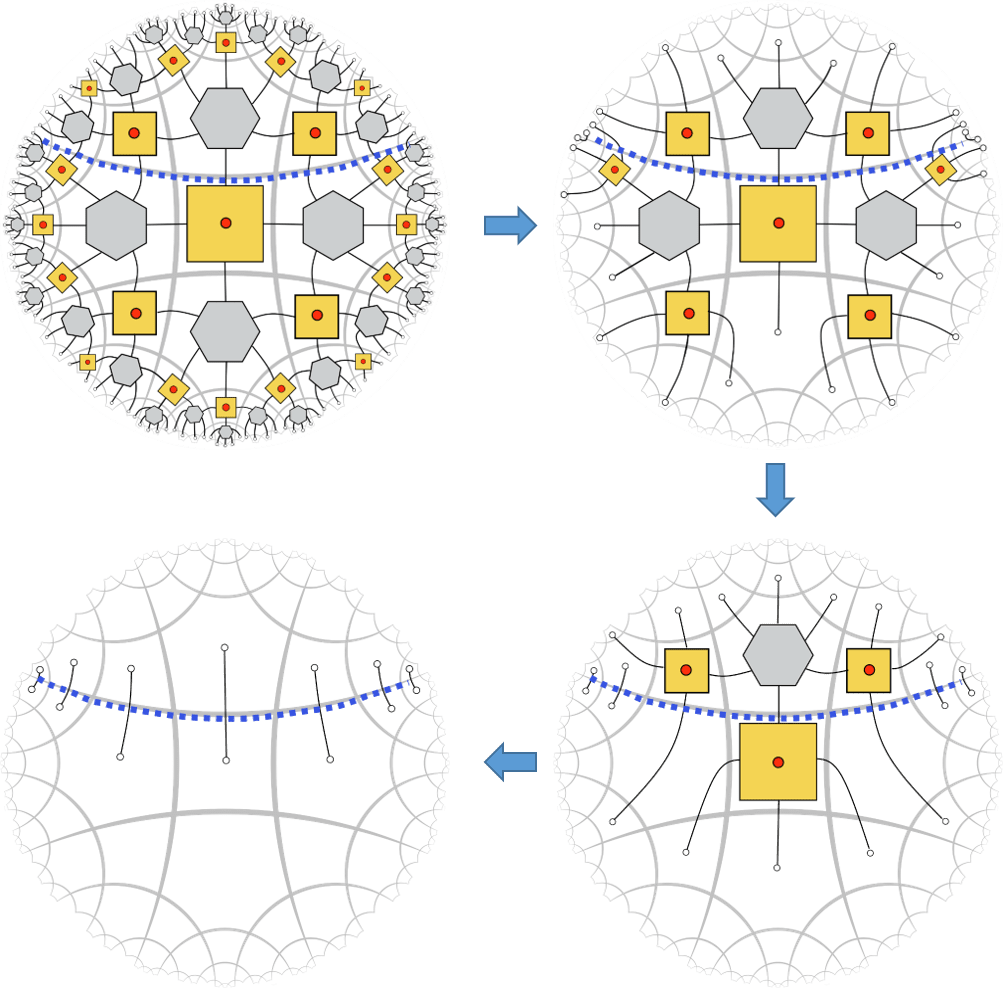}
    \caption{Distilling entanglement across the bipartitions using local moves. }
    \label{fig:decodestep1}
\end{figure}

\begin{figure}[ht]
    \centering
    \includegraphics[width=0.8\textwidth]{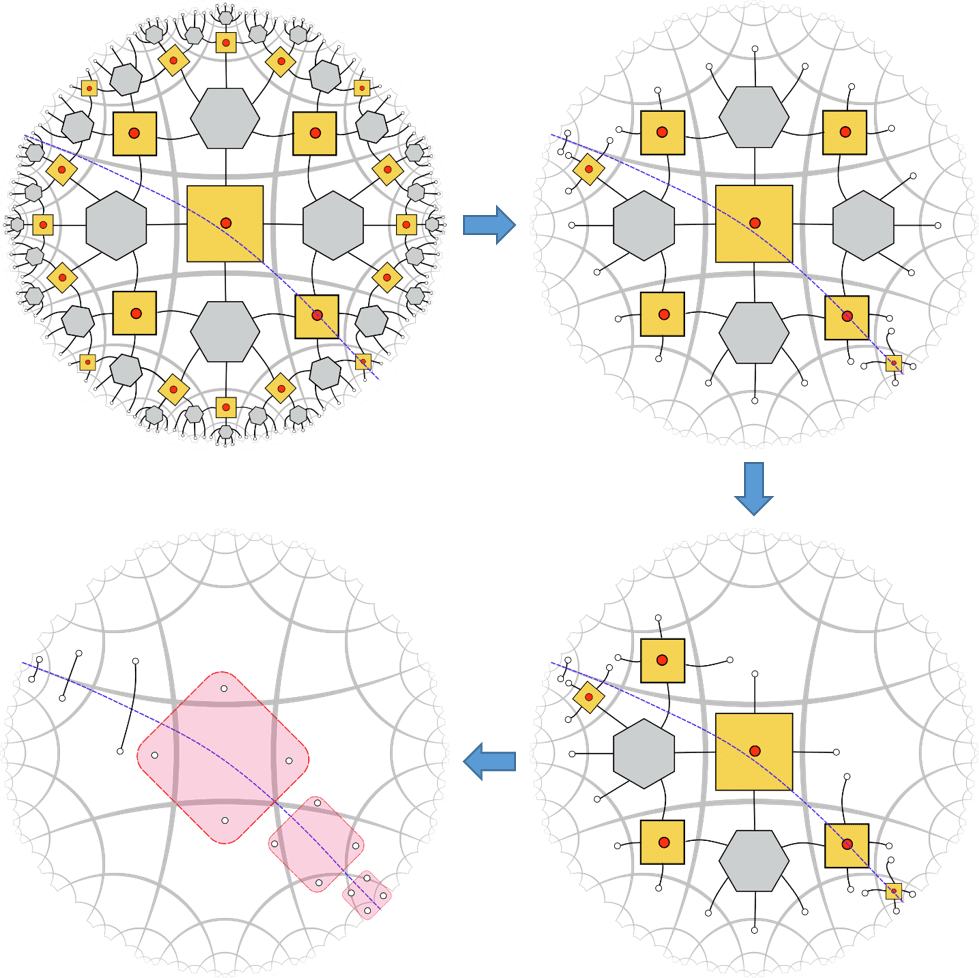}
    \caption{A bipartition is marked by the blue dashed line. The boundary degrees of freedom above and below the line are divided into two complementary sets. The entanglement can be distilled to Bell states (edges) and GHZ like states (hyperedges). }
    \label{fig:decodestep2}
\end{figure}

Now we apply these local moves and local unitary transformations to decode the information in the Greedy wedge following the Greedy algorithm.  Some examples are shown in Figures \ref{fig:decodestep1} and \ref{fig:decodestep2}. In Figure \ref{fig:decodestep1}, the geodesic cuts through each BST on exactly one edge. However, it is also possible to have bipartitions where the Greedy wedge contains two such edges in the BST (Figure~\ref{fig:decodestep2}). Note that although in the $ZZ=+1$ gauge, a single copy allows decoding and distillation into Bell states having control over only 2 qubits for suitably chosen states and bipartitions, this is not always possible for general multi-copies. In such cases, a part of the RT entanglement can remain GHZ-like. This is because we only need to decode one copy and the remaining copies can remain in a GHZ state (see (\ref{eqn:bs4ZZgauge}) in Section \ref{subsec:multic}). Therefore they are represented by hyperedges. 

 As such, we can distill the entanglement across the complementary regions into Bell-like or GHZ-like states $|\chi\rangle$, so that one such state can be assigned to each hyperedge that intersects the cut. For the sake of clarity, we will restrict ourselves to work for logical states initialized in computational basis states in a multi-copy code unless otherwise specified, so that the bulk contribution to entanglement entropy vanishes.

The above distillation procedure also generates a local circuit that decodes the logical information in the Greedy wedge from a subsystem. As we have seen from Figure~\ref{fig:local_decode}, for each local move in Figure~\ref{fig:local_moves}, we can associate a unitary that acts on the subsystem $A$. The steps to writing down the circuit are simple: we replace the isometries by their corresponding local decoding unitaries that perform the distillation/decoding. As shown in Figure~\ref{fig:local_decode}, where we denote the relevant gates as $U(Tk), U(B3)$. These are Clifford unitaries because they are derived from encoding unitaries of stabilizer codes. Therefore they allow further decompositions into 2-local Clifford gates. 
Because of the network structure, the number of layers of tensors scales logarithmically with the system size, yielding a log-depth circuit.

As an example, we can easily write down a local decoding circuit for the following tensor network
\begin{figure}[ht]
    \centering
    \includegraphics[width=0.5\textwidth]{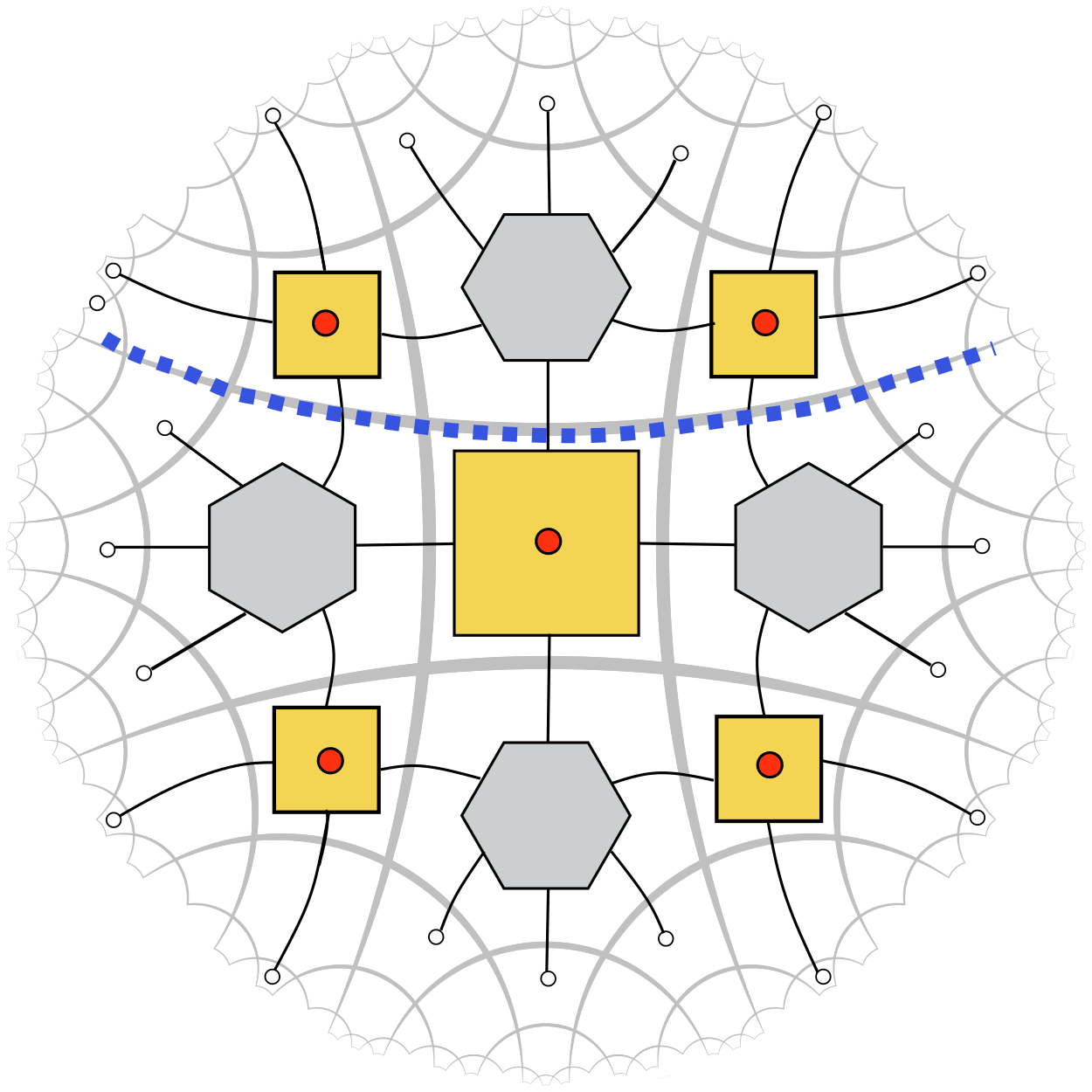}
    \caption{A two-layer hybrid tensor network.}
    \label{fig:shallow0}
\end{figure}
with access to the subsystem $A$ below the dotted line. 
Following the Greedy algorithm, for each local move we apply a gate element. Figures~\ref{fig:circuitgen1},\ref{fig:circuitgen2} show how such decoding circuits are constructed from local elements. The different colour shaded areas in the Greedy wedge indicate different steps in the local move. These mark the different layers in the circuit, which are shaded by the same colour. 

\begin{figure}[ht]
    \centering
    \includegraphics[width=\textwidth]{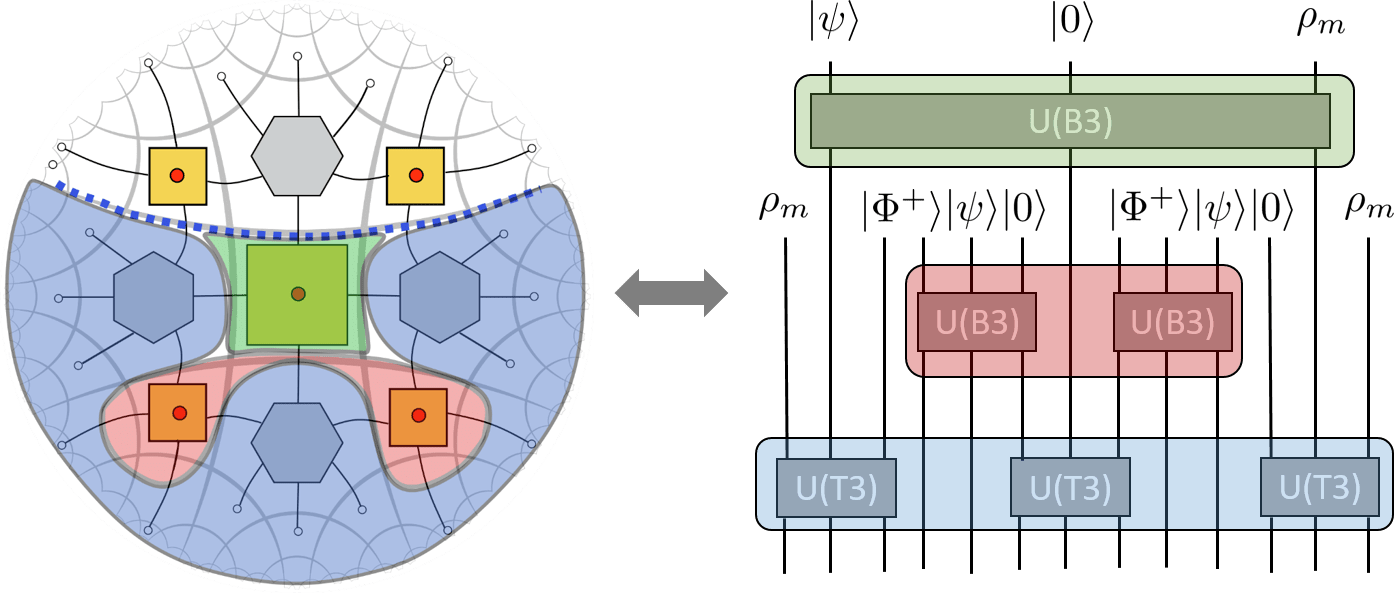}
    \caption{Decoding circuit for the bottom subsystem $A$ below the dotted curve. $\rho_m$ are maximally mixed states and $|\psi\rangle$ denote logical information. They need not be the same information. Time runs up during decoding and down during encoding. }
    \label{fig:circuitgen1}
\end{figure}

The decoding circuit is not uniquely constructed from local moves. For instance, we can consider a slightly different sequence of moves which produces a different circuit in Figure~\ref{fig:circuitgen2}.
Here $\rho_m$ is some mixed state which represents entanglement across the cut. For the examples we considered here, it is maximally mixed where $S(\rho_m)=N$ in a multi-copy construction. 

\begin{figure}[ht]
    \centering
    \includegraphics[width=\textwidth]{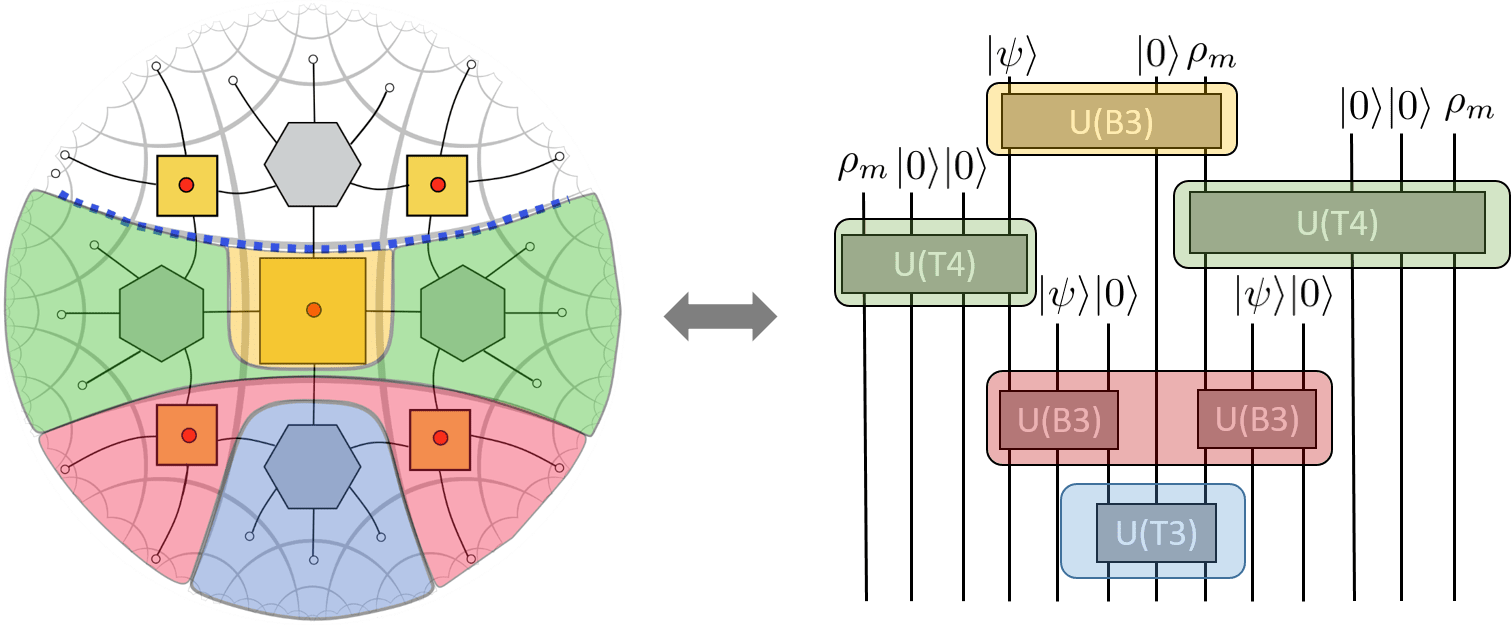}
    \caption{We can choose to decode the blue and red regions before the green region, whereas the previous construction corresponds to decoding the green and blue regions first.}
    \label{fig:circuitgen2}
\end{figure}

Globally, one can generate the decoding circuit in the same way by moving symmetrically from the outer layer to the inner layer. This yields a MERA-like 2-local log-depth circuit where each gate is a Clifford unitary. In particular, the Greedy algorithm that generates the local decoding circuit also relates the holographic tensor network, which corresponds to a slice of AdS, to a quantum circuit with a causal structure similar to that of the de~Sitter space\cite{Beny:2011vh,Czech:2015kbp,Bao:2017qmt,Milsted:2018san}.

\subsection{Entanglement Properties and the RT formula}
\label{subsec:exactEERT}
The entanglement properties of the BST heavily depends on the representative we take from the equivalence class. For instance, the $XX=+1$ gauge is completely dominated by bipartite entanglement for the same logical information $|\bar{0}\rangle, |\bar{1}\rangle$, whereas in $ZZ=+1$ gauge the entanglement is completely multipartite in the form of the GHZ state. There is also a high degree of flexibility to maneuver between bipartite vs multipartite entanglement by tuning the state of the gauge qubit and choosing a suitable representative. Therefore, the graph $G^*$ tied to the tensor network is not necessarily the best representation of its underlying entanglement structure due to this non-uniqueness. In this work, we will not characterize the general entanglement structure of all these possibilities, instead, we will focus on a particular example for the double-copy code in the $ZZ=+1$ gauge for simplicity.

As an exact error correction code with a factorizable boundary Hilbert space and a non-trivial center for the code subalgebra, the hybrid code obeys Theorem 5.2 in \cite{Harlow:2016vwg}, which naturally implies that the code satisfies a version of the RT/FLM formula. Recall that it says for a subregion $A$ on the boundary CFT, to the order of $G^0$,

\begin{equation}
    S(A) = \frac{A}{4G} + S_{\rm bulk}(\Sigma)
\end{equation}
where $\Sigma$ is the bulk region encircled by the boundary anchored minimal surface and the boundary subregion $A$.

We have shown previously that in the case of single interval boundary subregions with complementary recovery, the Greedy algorithm finds the global minimum cut in the dual graph. In fact, it is more accurate to represent the $ZZ=+1$ gauge entanglement using a hypergraph when the bulk states are initialized in the computational basis states.
\begin{figure}[ht]
    \centering
    \includegraphics[width=0.8\textwidth]{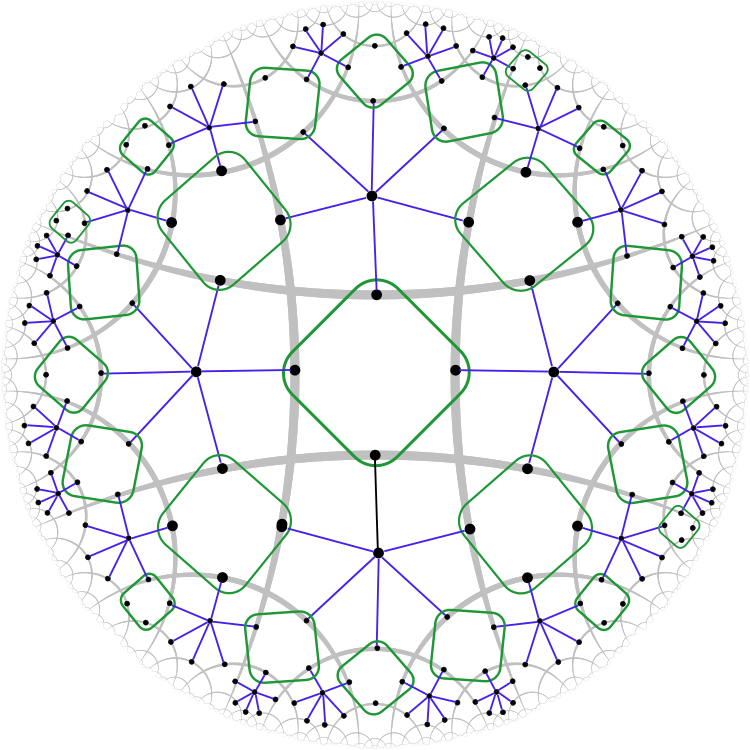}
    \caption{The hypergraph}
    \label{fig:hypergraph}
\end{figure}
In this hypergraph, we assign a degree 4 hyperedge to each square instead of the star graph in $G^*$. This is because the BST is in a GHZ state, such that any cut or bipartition of its degrees of freedom yields constant entanglement, as opposed to cuts in the perfect tensor\footnote{A similar construction that connects hypergraph representation and stabilizer states is discussed in \cite{Bao:2020zgx}.}. 
We will show that by assigning the proper amount of edge weights to the hyperedges according to the code, the entanglement entropy of a bipartition that leads to complementary recovery using the Greedy algorithm is equal to the minimal cut in the graph, thus proving the RT-FLM formula in these scenarios.

\begin{definition}
Given a bipartition of the boundary into $A$ and $A^c$, where both intervals are non-disjoint, a \textit{boundary-anchored minimal cut} $\tilde{\gamma}^*_A$ in the hypergraph $\tilde{G}$ is the cut that intersects the minimal number of edges which also bipartitions the boundary vertices into $A$ and $A^c$.
\end{definition}

There may be multiple minimal cuts for a particular bipartition, e.g. Figure \ref{fig:noncompdecodestep}, when the recovery is not complementary. In such cases, the Greedy algorithm terminates at a local minimum.

When the recovery is complementary, Greedy geodesics $\gamma_A^*,\gamma_{A^c}^*$ coincide, thus it is also the global minimal cut through the dual graph $G^*$. 
However, there are still multiple minimal cuts $\tilde{\gamma}_A^*$ through the graph $\tilde{G}$ that are consistent with the unique Greedy geodesic. This is because one can choose to cut through either an edge or an adjacent hyperedge of degree 4. To remove such ambiguity, we can consider minimal cuts that preferentially cut through the hyperedges. Indeed, to select the unique cut $\tilde{\gamma}_A^*$ in the hypergraph $\tilde{G}$, we can perform local deformation to $\gamma_A^*$ such that it preserves the Greedy wedge while preferentially cutting through the hyperedges.

Let us also assign edge weight $w_e$ to each edge in the hypergraph. 
\begin{definition}
The size of the cut $|\tilde{\gamma}|$ is given by the sum over the edge weights along the cut,

\begin{equation}
    |\tilde{\gamma}| = \sum_{e\in C_{\tilde{\gamma}}} w_e
\end{equation}
where $C_{\tilde{\gamma}}$ is the set of edges that the cut $\tilde{\gamma}$ intersects.
\end{definition}

For this section, we assign constant edge weight to each edge and hyperedge. We will see in Sec~\ref{subsec:SEERT} that these weights are state-dependent in the skewed code.

\begin{theorem}
Given a bipartition of the boundary vertices into $A$ and $A^c$, such that $A$ (and $A^c$) are non-disjoint intervals, the von Neumann entropy $S(A)$ of the multi-copy tensor network satisfies the RT/FLM formula if the recovery is complementary using the Greedy algorithm. In particular, in the $ZZ=+1$ gauge, the RT-like entanglement contribution is proportional to the size of the boundary-anchored minimal cut $\tilde{\gamma}_A^*$.
\end{theorem}

\begin{hproof}
First consider bipartitions of the boundary degrees of freedom into $A, A^c$. Complementary recovery implies the existence of decoding unitary $U_A, U_{A^c}$ that only act non-trivially on $A$ and $A^c$ respectively, such that for any basis $span \{|\widetilde{\alpha, ij}\rangle\} = \mathcal{C}\subset \mathcal{H}_{\rm phys}$

\begin{equation}
    U_AU_{{A^c}}|\widetilde{\alpha,ij}\rangle = |\alpha, i\rangle_{A_{1^{\alpha}}}|\alpha, j\rangle_{{A}_{1^{\alpha}}^c} |\chi_{\alpha}\rangle_{A_2^{\alpha}{A}_{2^{\alpha}}^c}
\end{equation}

Using the results from section 5.2 of \cite{Harlow:2016vwg}, we conclude that for any state $\tilde{\rho}$ in the code subspace, the entanglement entropy of is given by \ref{eqn:HarlowFLM}, which we reproduce below 

\begin{equation}
    S(\tilde{\rho}_A) = \Tr[\tilde{\rho}\mathcal{L}_A] + S(\tilde{\rho},M)=\sum_{\alpha}p_{\alpha}S(\chi_{\alpha}) + S(\tilde{\rho},M)
    \label{eqn:harlowFLM5}
\end{equation}
where the area operator
\begin{equation}
    \mathcal{L}_A = \oplus_{\alpha} S(\chi_{\alpha})I_{a_{\alpha}\bar{a}_{\alpha}}
\end{equation}
is in the center of the code subalgebra. The entropy contribution to the RT surface is given by $\sum p_{\alpha}S(\chi_{\alpha})$ for $\chi_{\alpha} = \Tr_{A_2^c}|\chi_{\alpha}\rangle\langle \chi_{\alpha}|$. 
This is an RT/FLM formula.

Now we show that the RT contribution $\sum p_{\alpha} S(\chi_{\alpha})$ is indeed proportional to the minimal cut in the hypergraph (Figure \ref{fig:hypergraph}) by computing the contribution of $|\chi_{\alpha}\rangle$ in this tensor network. To obtain $|\chi_{\alpha}\rangle$, we use the local moves from (Figure~\ref{fig:local_decode}) to decode the information in conjunction with the Greedy algorithm.

If the cut goes through a boundary leg, we add BST on square tiles as needed so the cut can be deformed through a square tile. The modified network now terminates on BSTs with the new boundary subregion containing all new BST boundary legs that emerged from the original boundary tensors. Modulo this modification, we assume that a cut can always be deformed such that it goes through BSTs on square tiles. 

Now for each cut in the bulk interior through the square tile/hyperedge, it may cut through 1 or 2 legs of the BST according to our local moves. 
If $\gamma$ cuts through only one leg of a BST, then either $W(A)$ or $W(A^c)$ must access 3 legs of the BST, therefore the bulk qubit must be recoverable in either the Greedy wedge or its complement. This implies that actions on $A^c$ or $A$ by decoding unitary can distill the entanglement across the cut to a Bell-pair or Bell pairs, thus adding one unit (or $N$ units in a multi-copy) of entanglement to this particular hyperedge that is cut. If 2 legs are contained  in  the wedge, then either wedge can recover the logical $Z$ subalgebra. Because we have initialized the states in the logical computational basis states, both wedges can decode and distill a state that is entangled across the bipartition with $N$ units for a multi-copy code. However, these states may be in the form of Bell pairs and GHZ states\footnote{For a single copy, the entropy may or may not be attributed to the RT term depending on the cut, thus we will focus on examples where this is not an issue. }. As such, we can attach $N$ units of entanglement to each edge or hyperedge. 
The total RT entanglement across the cut is therefore the sum over the entanglement across each such edge, which is proportional to the number of hyperedge cuts. 
\end{hproof}
If we assign unit weight to each edge in the hypergraph ($w_e=1$), then the $N$-copy tensor network yields RT-contribution,

\begin{equation}
   S(\gamma_A) =N\sum_{\alpha}p_{\alpha}S(\chi_{\alpha}) = N |\gamma_{\rm A}|. 
\end{equation}

From section \ref{subsec:bs4}, it is clear that the RT term is independent of the logical state because $S(\chi_{0})=S(\chi_1)$ for the Bacon-Shor code. Therefore, the RT-FLM formula \ref{eqn:harlowFLM5} holds.

If the bipartition does not lead to complementary recovery, e.g. in Figure \ref{fig:noncompdecodestep}, then it is clear that the previous argument with the Greedy algorithm breaks down. In particular, we can no longer associate distilled Bell-states or GHZ states to the edges that are cut in a straightforward way. However, it is possible to show on a case-by-case basis that minimum cut through the graph can still produce the correct prediction for the entanglement entropy when the bulk state is a computation basis state. We thus conjecture that minimum cuts in the hypergraph correctly captures the entanglement structure of the tensor network in the $ZZ=+1$ gauge.

\section{Approximate Hybrid code and Gravity}
\label{sec:6apphybrid}
While the QECC constructions~\cite{Pastawski:2015qua, Donnelly:2016qqt, Kohler:2018kqk}, including the one above, capture the features of bulk effective field theories on a fixed background, the actual bulk theory in AdS/CFT contains gravitational interactions as well. A defining feature of such interactions is back-reaction, where bulk matter content should ultimately affect the background geometry according to Einstein's equations or other generalizations. In particular, the quantum extremal surface \cite{Engelhardt:2014gca} will shift depending on the bulk matter configurations. This is a feature which is absent in the exact stabilizer code constructions, but can be found in the Renyi entropy calculations in \cite{Hayden:2016cfa}, where injection of bulk matter shifts the domain wall which computes the Renyi-2 entropy. 

In this section, we construct an approximate model of the hybrid holographic Bacon-Shor tensor network. We show that this tensor network exhibit features that may be interpreted as back-reaction\footnote{Note that we do not claim such back-reaction is necessarily related to Einstein gravity, which will likely require more fine-tuning than having a generic model.} both from the point of view of entanglement entropy and that of quantum error correction. Just like inserting massive objects into the bulk will shift both the boundary anchored minimal surface and the associated entanglement wedge, we show that 
\begin{itemize}
   \item Exact complementary recovery from subregions is generically broken in the approximate code, as we expect for gravitational interactions. Depending on the bulk state, there can be a ``no man's land'' that is not exactly accessible from either boundary subregion. 
   
   \item Although exact recovery of certain bulk/logical information is impossible from a subregion, it is possible to recover a subspace/subalgebra of the original code exactly.

   \item The exact recoverability of a bulk subalgebra is state or subspace-dependent. In other words, the Greedy wedge, which serves as an analog of the entanglement wedge, can depend on the bulk/logical state. Consequently, the Greedy geodesic, playing the role of minimal surface also shifts depending on the bulk state. Similarly, the representation of a bulk operator on the boundary will also depend on the bulk state.

    \item The von Neumann entropy associated with the RT surface as previously defined can depend on the bulk quantum information. Because these entropies are tied to minimal surface areas, this also implies changes in the background geometry depend on the bulk state, similar to gravitational back-reactions. 

\end{itemize}

We will also discuss how the approximate code exhibit a few other potential connections to gravity and gauge theory. 
We find that for generic approximate constructions, the bulk degrees of freedoms are weakly interdependent meaning the bulk degrees of freedom are not truly localized. However, it is possible to find states or subspaces in which the localization of some logical degrees of freedom can be made exact. This is similar to the consequence of coupling a local bulk effective field theory to gravity, where the locality of bulk degrees of freedom are approximate and state-dependent. The degree of such interdependence, the strength of back-reaction, and modifications to subregion duality as well as complementary recovery are all controlled by the level to which the approximate code deviates from the reference code.  We label it as a skewing (or noise) parameter $\epsilon$, which seemingly plays the role of $G_N$. In the limit of $\epsilon\rightarrow 0$, we recover the hybrid Bacon-Shor code in section~\ref{sec:5hybrid} where all the above effects vanish. This is analogous to the $G_N\rightarrow 0$ limit in which we simply recover a local quantum field theory on curved spacetime where the fields are decoupled from gravity.

Lastly, we construct a simple toy example using the approximate code that supports power-law correlation in connected two-point functions without injecting correlations in the bulk state. 

\subsection{General Constructions}
\label{subsec:skewedhybrid}
For the approximate code, we can replace the exact Bacon-Shor encoding isometries $W_{\stile}$ with its skewed counterparts $\eBST_{\stile}$ on the square tiles. Recall that this can be regarded as an approximate Bacon-Shor code whose encoding map is the superposition of exact stabilizer encoding maps with coefficients $\{\alpha_i\}$. 

\begin{equation}
    \eBST_{\stile} = \sum_{i_{\stile}=0}\alpha_{i_{\stile}}W_{i_{\stile}}
\end{equation}

More generally, the perfect tensor can also be replaced by its skewed counterpart

\begin{equation}
  |\ePT_{\htile}\rangle=\sum_{i_{\htile}=0}\beta_{i_{\htile}} |T_{i_{\htile}}\rangle,
\end{equation}
where $|\beta_{i_{\htile}\ne 0}|\ll |\beta_{0}|$ for small noises and $|T_0\rangle = |T\rangle$ is the original perfect tensor. The coefficients $\alpha_{i_{\stile}},\beta_{i_{\htile}}$ need not be the same for different tiles.  Without loss of generality, we take $|T_{i_{\htile}\ne 0}\rangle$ to be states orthogonal to the perfect tensor state. 

At each square tile $\stile$ and hexagonal tile $\htile$, the tensors can be replaced by its approximate version. Then the tensor network describing this approximate quantum error correction code is 

\begin{align}
    V&=\mathcal{N} (\bigotimes_E \langle \Phi^+_E|) \bigotimes_{\stile,\htile\in D}(\eBST_{\stile}\otimes |\ePT_{\htile}\rangle )\\
    &= \mathcal{N} \sum_{C} \prod_{\stile, \htile\in D}(\alpha_{i_{\stile}} \beta_{j_{\htile}}) \langle E| \bigotimes_{\stile,\htile\in D} W_{i_{\stile}}\otimes |T_{j_{\htile}}\rangle\\
    &=\sum_{c\in C} \Lambda_c V_c
    \label{eqn:superTN}
\end{align}
where we use $D$ to denote the set of tiles of the tessellation with an IR cut-off. $C$ denotes the different configurations of contracting the codes. For example, for codes where each tensor consists of two terms in the superposition $i_{\htile},i_{\stile} \in\{0,1\}$ then $C$ is the set of all binary strings with length $|D|$. In the last line, we can repackage the sum as a superposition of tensor networks with encoding isometry $V_c$, each of which is an exact hybrid Bacon-Shor code with semi-regular tiling. However, each has a different choice of the code subspace, or equivalently, a different choice of stabilizer and gauge elements. However, this decomposition is highly non-unique. It is possible to decompose the skewed code into other types of code that have completely different tensor network representations.

\begin{figure}[ht]
    \centering
    \includegraphics[width=0.8\textwidth]{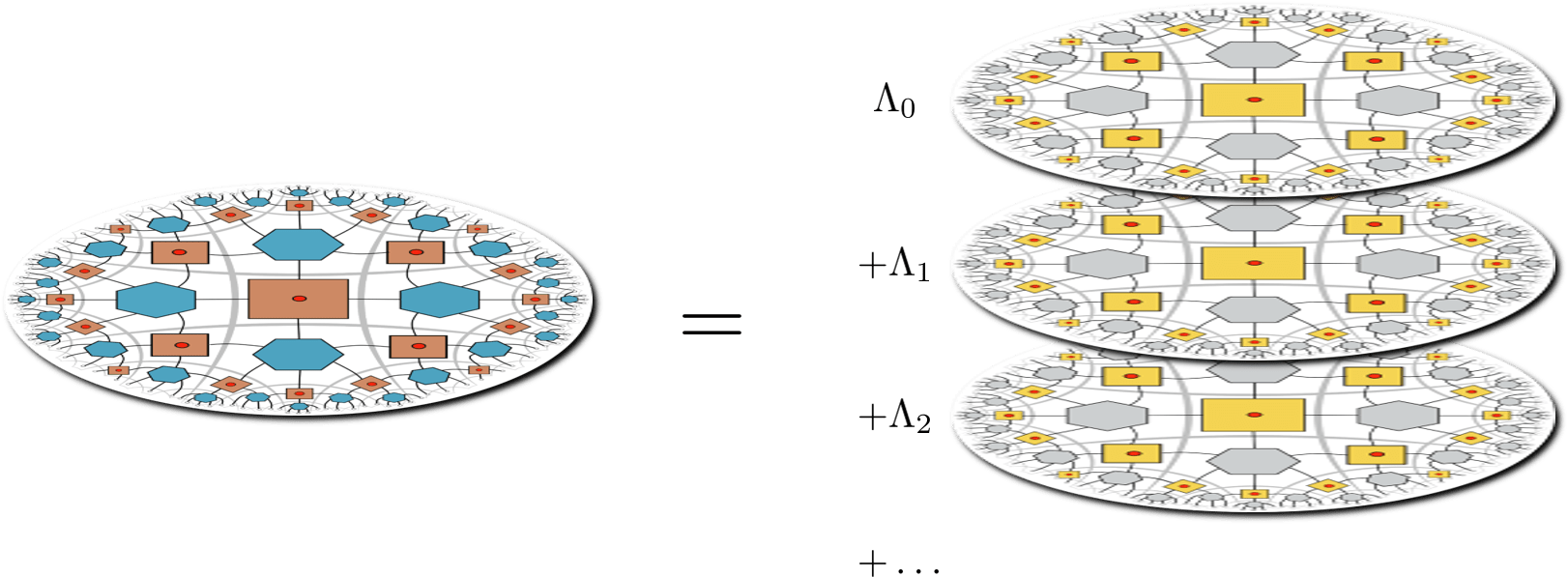}
    \caption{The skewed hybrid tensor network (left) can be thought of as a ``super tensor network''. It is a superposition of exact hybrid tensor networks, but each network can have a different set of stabilizers and gauge operators, hence pushing rules. The skewed tensors are coloured in blue or brown whereas the exact ones are coloured yellow or grey. This decomposition is non-unique, and the tensor network in each term of the superposition need not be the same.}
    \label{fig:superTN}
\end{figure}

It is possible to understand bulk-to-boundary operator pushing term by term in the superposition, where the pushing of each term is as given in section \ref{subsec:hybridpushing}. 
However, for such constructions, the resulting boundary operator for some localized bulk operator is global because of the projection operators, even though its counterpart in the reference code has support over only a subregion. 

Similarly, it is also useful to consider reference code representation when $|\alpha_i|, |\beta_j| \ll |\alpha_{0}|,|\beta_0|$ (for $i,j\not= 0$), where superposition has a most dominant exact hybrid code contribution. We take that to be the reference code, denoting the encoding map as $V_0$ and $\Lambda_0\approx 1$. Here let us assume that the other coefficients are nonzero but
\begin{equation}
    \sum_{c\ne 0}|\Lambda_{c}|<\epsilon
\end{equation}
are bounded by some small number $\epsilon$. This will be the noise or skewing parameter we use to control the degree of approximation in the skewed code. Then, logical operators of the reference code only preserves the code subspace approximately such that 

\begin{equation}
    V\bar{O} = O_0 V'\approx O_0 V.
    \label{eqn:skewTNpush}
\end{equation}
In such a case, all the results from the previous section on subregion duality still applies, albeit at a cost of a small error of order $\epsilon$. Hence, even if the exact bulk operator is global, we can still obtain an approximate logical operator which has support over a subregion using the reference code.
Relatedly, we may also use the reference code decoding circuit to extract encoded information. 

More explicitly, if $V$ is an encoding isometry such that $V^{\dagger}V=I$, then one can also derive representations of the boundary operator which manifestly reduce to the reference code operator in the $\epsilon\rightarrow 0$ limit,
\begin{align}
    V\bar{O} = V\bar{O}V^{\dagger}V = O'V,
\end{align}
where 
\begin{equation}
    O' = V\bar{O}V^{\dagger}= |\Lambda_0|^2 V_0 \bar{O}V_0^{\dagger}+\sum_{(c,d)\ne (0,0)} \Lambda_c\Lambda_d^* V_c \bar{O}V_d^{\dagger} = |\Lambda_0|^2 O_0 + \sum_{(c,d)\ne (0,0)} \Lambda_c\Lambda_d^* O_{cd}.
\end{equation}
If $|\Lambda_0|^2\gg |\sum \Lambda_c\Lambda_d^*|\sim O(\epsilon)$, then $O'\approx O_0$ at leading order, where we recover the reference operator. For each term, one can generate the physical operator using the exact hybrid code pushing rule to move $\bar{O}$ past $V_c$ isometry. The resulting $O'$ can then be construed as a superposition of operators where each term has a different support and representation on the boundary (physical Hilbert space). Clearly, $O'$ is supported on the union of boundary subregions over which $O_0, O_{cd}$ are supported.

Note that we must also be careful about the size of such errors incurred by the reference operator, which is related to terms in the superposition where $c\ne 0$. Suppose each skewed tensor is related to their exact counterpart by $\delta$, then each term with $\Lambda_{c\ne 0}\sim O(\delta)$ is of at most order $\delta$. However, the cumulative noise terms can be of order 1 by having a large number of contracted approximate tensors. To ensure the overall tensor network is close to the reference code, we need $\delta n\ll 1$ where $n\sim |D|$ is the total number of approximate tensors. 

Although we anticipate the subregion duality in the exact hybrid code to be generically broken by the skewed hybrid tensor network because of additional terms in the superposition, it can be controlled by the size of $\epsilon$. In the limit of $\epsilon\rightarrow 0$ we restore the original hybrid tensor network and its properties. 

We will pause the discussion of generic skewed code. In the following sections, we will consider specific constructions for the sake of building intuition.

\subsection{Breakdown of complementary recovery and Exact Subspace Recovery}
\label{subsec:subspacerec}
\subsubsection{Code Properties}
For concreteness, we again consider the example in (\ref{eqn:ghzcode}) where the skewed Bacon-Shor tensor is GHZ-like. Recall that for this tensor, bulk-to-boundary pushing is covered in the earlier sections, where the exact logical $\tilde{X}$-operator is now supported on all 4 legs and is no longer transversal while the reference logical $\tilde{X}$ incurs a small error penalty on the encoded state by taking it slightly out of the code subspace. On the other hand, logical $\tilde{Z}$ operators are identical to that of the reference code by construction, and are supported on any two legs of the skewed Bacon-Shor Tensor (SBST). The logical state $|\tilde{0}\rangle$, is stabilized by all Bacon-Shor code stabilizers, whereas other logical states are only stabilized by $Z$-type stabilizers.

\begin{figure}[ht]
    \centering
    \includegraphics[width=\textwidth]{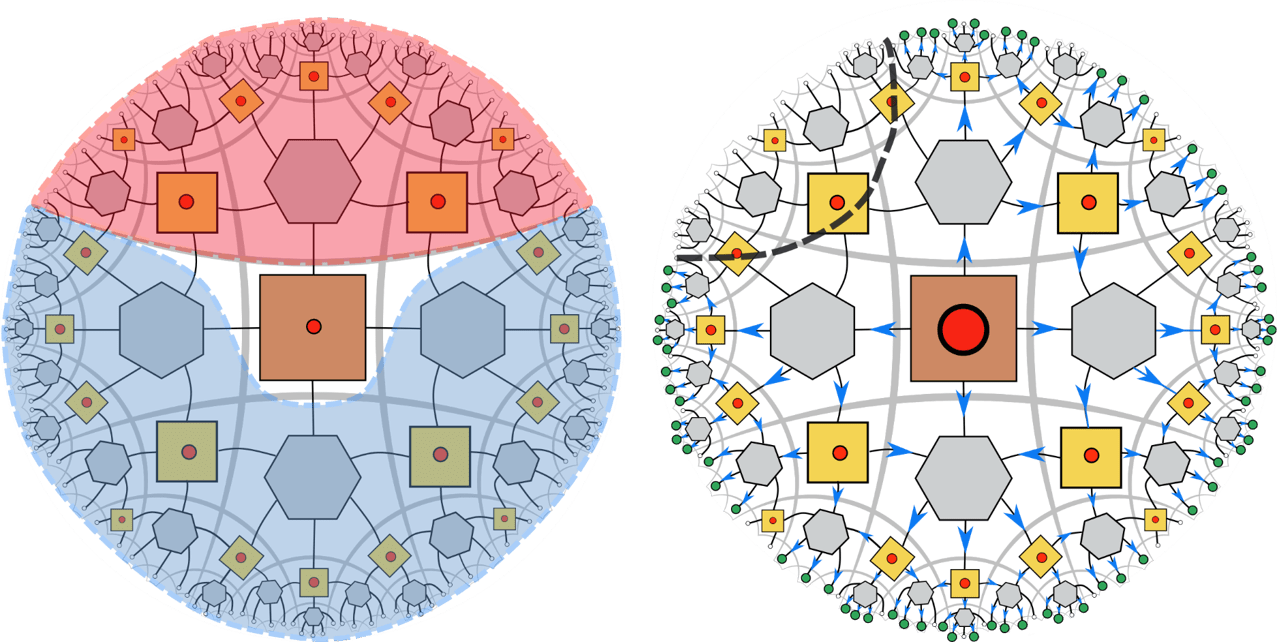}
    \caption{Left: When the central tile is approximate, its information cannot be recovered with perfect fidelity on either subregions. Greedy/entanglement wedges indicate where bulk information can be fully reconstructed. However, when the central qubit is in particular states, or equivalently, when the state of bottom wedge lives in a subspace $C_{a_i}$, then its information can be recovered on a subregion perfectly. Right: When we have enough support over the boundary, then the central qubit can be recovered exactly.}
    \label{fig:starlikeshadow}
\end{figure}

To start with a simple example, we take an exact hybrid tensor network but have replaced one BST by the SBST. The operator pushing \textit{from} this block is again given by Figure \ref{fig:weight4pushing}, and the right of Figure~\ref{fig:starlikeshadow}. The central bulk qubit now has support over a much larger boundary subregion. Thus it is clear that the bulk logical X cannot be exactly reconstructed by the subalgebra associated with the previous connected subregion (Figure \ref{fig:singlepush}). For the same bipartition in the left of Figure~\ref{fig:starlikeshadow}, this immediately leads to a breakdown of exact complementary recovery, which now only holds approximately, controlled by the degree of noise $\epsilon$. By contrast, the code's $Z$ subalgebra tied to the central tile can be pushed to the blue boundary subregion because its operator pushing was unchanged compared to the hybrid tensor network. 

Therefore, from the algebraic point of view, the physical degrees of freedom shaded in blue can access the full code subalgebra in the blue wedge as well as the $Z$ subalgebra of the unshaded region. They do not, however, access the full Pauli algebra of the unshaded region. This is  different from the exact hybrid tensor network construction because recovery is complementary (Figure~\ref{fig:singlepush}), and the bottom degrees of freedom also accesses the central bulk qubit fully. In some sense, in the bipartition where the red dashed line divides the top and bottom degrees of freedom, the bottom boundary subregion now only accesses a classical bit of the central tile exactly, whereas it accesses a full quantum bit in the hybrid tensor network. 

Equivalently, we can also see this from a state recovery point of view. It is straightforward to show that, using the Greedy algorithm, the decoding unitary for the original hybrid tensor network can extract the logical information of the central qubit when it is in the state of $|\tilde{0}\rangle, |\tilde{1}\rangle$ but not their arbitrary superposition\footnote{By extension, one can also recovery thermal states diagonal in the computational basis.}. This is because the decoding unitary for the BST also recovers the encoded information of the SBST when its logical states are 0 or 1, as shown in section~\ref{subsec:SBSdecode}. Therefore, we can still recover the encoded information in a complementary way provided that the SBST is in specific states.

We can summarize these observations as follows: although having access to a boundary subregion does not allow a universal reconstruction of the bulk code space $\mathcal{C}$ to perfect fidelity, it does allow us to reconstruct a subspace of $\mathcal{C}$. This resembles a version of the subspace recovery in \cite{HaydenPenington2017}, although here we do not show the reconstruction is universal. 

Let us explicitly construct these exactly recoverable subspaces. Let $C_a\subset \mathcal{C}$ be the subspace of the code subspace $\mathcal{C}$ such that it contains all the logical qubits in the complement of the red wedge. We can further decompose $C_a$ into two subspaces

\begin{equation}
   C_a = C_{a_0}\oplus C_{a_1}
\end{equation}
where the subspaces are defined as

\begin{equation}
    C_{a_i} = \mathrm{span} \{|\tilde{i}\rangle_{\rm SBST}\}\otimes C_{\mathcal{W}_{a\setminus o}}\} \text{ (for $\tilde{i} = 0,1$).}
    \label{eqn:recvsubsp}
\end{equation}

Here we use $C_{\mathcal{W}_{a\setminus o}}$ to denote the bulk subspace in the entanglement wedge $W_a$ of $A$ excluding the qubit associated with the SBST. Logical information is exactly recoverable when it lies in either subspace $C_{a_i}$.

Geometrically, because neither wedge can recover the logical information with perfect fidelity when the SBST is initialized. This yields a ``no-man's land'' in the central region. Therefore, generic complementary recovery is manifestly broken. However, if the central qubit is initialized in 0 or 1, which ensures the logical information of the bottom wedge to be in either of $C_{a_i}$, then the full information of the wedge can be recovered. In a way, here we have a limited form of complementary recovery, which we call \textit{state-dependent complementary recovery.} We will further discuss this property in section~\ref{subsec:SEERT}. At the same time, we see that the shape of these Greedy (entanglement) wedges are also state-dependent. The wedge contains the unshaded region when SBST is 0 or 1, but excludes it for other bulk states. Thus the overall Greedy wedge in this case (blue) is nothing but the intersection of all Greedy wedges that correspond to different bulk states\cite{Akers:2019wxj}.

Now, suppose we have access to a boundary subregion that is larger than the blue shaded subregion, then we can also recover the full central qubit or its Pauli algebra when that accessible region is large enough. For example, when we access a boundary subregion that is bounded by the black dashed curve below and to the right in Figure~\ref{fig:starlikeshadow}. This is because the region now has access to the prohibiting $\tilde{X}$ operator which has support over 4 legs. 

Thus far, we have been discussing the exact retrieval of encoded logical information. However, in the case where skewing is small, $|\alpha-\beta|<\epsilon$, it is clear that the reference code decoding unitary and logical operations have approximate complementary access to the encoded information. In this case, the logical $\tilde{X}\approx \tilde{X}_0$ operator is supported in the blue region up to errors of order $\epsilon$. Similarly, the central bulk qubit's information can be decoded approximately up to a small error $\epsilon$ regardless of the state. In the limit of $\epsilon\rightarrow 0$, we see that the state-dependence completely vanishes as we approach the reference code.

\begin{figure}[ht]
    \centering
    \includegraphics[width=\textwidth]{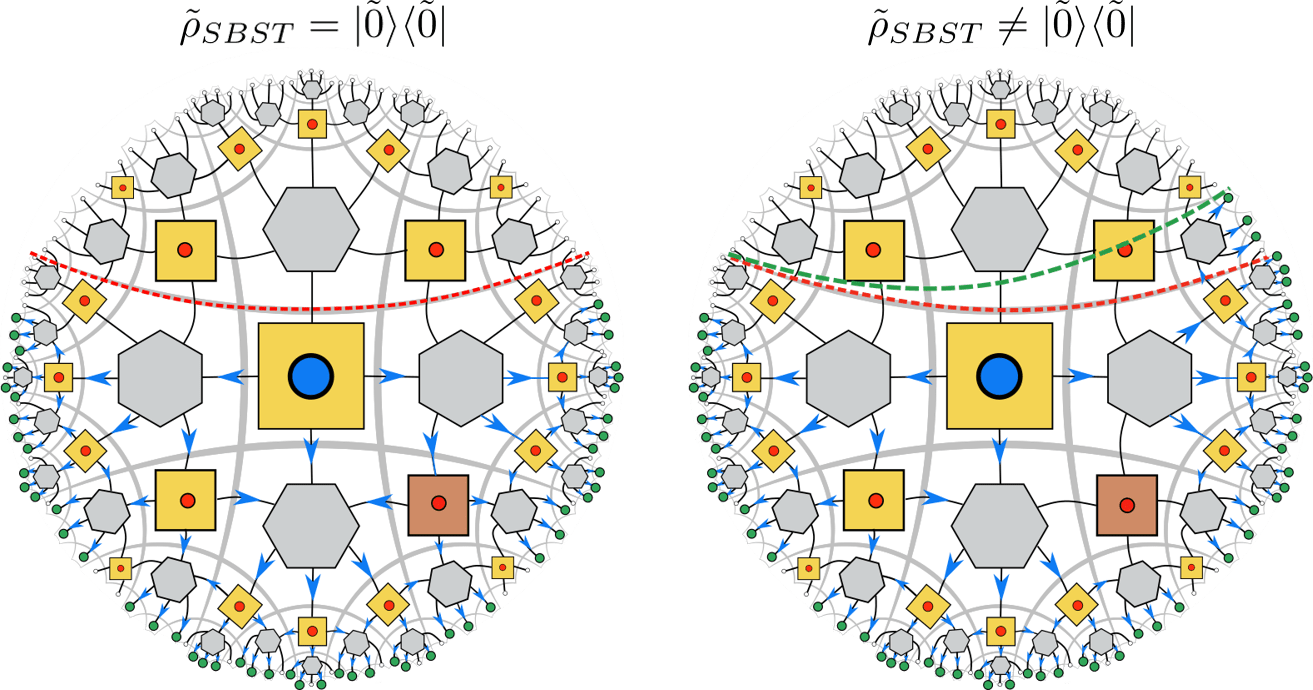}
    \caption{The darkened tensor correspond to the insertion of SBST. Operator pushing rules are different depending on the state. The central qubit can be fully accessed in the subregion below the red curve when the marked tile is in the 0 state. However, a larger subregion bounded by the green curve is needed, if the marked tile is in an another state.}
    \label{fig:shadow}
\end{figure}

 Let us supplement these observations with yet another example where we have the same bipartition of the boundary degrees of freedom in Figure \ref{fig:starlikeshadow}. However, instead of inserting the SBST in the central tile, we can insert it at the tile on the diagonal (Figure \ref{fig:shadow}). By symmetry, this is related to our previous configuration by passively moving the tessellation along the diagonal. We then consider the behaviour of the operator of the central BST when it is pushed to the boundary. Although naively it would seem like the physical representation of its code subalgebra on the boundary should not be affected, because its $\tilde{X}$ operator is not globally supported (on all 4 legs), it will be affected if we again restrict ourselves to certain boundary subregions. 

Again, consider bulk degrees of freedom below the red curve. Let us define subspaces $C_{a_i}$ in the same way for the bulk qubits in this region (\ref{eqn:recvsubsp}). Now suppose we want to push any bulk logical operator on the central tile to the boundary subregion below the red dotted curve, call it $A$, the flow of operator pushing needs to go through the marked tile. For states in the subspace $C_{a_0}$, because both $XXXX$ and $ZZZZ$ stabilize $|\tilde{0}\rangle$ of the SBST, by the pushing rules we discussed earlier, any single-site operator can be pushed across this tensor while preserving the logical information. Thus we find that any logical operator on the central tile will have support over the subregion $A$ (left of Figure~\ref{fig:shadow}). 

On the other hand, if the state is in the logical subspace $C_{a_1}$, then such pushing is no longer possible without errors\footnote{It may be possible to push through the marked tile if the single input is a $Z$ type operator. However, this requires the adjacent perfect tensor to be arranged such that it outputs only $Z$ or $I$ for any input from the central tile. This requires us to fine-tune the model by carefully choosing orientations of, for instance, the central tensor, which we do not consider here.}. The flow of operators must circumvent the marked tile, which results in a larger subregion of support. It is possible, however, to push through the marked tile regardless, but at the expense of error. If the error is sufficiently small, then the central operator can still be supported over $A$ but only approximately. Therefore, if the marked tile is in an arbitrary state, e.g. a superposition $a|\bar{0}\rangle+b|\bar{1}\rangle$ or a mixed state, then the above pushing flow for the purple bulk operator must avoid the X-marked tile unless the inputs are strictly $Z$.

We see that although the logical algebra of the central tile, which is supported on a BST, can be accessed on the boundary, their physical representations on a subregion can be very different depending on the state of the SBST. Indeed, like the previous example, different bulk information can lead to different accessibility on the boundary subregion. In this case, the region $A$ only accesses the full Pauli algebra of the central bulk qubit when the SBST is in the 0 state. In other words, given subregion $A$ on the boundary, we only recover the information in $C_{a_0}$. 
On the other hand, the latter example is also slightly different. The boundary representation of the \textit{same} bulk operator can be completely different depending on the state of the SBST \textit{somewhere else} because they need to be pushed differently.

\subsubsection{Connection with Holography}
In holography, having a massive object in the bulk will also deform the entanglement wedges and minimal surfaces, similar to our above examples when the SBST bulk qubit is initialized in a generic state.
As gravity deforms the spacetime around a massive object, the minimal surfaces (in this case geodesics) in general skirt around the object. Geodesics anchored to the same boundary points no longer coincide. As these surfaces bound the entanglement wedges, it creates a no man's land in the bulk, where a region lies in neither of the entanglement wedges (Figure \ref{fig:starlikeshadow}). 

However, when the bulk state lies in a particular subspace of $\mathcal{C}$, e.g. when there is no bulk matter or the mass of the matter is small, its Greedy wedge is consistent with the vacuum entanglement wedge configuration. Indeed, because of the state-dependence of such entanglement wedge, the Greedy wedge for a subregion is the \textit{intersection} of all Greedy wedges with different bulk states\cite{Akers:2019wxj}. This is precisely the same in holography, where we replace Greedy wedge with entanglement wedge\cite{Pastawski:2015qua, Dong:2016eik}. 

By analogy, the $\tilde{X}=\tilde{X}_0+O(\epsilon)$ operator may be interpreted as injecting matter into the bulk, similar to one that creates a gravitational charge; see for instance (\ref{eqn:exactSBSTpushing}) for a such a breakdown. While $\tilde{X}$ has support over a much larger boundary subregion, the bare operator $\tilde{X}_0$ can be recovered up to $\epsilon$ error in a smaller region consistent with the reference code entanglement wedge, where $\epsilon \sim G_N$ is zero. The residual terms in $O(\epsilon)$ that contributed to this error, are also responsible for $\tilde{X}$ having a larger support over the boundary. These order $\epsilon$ correction terms were not included in the exact QEC limit where there is no ``back-reaction''. We may treat them as ``dressing'' for the bare operator. Here $\epsilon$ plays the role of $G_N$, while a gauge invariant operation is analogous to one that preserves the code subspace.

Taking this analogy further, we can interpret this skewed code subspace to model a bulk low energy subspace that also includes different mass distributions on the central tile. Here, $\tilde{X}$ is the dressed operator that inserts a mass in the central tile. Given the particular subregion, we can not access the full Pauli algebra because the actual mass operator has Wilson lines that extend to the complementary subregion. As a result, the Greedy/entanglement wedge does not contain this region. On the state/subspace level, if the central tile is in an arbitrary unknown state, then the Greedy wedge does not contain this region either, similar to the case of a mixed state black hole, where the minimal surfaces skirt around the black hole horizon. However, if we are promised that the central tile is in a particular state, then it is possible to recover the information in this region, similar to being promised a particular black-hole microstate, such that the black hole is now contained in the entanglement wedge~\cite{Hayden:2018khn}. In our case, the role of thermal state black holes can be played by generic thermal states that are not diagonal in the computational basis or states in a superposition of $|\tilde{0}\rangle$ and $|\tilde{1}\rangle$.\footnote{ For a fixed entanglement entropy of the SBST, there can be multiple logical states that are consistent with such RT entropy. Hence from an outside observer point of view, the exact microstate that generated this entanglement geometry is unknown. However, $|\tilde{0}\rangle, |\tilde{1}\rangle$ are the only states where the entanglement-geometry uniquely determines the microstate itself.} While it may seem strange that a pure state also causes the surfaces to diverge, we should recall that for this code, any subregion can only access at most the code's $Z$ subalgebra. The superposition of 0 and 1 reduces to a mixed state on this code subalgebra.

The observation of exact subspace recovery also has its analogue in holography. In some sense, the code subspace here is ``too large'' in that it contains states with different bulk mass distributions and hence different geometries. Because for some fixed bipartition, these states have different entanglement wedges, a given boundary subregion only recovers the intersection of these entanglement wedges. However, if we shrink the code subspace by considering only a subspace, e.g. $C_{a_0}$, that contains only one background geometry, say that of the AdS vacuum, then bulk information can be accessed in the same way as the exact holographic quantum error correction codes where $\epsilon\sim G_N=0$.

In the second example, we see that it reflects many of the similar properties of the first example, such as the state-dependence of the entanglement wedge and recoverability. Moreover, we explicitly see that the same logical operator has (bulk) state-dependent boundary representations because of the SBST insertion. 
Thus, the SBST can also create a ``shadow region'', such that operator pushing of the nearby tiles also depend on the SBST bulk state. This again is similar to what we expect as gravitational back-reaction when there is a massive object such as a star or black hole. In order to reach the boundary, the bulk operator has to be pushed around this inaccessible region created by the SBST. This is a direct consequence of the changes in entanglement wedges for different subregions when the SBST is initialized in different bulk states.

\subsection{Towards More General Approximate Codes}

So far, we see that there is a clear case of state dependence when we move from exact error correction codes to approximate ones. However, in the previous example, we only allow one skewed tile to ``back-react''. More generally, we can replace more BSTs with SBSTs.

\begin{figure}[ht]
    \centering
    \includegraphics[width=\textwidth]{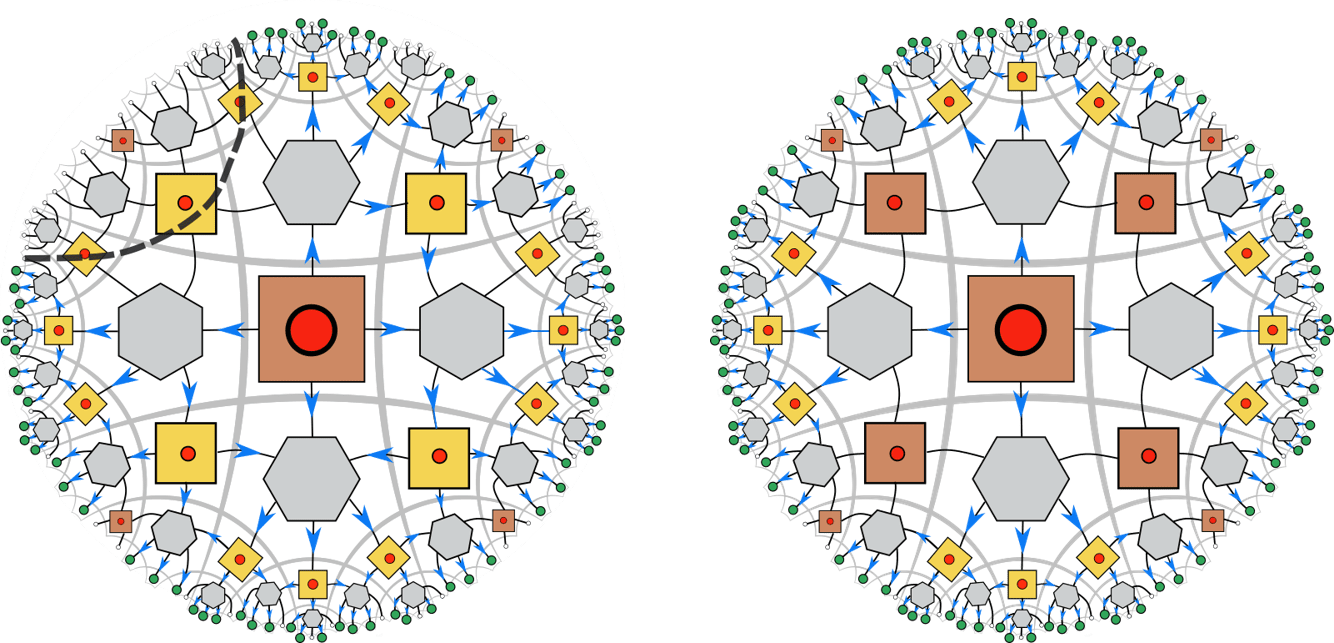}
    \caption{Inserting SBSTs sparsely into the tensor network can still allow operators to be pushed to the boundary independently. Therefore, each bulk qubit can be accessed independently from the boundary. However, more SBST imposes more obstruction for pushing. Left: operator can be pushed to a boundary subregion. Right: the operator can be accessed globally, but not in the previous subregion. The pushing assumes SBSTs are intialized in non-zero logical states.}
    \label{fig:doping}
\end{figure}

For example, consider a construction where more SBSTs are added. If they are placed sparsely in the bulk, then it is easy to check that all of the bulk logical operators, including those of the SBSTs, can be pushed to the boundary while preserving the logical state. This implies the apparent bulk qubits, as shown in the tensor network, are indeed independent degrees of freedom that live in tensor factors of the bulk Hilbert space, which are accessible from the boundary. In other words, one can perform logical operations on one encoded qubit without affecting another if one has access to the global boundary. One such construction is shown in Figure \ref{fig:doping}. BSTs are coloured in yellow, while SBSTs are brown. The perfect tensors are exact (unskewed).

Each of the added SBSTs here will contribute to a higher degree of state-dependence when it comes to bulk reconstructions. This is by extending the same reasoning from the previous section, where any single site operator can be pushed through the SBST if it is initialized in the logical 0 state, whereas operators (except for $Z$) must be pushed around it. Clearly, having more SBSTs in the tensor network creates more state-dependent obstructions for bulk operators to reach a particular designated boundary subregion, thus impacting their boundary representations.

\begin{figure}
\centering
\begin{subfigure}[b]{0.55\textwidth}
   \includegraphics[width=1\linewidth]{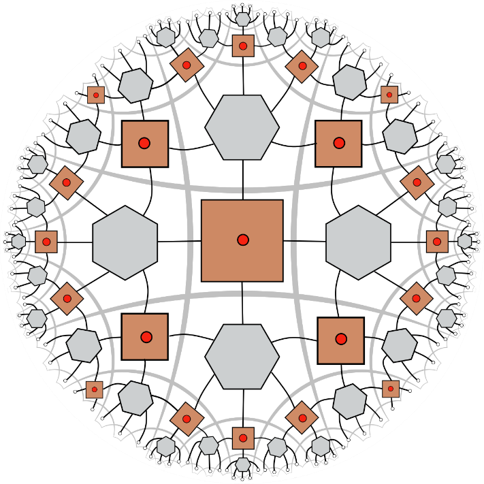}
   \caption{}
   \label{fig:allSBSTTN1} 
\end{subfigure}

\begin{subfigure}[b]{0.55\textwidth}
   \includegraphics[width=1\linewidth]{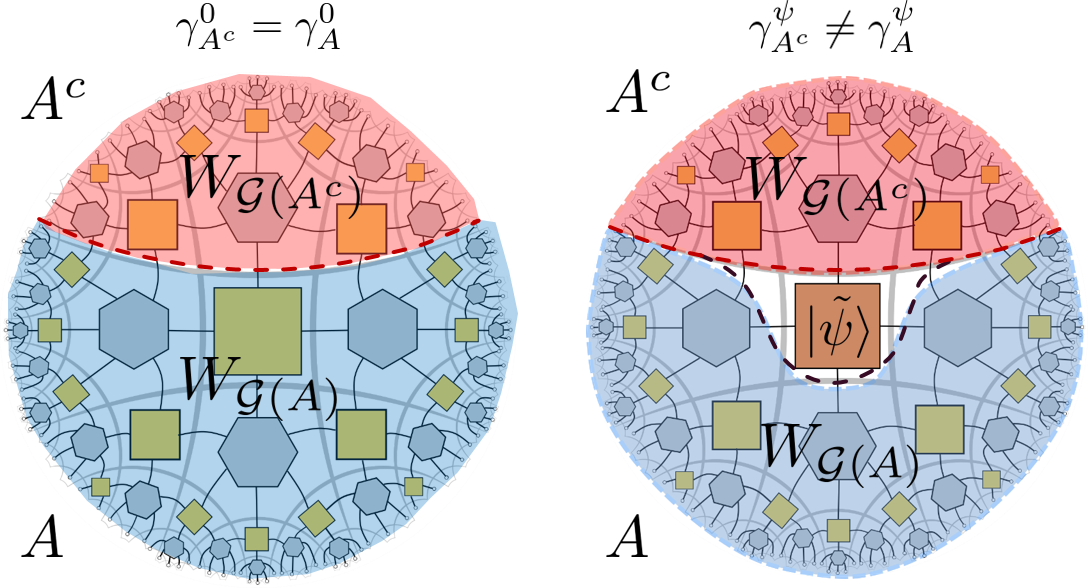}
   \caption{}
   \label{fig:allSBSTTN}
\end{subfigure}

\caption[Two numerical solutions]{(a) Noisy hybrid code with skewed Bacon-Shor tensors (brown). (b) Dashed curves $\gamma_{A,A^c}^{0, \psi}$ are ``geodesics'' that are anchored on the endpoints of boundary regions $A$, $A^c$. Yellow squares denote the $|\tilde{0}\rangle$ logical state whereas a brown square denotes some superposition $|\tilde{\psi}\rangle=c|\tilde{0}\rangle+d|\tilde{1}\rangle$ or a generic mixed state. The wedges in the left figure also hold for a central tile in the $|\tilde{1}\rangle$ state.}
\end{figure}

When the SBSTs are sufficiently dense in the tensor network, it will no longer be possible to recover the ``bulk information'' even globally.  
For example, consider a scenario where all BSTs are replaced by their skewed counterparts (Figure~\ref{fig:allSBSTTN}a).
Intuitively, the bulk subspace now contains states that correspond to a large number of different background ``geometries'', whereas we were limited to 2 in the previous section. This is visible from both the Greedy wedge and the RT contribution to the entanglement entropy, which we will discuss below and in section \ref{subsec:SEERT}. 


For example, the analogous vacuum AdS configuration is now captured by the all zero logical state $|\tilde{0}\rangle^{\otimes k}$ where $k$ is the number of encoded qubits. The analogous entanglement wedge reconstruction (Figure~\ref{fig:allSBSTTN}b left) recovers what we expect for the usual non-backreacting configuration. However, these entanglement wedges clearly depend on the bulk state (Figure~\ref{fig:allSBSTTN}b right), where a generic state leads to a deformed entanglement wedge bounded by geodesics that skirts around the ``mass'' inserted in the bulk central region. 

One key difference from holography is that such kind of ``back-reaction'' also seems to happen for pure states, whereas in holography it is not the case. This difference is in part because of the simplicity of the model where at best only the Z subalgebra is supported on any proper subregion of the skewed Bacon-Shor code as opposed to a full quantum degree of freedom. ``Back-reaction'' requires that different code subalgebras to have different degree of support on the physical qubits. In the skewed Bacon-Shor construction, it is manifested through the X code subalgebra being supported on 4 legs while Z code subalgebra is supported on 2. Recovery of the bulk information is only complementary on $A$ and $A^c$ when the logical state we consider also reduces to a pure state in the code subalgebra supported on a subregion $A$. Therefore, in this model, only the two ``classical states'' $|\tilde{0}\rangle,|\tilde{1}\rangle$ satisfy such purity conditions and are effectively functioning as the pure state counterparts in holography. On the other hand, a pure state like $|\tilde{\psi}\rangle$ is behaving like a mixed state in holography. This is because when restricted to a subregion $A$, we must consider its representation on the $Z$ subalgebra. Explicit examples can be found in Sections~\ref{sec:3baconshor} and \ref{sec:4aqec}, where even for unentangled pure states like $|\tilde{\psi}\rangle$, there is entropy contribution from bulk ``matter''. This is because $|\tilde{\psi}\rangle$ when reduced to the $Z$-subalgebra is mixed. A model more similar to what we see in holography can likely be produced by considering more sophisticated versions of such models where there is a larger code subalgebra.

Note, in particular, that the no man's land in the central region is not because of the failure of the Greedy algorithm, but rather a real physical effect because it is a fundamental obstruction for recovery created by the approximate Bacon-Shor code. Indeed, as we have shown that the approximate 4 qubit code only corrects an erasure approximately, it is impossible to recover the any logical information from any 3 qubits with perfect fidelity\footnote{, Recall that although the skewed Bacon-Shor tensor can recover the encoded information from any 3 legs when the logical state is $|\tilde{0}\rangle$ or  $|\tilde{1}\rangle$, it does not do so with perfect fidelity for other states. Therefore it corrects one erasure approximately.}. The situation here with the larger tensor network is identical --- because all surrounding tensors are isometries, subregion $A$ fully supports 3 legs of the tensor on the central tile while $A^c$ supports 1. Therefore the failure to recover the central region is directly inherited from the approximate error correction property of the approximate Bacon-Shor code.

One can repeat this experiment for other states by replacing the yellow square tensors with brown tensors, thereby obtaining different entanglement wedges and Greedy geodesics for each such state. Therefore, these different states are taken to be different geometries. 
Clearly, state dependence holds to a much greater extent where operator pushing and, equivalently, information recovery depends on the state of every single SBST. For example, the operator pushing from a central tile to a boundary subregion is the same as before if all other states are initialized in 0. However, all surrounding SBSTs can form obstacles to pushing if they are in other states. Thus, instead of depending on the state of a single bulk tile, now operator pushing depends on each square tile. We also expect the RT entanglement for different logical computational basis states to be different, because we are tracing together SBSTs that have different RT entanglements in different states. 

There are also richer structures that were absent in the previous examples. We make comments on three major properties.
\begin{enumerate}
    \item Interdependence of bulk qubits that appear independent in the tensor network geometry. As an important corollary, state or subspace dependence of operator reconstruction now can occur even when we have global access to the boundary.
    \item Existence and identification of subspaces where the bulk degrees of freedom are independent and exactly recoverable. This is a generalization of the statement we made in the previous section. 
    \item The state-dependent obstruction to operator reconstruction where bulk ``charges'' can shield the interior information from reaching the boundary via error accumulation. More specifically, decoding the bulk qubits that are closer to the center of the tensor network, i.e., logical qubits that have a larger distance will incur a larger error if the surrounding states are not $|\tilde{0}\rangle$. 
\end{enumerate}

These observations have analogues in gravity, where the locality of bulk degrees of freedom is approximate thanks to the gravitational dressing. The degree of locality, however, can be made precise when we turn off gravity, i.e., in the limit where $G_N\rightarrow 0$, which is similar to the limit of $\epsilon\rightarrow 0$ in the reference code limit.

\begin{figure}[ht]
    \centering
    \includegraphics[width=\textwidth]{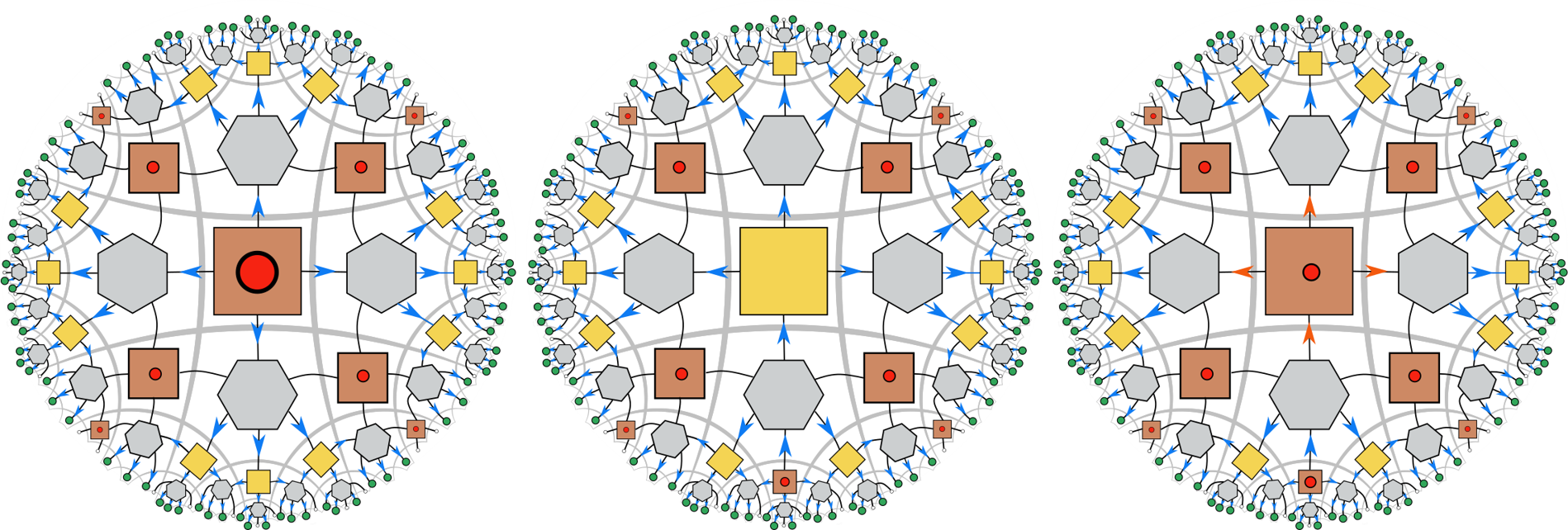}
    \caption{We can identify different bulk subspaces that are recoverable from the boundary. Each individual subspace (left, center) is exactly recoverable, however their union (right) is not.}
    \label{fig:dense_global}
\end{figure}

For an all $|\tilde{0}\rangle^{\otimes N}$ bulk logical state, the bulk information can clearly be recovered from the reference code decoding unitary because the codewords from our approximate code and that of the reference code are identical. There is also no obstruction to pushing operators from a single bulk site to the boundary. In other words,  Pauli algebra of a bulk qubit can be accessed globally on the boundary and on certain boundary subregions as allowed by the operator pushing rules and associated subregion duality. 
In fact, we can also access some of the bulk $\tilde{X}$ operators provided they are sparsely distributed so that these operators can be pushed independently to the boundary without interfering with each other. This is equivalent to what is shown in Figure~\ref{fig:doping}, except the 0 states in the SBSTs are playing the roles of BSTs. Of course, with regard to operator pushing, SBST only behaves like BSTs in a state-dependent manner. As such, we denote the 0 state of an SBST as a yellow square without the bulk leg to indicate that it is a state (Figure~\ref{fig:dense_global}).
Therefore, as long as sufficiently many bulk states are initialized in $|\tilde{0}\rangle$, it is possible to access the full Pauli algebra of some of the bulk qubits exactly. However, the boundary subregions on which they are supported can differ drastically from the reference code. Equivalently, for any fixed boundary subregion, the corresponding Greedy wedge can differ from that of the reference code by some region of the AdS scale. 

Interestingly, as we flip more states on the SBST from 0 to some other state, the global reconstruction of some bulk operator can also be state-dependent. For example, compare the left and right figure of \ref{fig:dense_global}. Suppose we wish to represent the central bulk operator on the boundary, there exists an exact representation on the left figure when some of SBSTs are initialized in the 0 state. However, this is no longer possible when one of the SBST 0 states are rotated into an arbitrary state (right figure \ref{fig:dense_global}). 

By now, it should be clear that in the tensor network where all BST are replaced by SBSTs, the bulk degrees of freedom are actually inter-dependent. They no longer reside in independent tensor factors of the code subspace and hence ``non-local''. This can be seen from our analysis of state dependence, where accessibility of certain bulk qubits will depend on others and vice versa.
Indeed, although any bulk qubit can appear ``localized'' to a bulk tile in the tensor network, this ``locality'' is only exact in the reference code limit with no back-reaction or if we choose some subspace in which the SBSTs in nonzero states are sufficiently sparse. If all of the bulk qubits are initialized in a non-zero state for the dense SBST model we discuss here, then all operator pushing have to go through some SBSTs to reach the boundary. As it is generically impossible to ``push'' a single site operator across the SBSTs while leaving the state invariant. This means that we can no longer find a boundary representation of any non-trivial logical operator that exactly preserves the encoded information. As such, performing an apparent local operation on one of the bulk qubits will inevitably affect our access to other bulk qubits.

Physically, this is very similar to the observation that locality is approximate when a local field theory is coupled to gravity. For generic bulk states, the logical degrees of freedom we have in the skewed code are only local up to some order $G_N$ correction and have global influence over other parts of the spacetime. We can also conclude from the following analysis that there are different subspaces where exact independent access is possible for some of these apparent bulk degrees of freedom. This reduces the locality to a subspace-dependent statement.

Subspace recovery has a much richer structure than before, as we can construct different subspaces that are exactly recoverable on the boundary. Graphically, they are represented by configurations where different subsets of the bulk qubits are fixed to the 0 state. A trivial example is the set of subspaces in which all but one SBSTs are fixed in the zero state. 
For another example, consider recovering bulk information from the entire boundary in Figure~\ref{fig:dense_global}. We see that it is possible to recover the bulk degrees of freedom if the yellow tiles are fixed in the 0 state. This is only a subspace of the global code subspace. However, there are more than one choice of such subspaces where different bulk qubits in SBSTs are independently accessible (left vs center). It is also clear that we can identify a number of similar subspaces by rotation, or by shifting the location of the brown tile in Figure~\ref{fig:dense_global} (center). In the gravity language, this means the bulk degrees of freedom are ``local'' in this subspace. 

Of course, these separately recoverable bulk qubits are not simultaneously recoverable from the boundary without error. This can be seen by taking both the central tile and one of the bottom SBSTs to be in arbitrary state (right of Figure~\ref{fig:dense_global}). The bulk operators cannot be pushed to the boundary without passing through a SBST, which incurs error (orange arrows). 

Interestingly, SBSTs that are not in the logical 0 state also act like ``charges'' that shields other bulk information from being recovered. This is because for each SBST not in the zero state, we can use the reference code stabilizer for pushing, which comes at a cost. Any operator that is ``pushed'' this way is only approximate, and adds some noise to the original quantum operation. The more SBST that the operator pushes through, the more noise it will incur by the time it reaches the boundary. Hence the level of accuracy at which the physical reference code operator acts on some particular bulk qubit depends on the minimal number of SBSTs the bulk operator has to pass through in order to reach the boundary. Because bulk qubits closer to the center have to pass through more layers of SBSTs to reach the boundary, they are also less accessible than the qubits near the boundary.
Because access to $\tilde{X}$ operators can create non-zero states, they behave like ``charge-creation operators'' by this analogy.

\begin{figure}[h]
    \centering
    \includegraphics[width=0.7\textwidth]{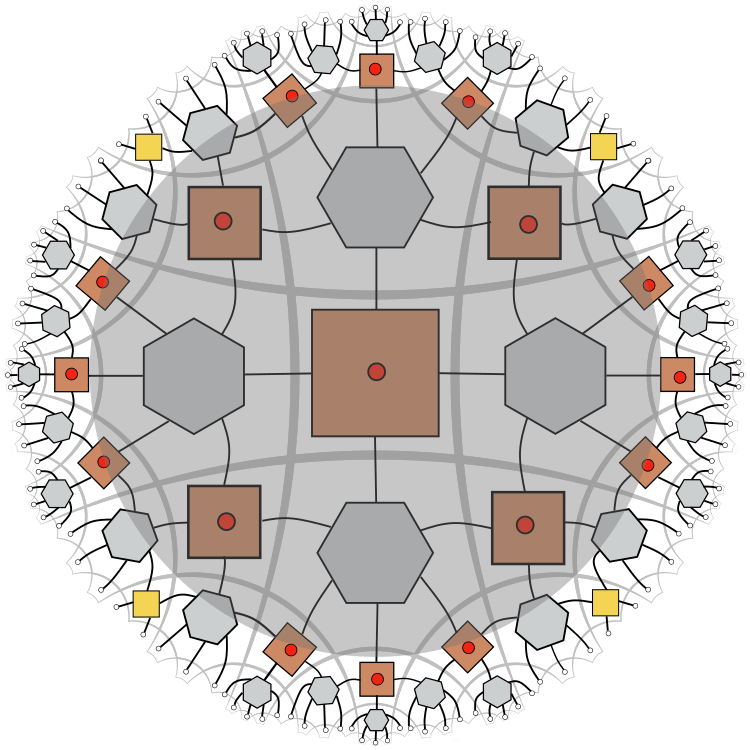}
    \caption{If the darker tiles have states in a non-zero state, then one can only access the $Z$ subalgebras on the boundary tiles of the shaded region. Interior of the shaded region cannot be accessed exactly. All other states are in the 0 state. The central qubit is ``shielded'' from the boundary as it has to go through SBSTs to reach the boundary. }
    \label{fig:not_bh}
\end{figure}

For example, consider Figure \ref{fig:not_bh}, where all the marked tiles are not in the vacuum state.  Indeed, the full Pauli algebra of these qubits cannot be accessed exactly even from the entire boundary. However, it is possible to access the $Z$ subalgebras exactly for those tiles that are on the boundary of the shaded region. In this way, the qubits deeper in the bulk are shielded from the exterior. These marked tiles represent ``charge excitations'', such that no information bounded by the shaded region can be accessed exactly, forming an ``exact-access horizon''. Nevertheless, the information can be accessed approximately using the reference code decoding, but such recovery procedure leads to a larger error for the bulk information closer to the center, as they are screened by more SBSTs, each of which introduces errors to the information recovered. 

In summary, when bulk states are initialized in 0 states, they behave like BST tiles with respect to operator pushing. However, a generic state of an SBST is a form of obstruction to operator pushing. Therefore, each SBST introduces additional state-dependence in this tensor network. While certain bulk degrees of freedom can be accessed independently when other bulk states are fixed to be 0, the same cannot be said when the latter apparent degrees of freedom are initialized in other states. As a result, it is impossible to change one apparent bulk degree of freedom without influencing another. This implies that even global reconstructions of certain bulk operators can heavily depend on the bulk state. In general, the bulk operators are approximately local, in that they commute up to $O(\epsilon)$ corrections, (as indicated by the reference code). Interestingly, although the apparent bulk degrees of freedom are not exactly independent, it is possible to access different subsets of them independently when we restrict ourselves to a subspace of the bulk information by fixing some tensors to the 0 state. This subspace corresponds to a configuration in which mass insertions are sparse and the general geometry is close to the vacuum state. We will refer to this as a subspace-dependent locality. From the Hilbert space point of view, it corresponds to identifying a subspace of the code subspace (e.g. an $\alpha$-block) where certain bulk degrees of freedom are in a tensor product even though the full code subspace may not be factorizable. For example, this is consistent with a code subspace of the form 
\begin{equation}
    \mathcal{C} = \bigoplus_{\alpha} C_{\alpha}
\end{equation}
where 
\begin{equation}
    C_{\alpha} = \bigotimes_{i_{\alpha}}^{N_{\alpha}} \mathcal{C}_{i_{\alpha}},
\end{equation}
similar to (\ref{eqn:MCdecomp}).

\subsection{Entanglement Entropy and the RT formula}
\label{subsec:SEERT}
We see that generally because $0$ and $1$ states in the SBST have distinct RT entanglement, the entanglement patterns of the tensor network will also depend on the bulk state. For example, by tracing together SBSTs in different bulk states, we also obtain RT entanglement that is state-dependent.  This dependence can also be made more explicit for codes (or subspaces) whose bulk degrees of freedom are independent (for instance, in Figure \ref{fig:doping}) using the Greedy algorithm.

Note that in \cite{Harlow:2016vwg}, the area operator defined in the RT/FLM formula depended on having complementary recovery in the error correction code. Here we only focus on the cases with a somewhat weaker form of complementary recovery. The reason for this requirement is mostly technical. While there may be other ways to define the RT entropy contribution, we follow the most straightforward path and define RT contribution the same way. Furthermore, we use the same argument in Sec~\ref{subsec:exactEERT} to prove a similar but state-dependent version of the RT/FLM formula for this tensor network, which uses complementary recovery in the assumption.

In this SBST construction, exact complementary recovery breaks down in general, especially when the SBSTs are initialized in arbitrary states. 
However, all is not lost, as there are subspaces in which information can be retrieved in a complementary manner. We have seen simple forms of this in sections~\ref{subsec:SBSdecode} and \ref{subsec:subspacerec}. Here we define \textit{state-dependent complementary recovery} as follows. 

\begin{definition}
We say that the code $\mathcal{C}\subset\mathcal{H}$ has state-dependent complementary recovery in a given bipartition
\begin{equation}
    \mathcal{H} = \mathcal{H}_A\otimes \mathcal{H}_{A^c},
\end{equation}
if there exists any state $\tilde{\rho}\in L(\mathcal{C})$ such that all encoded information of this particular state can be decoded on either $A$ or $A^c$. 
\end{definition}

Let us assume that we have such a decomposition where there is state-dependent complementary recovery. Then there is at least one state $|\widetilde{\alpha,ij}\rangle\in \mathcal{C}$, where for some decomposition of the code subspace

\begin{equation}
    \mathcal{C} =\bigoplus_{\alpha} \mathcal{C}_{\alpha}=\bigoplus_{\alpha} \mathcal{C}_{a_{\alpha}}\otimes \mathcal{C}_{a^c_{\alpha}},
\end{equation}
there exists decoding unitaries acting only on $A, A^c$ where
\begin{equation}
    U_AU_{A^c} |\widetilde{\alpha,i,j}\rangle = |\alpha,i\rangle_{A_1^{\alpha}}|\alpha,j\rangle_{{A^c}_1^{\alpha}} |\chi_{\alpha}\rangle_{A_2^{\alpha}{A^c}_2^{\alpha}}.
\end{equation}

If there exists a set of such states $|\widetilde{\alpha, ij}\rangle$, then any state
\begin{equation}
    \tilde{\rho} = \bigoplus_{\alpha} p_{\alpha}\rho_{\alpha\alpha}
\end{equation}
where $\rho_{\alpha\alpha}=|\psi_{\alpha}\rangle\langle\psi_{\alpha}|$ and 
\begin{equation}
|\psi_{\alpha}\rangle = \sum_{ij} c_{ij}|\alpha, ij\rangle
\end{equation}
for some coefficients $c_{ij}$, would satisfy the same RT/FLM formula \ref{eqn:harlowFLM5}. This is because the same derivations in (\ref{eqn:HarlowFLM}) applies here also.
In the case that $|\widetilde{\alpha, ij}\rangle$ spans the entire code subspace, we recover the original definition of complementary recovery. 
 Therefore, the RT contribution to the entropy still takes on the same form  (\ref{eqn:RTareaEnt})
\begin{equation}
    S_{RT} = \sum_{\alpha} p_{\alpha} S(\chi_{\alpha}).
\end{equation}

The skewed Bacon-Shor code we constructed in Figure \ref{fig:starlikeshadow} is one such instance where for any bipartition, one can find decoding unitaries such that it has complementary recovery when the logical state is in either 0 or 1, and therefore also a thermal state diagonal in the computational basis. However, it is not complementary recoverable when the logical state is in an arbitrary superposition. Hence such a code has state-dependent complementary recovery. 

As our tensor networks reflect the basic properties of SBST, it is possible to compute the entanglement entropies for the RT contribution by isolating the state $|\chi_{\alpha}\rangle$ using the Greedy algorithm when the recovery is complementary. Whether the Greedy algorithm still achieves such complementary recovery depends heavily on the state. This is generally possible for the case where bulk states are mostly 0, and the non-zero states lie close to the Greedy geodesic\footnote{When the bulk state is an arbitrary superposition, it is no longer possible to have a decoding unitary that recovers the exact encoded information by acting only on a subsystem as we have previously discussed. At this point, we cannot naively apply Harlow's theorem, and thus we cannot meaningfully separate entanglement entropy into an RT contribution and the bulk contribution.}. 

To prove the RT formula, we reuse the same argument in section~\ref{subsec:exactEERT}, which obtains the states $\chi_{\alpha}$ through local moves, using the Greedy algorithm. Again, we focus on states where the bulk is in a computational basis state (i.e. each bulk qubit is either 0 or 1). The local moves apply because the decoding unitary acting on any 2 or 3 legs decodes the logical information in the same way except $\chi_{\alpha}$ is state-dependent. However, we have to be careful because if the local moves pass through a 1 state, one can not make any subsequent moves as $\chi_{1}$ is not maximally mixed. Therefore, the Greedy algorithm across SBSTs will terminate if the current move passes through a non-zero bulk state. 

\begin{figure}[ht]
    \centering
    \includegraphics[width=0.5\textwidth]{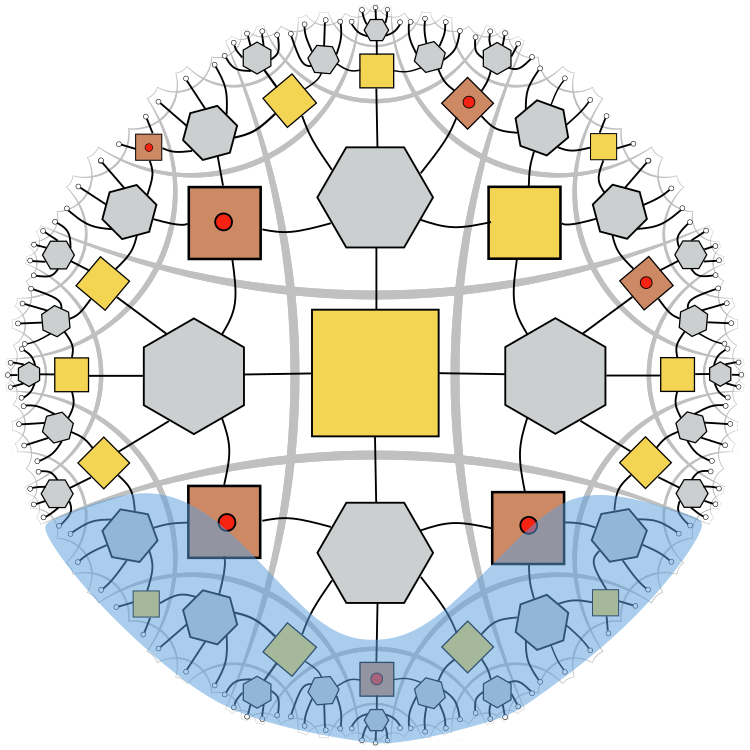}
    \caption{If X tiles are initialized in arbitrary states, then the Greedy wedge cannot move past the perfect tensor. This is true even if the information on these tiles are in $|\tilde{1}\rangle$ and can be recovered. Yellow region is the Greedy wedge if the marked tiles are initialized in 0.}
    \label{fig:brokenCR}
\end{figure}

An example is shown in (Figure~\ref{fig:brokenCR}). Even if the SBSTs are initialized in logical 1 states, which are individually recoverable from local moves on an SBST, the resulting  $\chi_1$ is not maximally mixed. This ruins the isometry property of the adjacent perfect tensor. Therefore the algorithm must terminate at the boundary marked by the blue wedge. We do not make any conclusive statement about entanglement entropies in these cases where the Greedy algorithm fails.

\begin{figure}
    \centering
    \includegraphics[width=\textwidth]{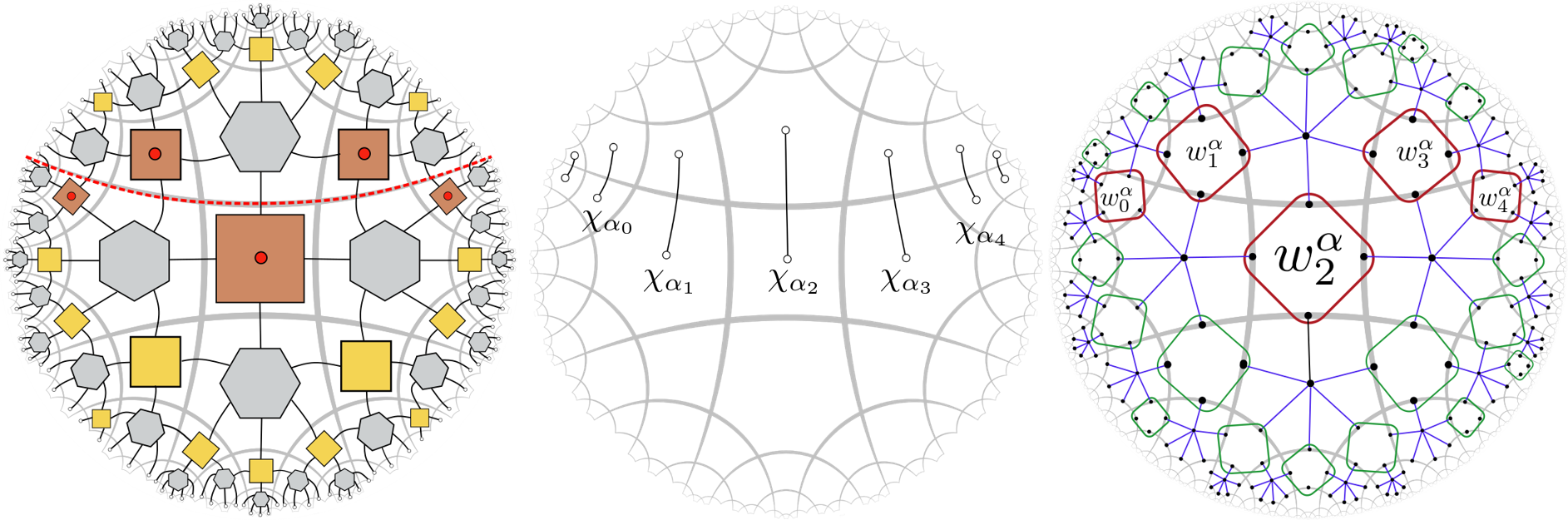}
    \caption{Going from left to right, we choose a bipartition such that the boundary degrees of freedom are divided into two complementary sets above and below the dashed line. Greedy algorithm ``distills'' the state $\chi_{\alpha}$, whose von Neumann entropies are then assigned as $\alpha$-dependent edge weights in the hypergraph. }
    \label{fig:SBST_RT}
\end{figure}

However, if we do have state-dependent complementary recovery (one example shown in Figure~\ref{fig:SBST_RT}), \textit{i.e.,} the Greedy algorithm does not terminate prematurely, then by the same distillation argument we can obtain the $\chi_{\alpha}$ states associated with the global minimum cut in the tensor network. In this case, 
\begin{equation}
    \chi_{\alpha} = \bigotimes_{T}\chi_{\alpha_{T}}^{SBST},
\end{equation}
where $\chi_{\alpha_{T}}^{SBST}$ are the distilled Bell-like or GHZ-like states across the bipartition (\ref{eqn:chi0},\ref{eqn:chi1}). Here $T$ represents the square tiles that are cut by the Greedy geodesic. For multi-copies, there will be $N$ such factors at each square tile $T$ associated with the SBSTs across the Greedy geodesic $\gamma_A^*$. Now its associated entropy $S(\chi_{\alpha})$ is clearly $\alpha$-dependent.

To derive the fact that the minimum cut in the hypergraph is proportional to the RT entropy $S(\chi_{\alpha})$, we simply assign state-dependent edge weights to the hypergraph~(Fig. \ref{fig:SBST_RT}). For the distilled state $\chi_{\alpha_T}^{SBST}$ associated with the square tile $T$ across the minimum cut, we assign weight $w_T^{\alpha}=S(\chi_{\alpha_T}^{SBST})$ to the corresponding hyperedge on that tile\footnote{The nearby degree-2 edge is assigned the same weight. }. Then by construction,
\begin{equation}
    S(\chi_{\alpha}) = \sum_{T} S(\chi_{\alpha_T}^{SBST})=\sum_{T} w_T^{\alpha}
\end{equation}
where $w_i^{\alpha}$ are the weights associated with the distilled states across $\gamma_A^*$. 
For example, in our construction for a single SBST double-copy tensor, it is a hypergraph with four vertices and one degree 4 hyperedge that connects all the vertices. The edge weights are then $w^{0}=S(\chi_0)=2$ whereas $w^1=S(\chi_1)<2$. 

Then the size of the graph cut is indeed proportional to the RT entropy $S(\chi_{\alpha})$
\begin{equation}
    \mathrm{Area}(\gamma_A^*)\propto|\gamma_A^*(\alpha)| = \sum_T w_T^{\alpha} = S(\chi_{\alpha}). 
\end{equation}

More generically for a bulk state $\tilde{\rho}$ that is complementarily recoverable (e.g. the central bulk qubit is in a mixed state diagonal in the computational basis and the rest of the SBST are in the logical 0 state), 
the RT entropy is proportional to a weighted sum over the Greedy geodesic lengths $\gamma_A^*(\alpha)$ defined by a weighted sum over edge cuts
\begin{equation}
    \mathrm{Area}(\gamma_A^*)\propto \sum_{\alpha}p_{\alpha}S(\chi_{\alpha}) \propto \sum_{\alpha}p_{\alpha}|\gamma_A^*(\alpha)|.
\end{equation}
The total entropy $S(A)$ now also includes a bulk term that is given by the Shannon entropy of $\{p_{\alpha}\}$.

Therefore, the approximate code also clearly obeys the RT/FLM formula  when the bulk states are chosen such that the recovery is complementary. The RT entropy contribution is equal to the sum over the edge weights of the hypergraph. 

Notably, the state-dependence of the RT entropy is yet another indicator of bulk back-reaction. For different bulk states that correspond to different ``matter contents'', the corresponding RT entropies also take on different values. This is synonymous to having background (entanglement) geometries that depend on the bulk content. This is also similar to having a code space that contains different semi-classical geometries each labelled by a different $\alpha$.

We emphasize the importance that the state-dependence of this entropy is associated purely with the RT term. It is otherwise straightforward to obtain tensor networks whose overall von Neumann entropies are state-dependent. For instance, this is easily achievable in \cite{Pastawski:2015qua,Hayden:2016cfa} by simply changing the entropy of the bulk state. However, the leading area contribution remains unchanged in these constructions.

\subsection{Power Law Correlations}
\label{subsec:powerlaw}

A challenge with exact stabilizer code construction of the holographic tensor network is that the two-point correlation functions on the boundary is either a phase or zero\cite{Pastawski:2015qua}, unless the logical states are initialized in a special way\cite{Gesteau:2020hoz}. For example, it is possible to have a power-law correlation for localized operators on the boundary as long as the bulk degrees of freedom have correlations that are exponentially decaying with bulk distance. More explicitly, let the bulk be initialized in some state $|\tilde{\psi}\rangle$ that have exponentially decaying bulk correlation with respect to bulk distance, and $O_A, O_B$ be two localized operators on the boundary separated by some distance $\delta x$ on the boundary. Suppose $O_A,O_B$ are logical operators, $\tilde{O}_a, \tilde{O}_b$, of the bulk degrees of freedom that are close to the boundary, then 

\begin{equation}
    \langle \tilde{\psi}|O_A O_B|\tilde{\psi}\rangle  = \langle \tilde{\psi}|\tilde{O}_a \tilde{O}_b|\tilde{\psi}\rangle \propto \exp(-\tilde{K} d(a,b))\approx \exp(-2\Delta \log(\delta x)) = \delta x^{-2\Delta},
\end{equation}
which is a power-law. 
Notably, by considering localized operators without bulk correlations, \cite{Jahn2020} shows that it is also possible for stabilizer holographic pentagon codes to support a power-law decaying correlation on a subset of the boundary sites.

However, we will show that a different source of power-law correlation is possible from the approximate hybrid Bacon-Shor code without using bulk correlations. This may be related to similar behaviour in~\cite{Hayden:2016cfa} in the finite bond dimension limit.

We explicitly construct skewed tensors that will give power-law correlation to $\langle X X\rangle$ correlators in a simple model. In particular, the scaling dimension $\Delta$ of this operator is larger for smaller skewing. This means in the limit of $\epsilon \ll 1$, where the noise terms are small, there is a large gap  in the scaling dimension \cite{Heemskerk:2009pn}, similar to what we observe in holography where $G_{N}\ll 1$. 

\begin{figure}[h]
    \centering
    \includegraphics[width=0.8\textwidth]{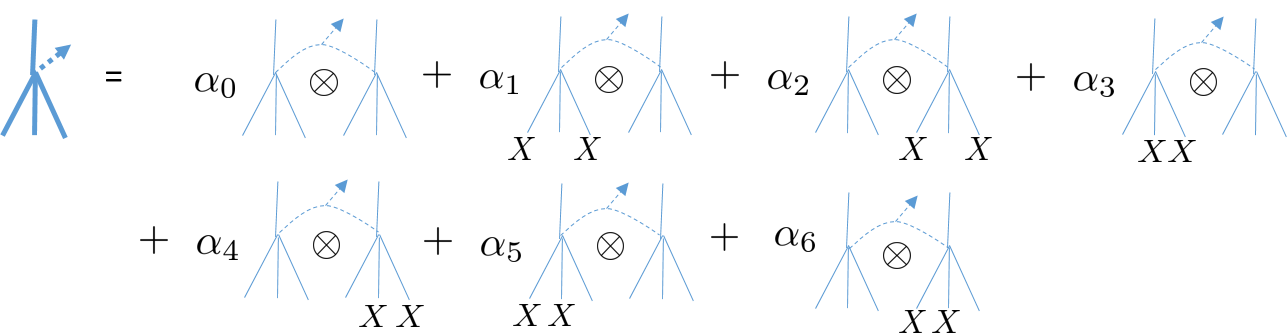}
    \caption{Consider approximate isometries constructed as such. Because no weight two $X$ type operator can be the logical operator in this concatenated code, they are detectable $X$ type errors, and are orthogonal states. Like before, the physical qubits of the second copy are collectively rotated by 90 degrees relative to the first. }
    \label{fig:abst_powerlaw}
\end{figure}

Consider skewed versions of the double-copy Bacon-Shor tensor which can be written as a superposition of exact double-copy Bacon-Shor codes, but are different in their choice of codewords:
\begin{equation}
    V= \sum_{i=0}^6\alpha_i W_i
\end{equation}
where each $W_i$ is an encoding isometry of the double-copy Bacon-Shor code in a particular gauge, e.g. the $ZZ=+1$ gauge. However, different $W_i$ are related by applying a single copy gauge or logical $\bar{X}$ operator.
Diagrammatically, this superposition is written as in Figure~\ref{fig:abst_powerlaw}, 
where each Bacon-Shor tensor is again an encoding map with 4 physical qubits marked by solid lines. The dashed line connecting the two tensors denote the concatenation by the repetition code $|\tilde{0}\rangle = |\bar{00}\rangle$ and $|\tilde{1}\rangle = |\bar{11}\rangle $, which outputs a single bulk qubit in the end. This is yet another example of the approximate double-copy Bacon-Shor code. 

We treat it as a superposition of the reference code $i=0$, and the set of codes related to the reference code by a 2-qubit $X$ error on a single copy. We can check that the noise terms are isotropic if each $i\not= 0$ has $\alpha_{i}$ all taking the same value.\footnote{This can be easily checked by rotating the SBST by 90 degrees. The rotated tensor is equal to the unrotated version. Note that some terms are related by stabilizer multiplication.} In the diagram, we have also chosen a particular orientation to represent the tensor, because we can also treat it as an isometry where we have a single input leg on top which points toward the bulk interior, and three output legs in the bottom that point toward the boundary. The bold tensor indicates a superposition of multiple double-copy Bacon-Shor tensors where one copy is rotated 90 degrees relative to the other. For the dominant reference term with amplitude $\alpha_0$ on the right-hand side, it is nothing but the original double-copy Bacon-Shor tensor. For the other terms, we produce a slightly different Bacon-Shor encoding isometry by acting Pauli $X$ operators on the qubits marked with $X$.

\begin{figure}[H]
    \centering
    \includegraphics[width=0.9\textwidth]{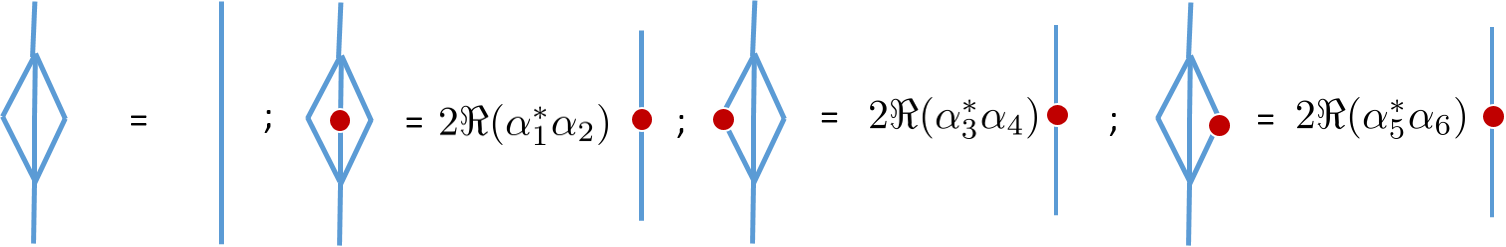}
    \caption{The tensor contracts as an isometry in the 1-3 split. Each red marker indicates an $X\otimes X$ operator insertion. We drop the bulk leg as we have fixed it to a specific state.}
    \label{fig:ABST_contraction}
\end{figure}

For simplicity, let us fix the bulk state to be $|\tilde{0}\rangle$, the double-copy tensor is again an isometry in a 3-1 split~(Figure \ref{fig:ABST_contraction}). This is because any one leg (or the two physical qubits associated with the leg) is (are) maximally entangled with the rest of the system. Alternatively one can check through explicit tensor contraction term by term. Details of the tensor contraction and their properties are further explained in Appendix~\ref{app:tensorcontract}. 

However, if we insert an $X\otimes X$ operator on a single leg\footnote{Each leg corresponds to two qubits because of the double copy. One $X$ is inserted for each copy.}, the tensor contracts to the same operator but with a multiplicative constant. For appropriate choices of $\alpha_i$, $X\otimes X$ is thus a single site scaling operator under this superoperator induced by the SBST isometry we constructed.

We can also skew the perfect tensor to make it an approximate stabilizer state, where it behaves like an isometry when we split the tensor into one input leg and 5 outputs legs. In particular, we want to construct the noisy components around the reference state such that the tensor contraction rule is similar to Figure \ref{fig:ABST_contraction} when an $X$ operator is inserted on an output leg. Then we can simply construct a double copy by taking the tensor product of two single copies. 

To prepare a single copy of such a tensor, we consider a superposition of the perfect tensor, which we call the reference perfect tensor, and those related to the perfect tensor by weight two $X$ errors. 
More explicitly, the single copy \textbf{imperfect tensor} is written as
\begin{equation}
    |\ePT\rangle = \beta_0|T_0\rangle+\sum_{<i,j>}\beta_{<i,j>} X_{i}X_j|T_0\rangle,
\end{equation}
where $|T_0\rangle$ is the reference perfect tensor. Each term in the summand is related to $|T_0\rangle$ by a weight two $X$ operator acting on two distinct sites $i,j$. We use $<i,j>$ to denote all distinct unordered pairs in 6 qubits, totalling $\binom{6}{2}=15$ terms. Similar to the the SBST, the imperfect tensor is isotropic when $\beta_{<i,j>}$ are equal. Because the perfect tensor is non-degenerate with a code of distance 4, all states in the sum are mutually orthogonal.

\begin{figure}[h]
    \centering
    \includegraphics[width=\textwidth]{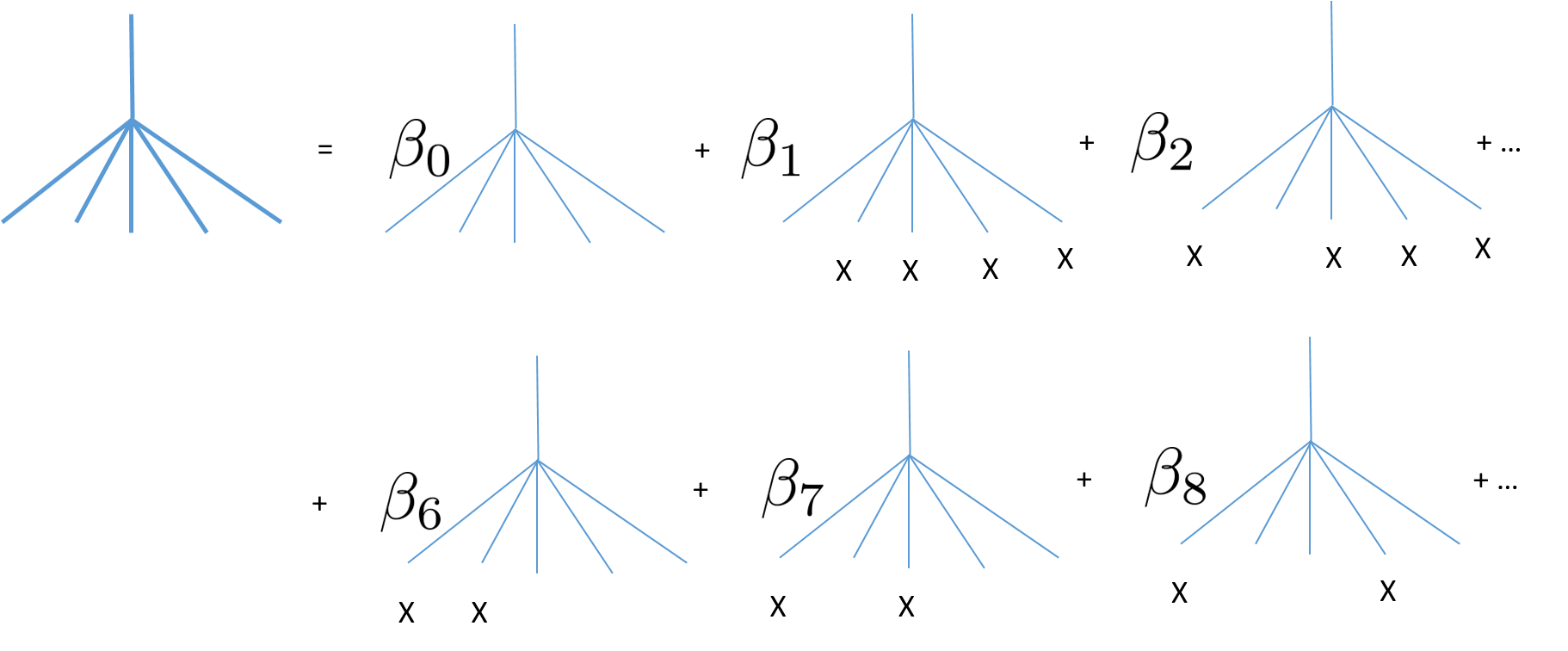}
    \caption{We construct an imperfect tensor by taking a superposition of the following tensors, where tensor without $X$ insertion is the reference perfect tensor. For small errors, $\beta_0\gg \beta_{i\ne 0}$. }
    \label{fig:imperfect_tensor}
\end{figure}

Diagrammatically, we can again re-express this imperfect tensor as superposition over different tensors (Figure \ref{fig:imperfect_tensor}). We can choose to express the tensor with one leg on top (towards the bulk interior) and the remaining in the bottom (towards the boundary). We can clear the top leg by multiplying by $XXXXXX$, which stabilizes $|T_0\rangle$, hence the weight four $X$-operator insertions. 

\begin{figure}[h]
    \centering
    \includegraphics[width=0.5\textwidth]{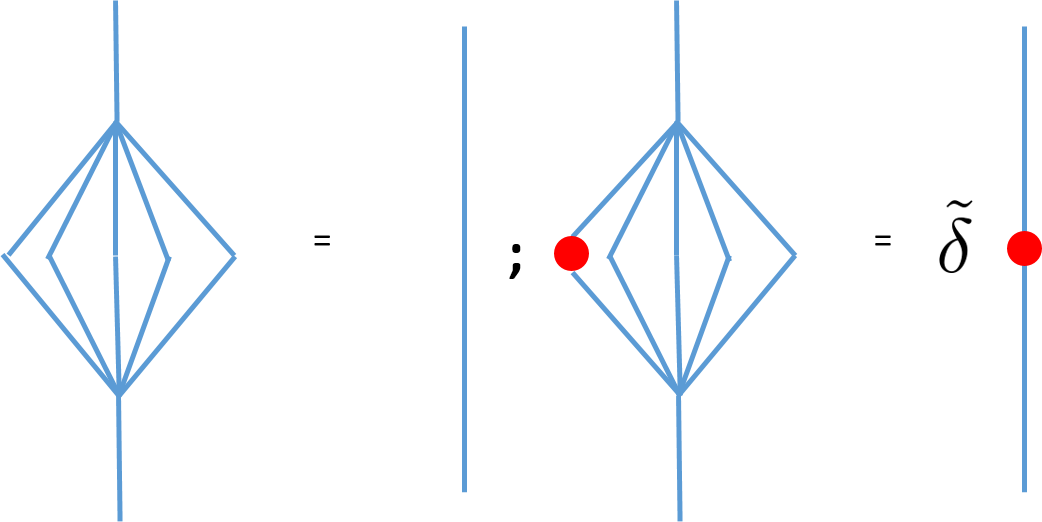}
    \caption{The imperfect tensor is any isometry with 1 input and 5 outputs. An $X$ operator insertion into this contracted isometry leads to another $X$ operator times some constant multiplicative factor $\tilde{\delta}$. }
    \label{fig:imperfect_tensor_contract}
\end{figure}

The overall pushing rule of the single $X$ insertion is shown in  Figure \ref{fig:imperfect_tensor_contract}, where $\tilde{\delta}$ depends on the noise coefficients $\beta_0,\beta_{<i,j>}$, $|\tilde{\delta}|<1$. Intuitively, every time we perform a tensor contraction with a single $X$ insertion, we can push the operator deeper into the bulk, then contract the tensor. In the meantime we must pick up a multiplicative factor $\tilde{\delta}$.

Although we list only one example in Figure \ref{fig:imperfect_tensor_contract}, it should be understood that all single-leg operator insertions will essentially be equivalent except the differences in the coefficients.
Because the double copy of the imperfect tensor would be a tensor product of itself, it suffices to derive the pushing rule for a single copy. The double copy network has the same pushing rule if we replace the single $X$ insertion with $X\otimes X$ in the double copy, and the coefficient $\tilde{\delta}\rightarrow\tilde{\delta}^2$.
 
\begin{figure}[H]
    \centering
    \includegraphics[width=0.8\textwidth]{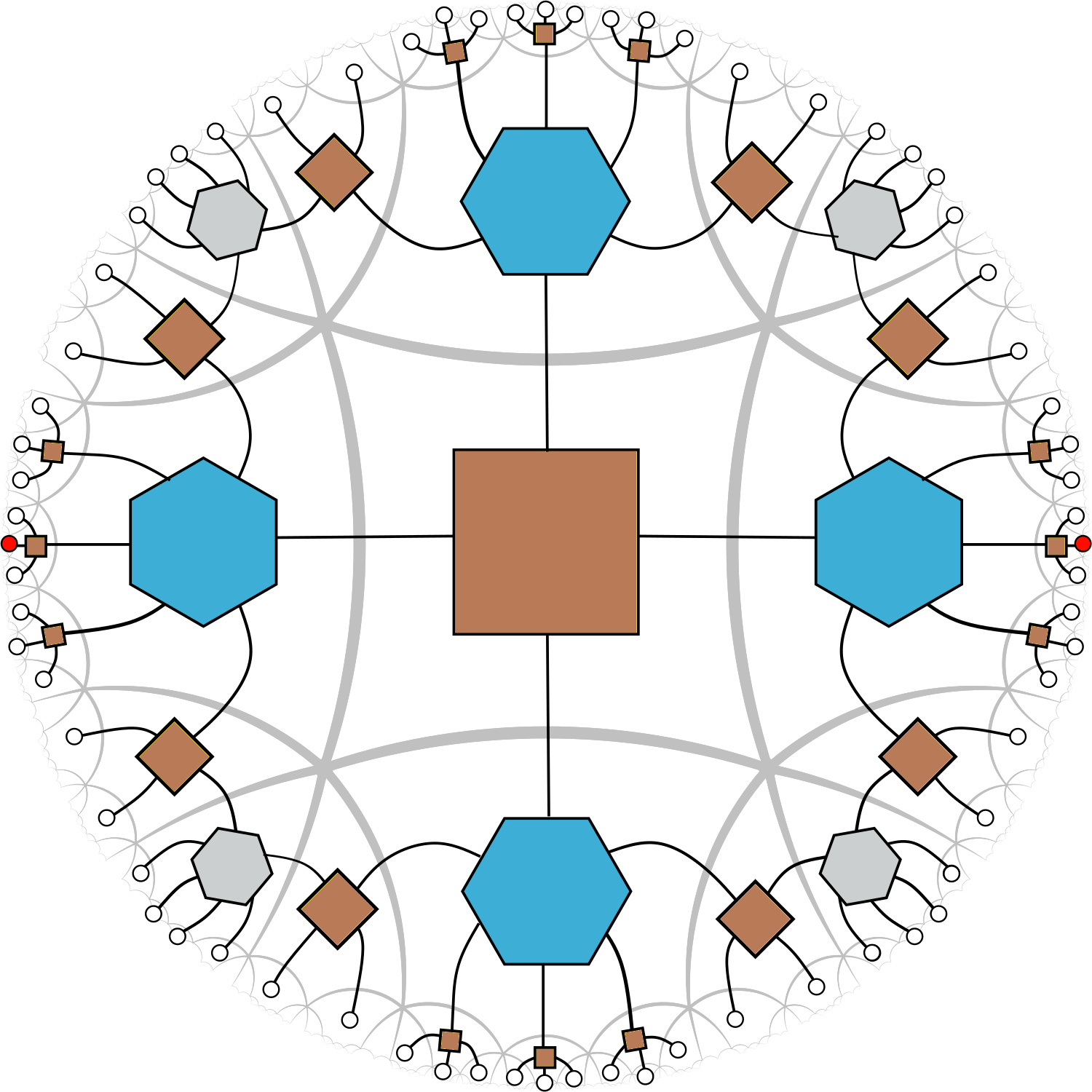}
    \caption{It is understood that the tensor network is the dual graph and red markers are $X\otimes X$ in the double copy. We can alternate the perfect tensor (grey) and imperfect tensor (blue) tiles such that the tensor network can be efficiently contracted while having power-law for certain two-point operator insertions.}
    \label{fig:hybrid_contraction}
\end{figure}

By the pushing rules, we can push the operator through as we contract the tensor network. In this example (Figure~\ref{fig:hybrid_contraction}), all squares are SBSTs (brown) initialized in the product of $|\tilde{0}\rangle$ state while we only intersperse imperfect tensors (blue). This construction is to ensure that we can efficiently contract these approximate tensors when evaluating the correlation function\footnote{If we replace all perfect tensors with imperfect tensors, then we no longer have isometric tensor contractions. Although a more natural tensor network to construct, the evaluation of correlation functions for such cases is much more complex. For clearer proof and illustration, we choose this simpler case.}.

For the sake of simplicity, we fix the amplitudes of the noise term for the double copy Bacon-Shor tensors to be $\alpha_{i\ne 0} =\gamma$ and $\tilde{\gamma} = 2|\gamma|^2$. Similarly, set $\beta_{<i,j>} = \beta$ and the constant factor $\tilde{\delta}(\beta)$ now is the same for operator insertions at all 5 legs. These tensors are also isotropic. 

Suppose the two insertions are at sites $i, j$ where $|i-j|\gg 1$, then by the tensor contraction rules, each iteration of the contraction moves the operators inward by one layer. The operators will take roughly $\log_{\eta}(|i-j|)$ contractions before the two insertions meet at the ``top'' tensor where the two operators are applied to the same tensor, be it an SBST or an imperfect tensor.

It is straightforward to check that a two-point insertion on SBST or an imperfect tensor, denoted as $C_{XX}$, is nonzero.  Therefore, the double-copy two-point correlator 

\begin{equation}
    \langle X^{\otimes 2}_i X^{\otimes 2}_j\rangle \approx C_{XX}(\tilde{\gamma}\tilde{\delta}^2)^{\log_{\eta}(|
    i-j|)}
    = \frac{C_{XX}}{|i-j|^{2\Delta}}
\end{equation}
where 
\begin{equation}
    \Delta = \frac{-\log (\tilde{\delta}^2\tilde{\gamma})}{2\log \eta}
\end{equation}
and $\eta$ is the approximate growth ratio for each layer. Each layer consists of hexagons and their immediate square tile neighbours, similar to the counting we use for the encoding rate in appendix~\ref{app:codeprop}. Indeed for $\tilde{\delta},\tilde{\gamma}\ll 1$, $\eta\approx 5.8$, the scaling dimension $\Delta >0$.

\begin{figure}[H]
    \centering
    \includegraphics[width=0.8\textwidth]{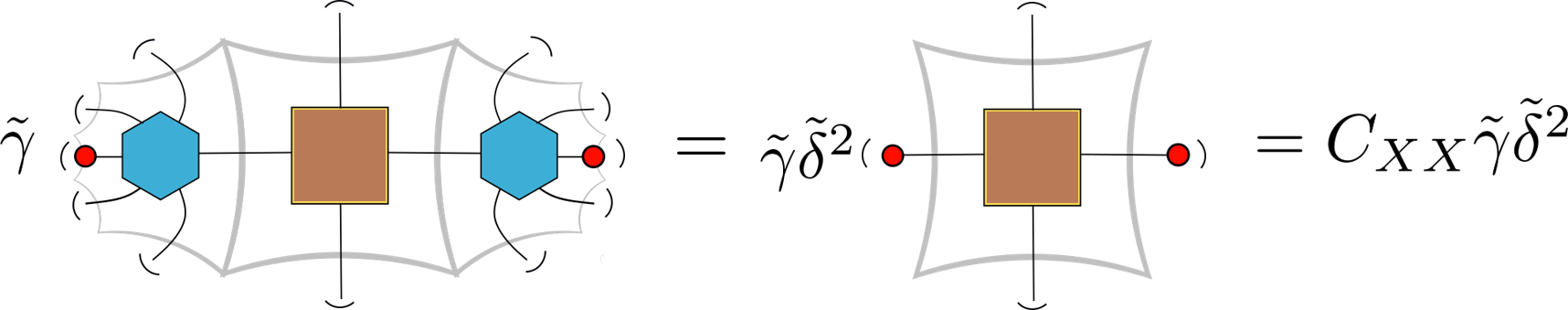}
    \caption{Tensor contraction for two-point function with operator insertions shown in Figure~\ref{fig:hybrid_contraction}. The arc denotes tensor contraction. $\tilde{\delta}^2$ because of the double copy.}
    \label{fig:APL_contract_steps}
\end{figure}

For example, for two point insertions shown in Figure~\ref{fig:hybrid_contraction}, the tensor network can be contracted as follows (Figure~\ref{fig:APL_contract_steps}) to evaluate the correlation function. 

The arcs denote tensor contraction as follows.

\begin{figure}[H]
    \centering
    \includegraphics[width=0.3\textwidth]{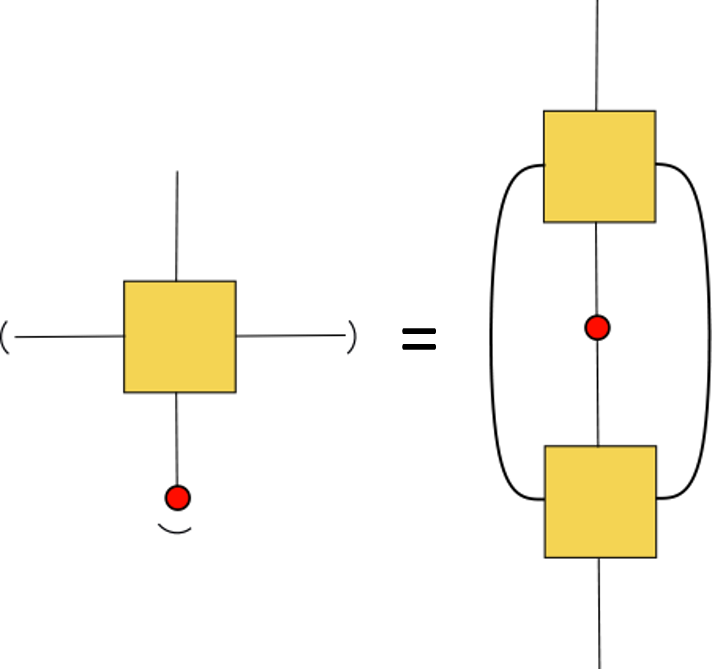}
    \caption{Arc at the end of a tensor leg denotes contracting with its conjugate transpose.}
    \label{fig:contraction_notation}
\end{figure}

\begin{figure}[H]
    \centering
    \includegraphics[width=0.6\textwidth]{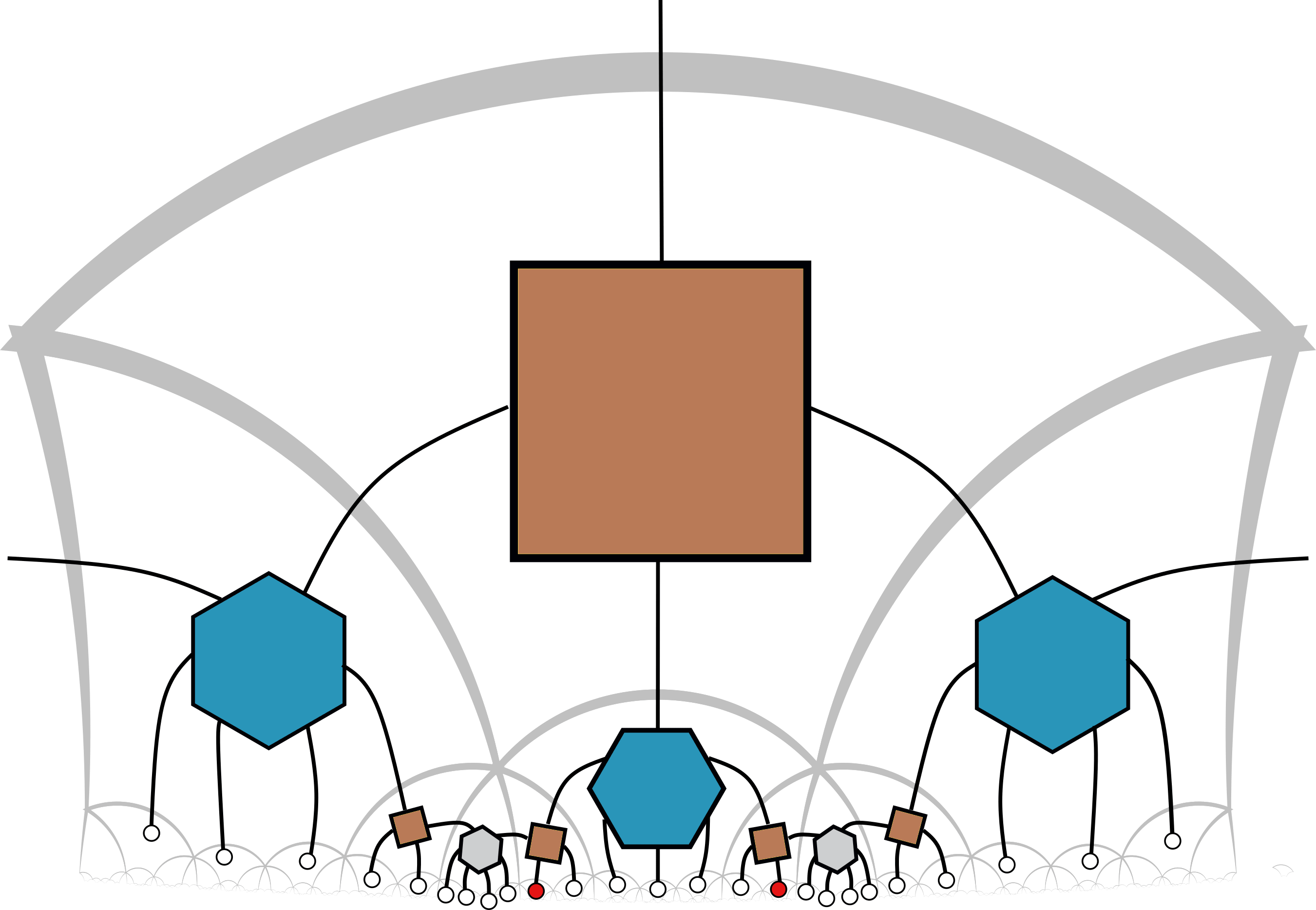}
    \caption{We can also consider correlation function where inserted operators are seprated by a shorter distance (marked red). In this case, we can check that the operators meet at the top tensor after only a single layer. The figures are zoomed into the bottom of the previous figure for clarity.}
    \label{fig:zoomed}
\end{figure}

In Figure~\ref{fig:zoomed}, we also show an example where inserted operators are inserted closer together where tensors closer to the boundary are drawn. This requires fewer steps of contraction (hence fewer powers of multiplicative constants) to evaluate the 2 point function.

Note that the connected 2 point function is precisely the two-point function in our example, because one-point functions of $X^{\otimes 2}$ vanish for this tensor network. We can see this because the expectation value of $\langle X^{\otimes 2}\rangle$ on either an SBST top tensor or an imperfect top tensor is zero. Indeed they correspond to evaluating $\Tr[X^{\otimes 2}]$ according to the pushing rule. Equivalently, we can prove it by contracting the approximate tensors term by term in Figure \ref{fig:abst_powerlaw}. We see that each contraction corresponds to the inner product between a logical state and a single error state, which are orthogonal. Similarly, if the top tensor is a hexagon, a single $X$ expectation value contains terms that are the perfect tensor expectation value of weight 1 or weight 3 transversal $X$ operators. Because of the code property, this expectation value must be zero.

Each tensor is isotropic in this skewed tensor network construction, which respects the discrete symmetries of the hyperbolic space, similar to \cite{Evenbly2017}. Unlike MERA \cite{Vidal2008}, it does not assume a special direction or a different top tensor. However, because we only make some (alternating) hexagons to be imperfect, certain correlators will still vanish; e.g. if the inserted operator is pushed across a perfect tensor. In this sense it is similar to \cite{Jahn2020}, but correlations are computed with actual weight two operators. 
Nevertheless, this simple example demonstrated that it is possible to construct a holographic tensor network that has power-law decaying correlations without bulk correlation using explicit approximate quantum error correction codes. Much of this observation remains to be explored and generalized, including whether it can be used as an efficient variational ansatz to reproduce the correct conformal spectrum by tuning the noise parameters $\alpha_i,\beta_i$. Some of us will discuss these topics in an upcoming work\cite{CaoPollackWang}.

\section{Discussion}
\label{sec:7disc}

In this work, we have constructed an explicit tensor network toy model using the Bacon-Shor code, perfect tensors, and their generalizations. In particular, we find that it is possible to realize a qubit code with a non-trivial center in the code subalgebra. The generalized (approximate) code exhibit many features similar to holography. In our construction, the code is a superposition of different stabilizer codes whose code subspace can contain codewords that correspond to different ``semi-classical'' geometries. The entanglement wedge, and relatedly, bulk operator reconstruction become state/subspace-dependent while the bulk degrees of freedom are only approximately local. The approximate code also shows flexibility to accommodate power-law correlations in the connected two-point function. Because different $\alpha$ blocks can correspond to different ``semi-classical'' geometries as defined by RT contribution to the entanglement entropy, the entanglement wedge, and operator reconstruction properties, the code reproduces some salient features of back-reaction where bulk state affects such geometries. Precise connections with gravity remain to be explored. One particular advantage of this construction of the approximate quantum error correction code is that the tensor constructions are explicit and well-controlled. Furthermore, the logical information can be decoded with small error using an explicit local log-depth reference code decoding circuit. In the limit of zero skewing, we also constructed a new type of hybrid subsystem/gauge code, where stabilizer and gauge operators can be explicitly constructed through operator pushing in the dual emergent network geometry. For a gauge fixed version, error detection and correction can be performed using the stabilizer formalism, which is well-known.  

We also comment on several connections with physics and potential future directions below.

\subsection{Fixed-Area States and Superposition of Codes}
We have shown that the exact stabilizer code has non-trivial centers in the code subalgebra, which allows us to identify the so-called ``$\alpha$-block'' structures \cite{Harlow:2016vwg, Akers:2018fow} in the Hilbert space decomposition. In the examples we constructed, bulk computational basis states correspond to basis vectors in different blocks. These are stabilizer states which clearly have a flat entanglement spectrum and can be considered as fixed-area states\cite{Akers:2018fow,Dong:2018seb}. 
Despite the presence of these $\alpha$-blocks, the dimensionality of each $\alpha$ is the same in the exact holographic hybrid code. For any bipartition, it is easy to check that each $\alpha$-block yields equal entanglement, and each $\chi_{\alpha}$ that gives the RT entanglement contribution has full rank and is maximally mixed. Therefore, the underlying entanglement geometries for different $\alpha$ blocks are identical.

It was argued in~\cite{Akers:2018fow,Dong:2018seb} that existing holographic QECC models, which have flat entanglement spectrum, correspond to fixed area states and not states with smooth geometries. However, these tensor network models do form an overcomplete basis of the Hilbert space, such that any smooth geometry can be constructed by the superposition of such codes. Indeed we find something similar here with the skewed holographic hybrid code construction, where it can be construed as a super tensor network~\cite{Akers:2018fow}. Such codes are built from a superposition of different stabilizer codes/states where each corresponds to a different ``fixed-area state''. For example, in section~\ref{sec:6apphybrid}, $\chi_{\alpha}$ takes on different forms in different $\alpha$-blocks. They clearly produce different RT contributions to entropy, and hence different entanglement geometries. Therefore, a generic state in the skewed code subspace is indeed a superposition over these fixed-area states and can be constructed to have a non-flat entanglement spectrum. 

Given that the relative amplitudes in this superposition can be tuned in the generalized code, it is worthwhile to find out how much flexibility do these degrees of freedom add to the tensor network's ability to fit the entanglement spectrum of, for instance, the vacuum state of a CFT. This may enable the construction of an AQECC tensor network which is also a variational ansatz that efficiently reproduces a CFT ground state.

\subsection{Holographic Renyi Entropy and the Large-N limit}

While we emphasized the flexibility in choosing $\chi_{\alpha}$ above, it is known that by requiring consistent FLM-like formulae for the Renyi entropies in the cosmic brane prescription\cite{Dong:2016fnf}, one should also restrict $\chi_{\alpha}$ of each block to be maximally mixed\cite{Akers:2018fow}. This corresponds to the choice where each $\alpha$-block is a stabilizer code, which serves as an additional constraint for the holographic skewed codes. 

For instance, consider a single-copy Bacon-Shor tensor shown in (\ref{eqn:ghzcode}). We can generate such codes by having $\alpha=1$ and $\beta=0$ in the SBST construction. In the case of a single copy, $\chi_{1}$ is trivial while $\chi_0$ is of rank 2 and maximally mixed. This construction is similar to a super-tensor network where we have a superposition of two $k=0$ stabilizer codes (or rather, stabilizer states), each being a fixed-area state whose $\chi_{\alpha}$ is maximally mixed over a zero and a two-dimensional Hilbert space respectively. Using this SBST as a basic component in the tensor network, we recover structures where the logical information has support over multiple $\alpha$-blocks where the entanglement across each block is maximal. 
In a multi-copy construction, there is more room in accommodating maximally mixed $\chi_{\alpha}$ that are of different ranks. For example, one can concatenate two copies where we choose $\beta$ to be 0 in one and  $\beta=1/\sqrt{2}$ in the other. The resulting $\chi_{\alpha}$ are maximally mixed with rank two and four respectively.

As we have discussed in section~\ref{sec:6apphybrid}, the differences in $\chi_{\alpha}$ lead to ``back-reactions'' where different bulk information modifies the underlying background geometry. We can heuristically define the strength of ``back-reaction'' as $\Delta S(\chi)/S(\chi)$, or the relative of changes in $S(\chi)$ one can have as a result of changing the bulk quantum state. It is clear that one can tune the relative size between $\alpha$ and $\beta$ in a generic skewed code to have arbitrarily small ``back-reactions''. However, if we impose the above constraint on $\chi_{\alpha}$ being maximally mixed on each $\alpha$-block, then back-reactions in a single copy code is of order 1.

 This strong back-reaction can be mitigated by adding more copies in the SBST. For example, we can generate an $N$-copy concatenated SBST where only one copy has state-dependent $\chi_{\alpha}$ entanglement. As such, the entanglement across for $|\bar{0}\rangle$ is $N$ while that for $|\bar{1}\rangle$ is $N-1$, yielding a back-reaction of order $1/N$. Thus by requiring the holographic Renyi entropy constraint and a weak back-reaction, the QECC model must have a large $N$ copy of codes, analogous to the large $N$ limit in holography. Also see \cite{Milekhin:2020zpg} for other connections between large N and QECC.

\subsection{Multipartite Entanglement}
The network supports multi-partite entanglement through its Bacon-Shor and perfect tensor components. The 4-qubit Bacon-Shor code supports entanglement that is GHZ-like or Bell-like, depending on the gauge qubit. It has been pointed out that multipartite (or rather, tripartite) entanglement plays an important role\cite{Akers:2019gcv} in understanding the nature of entanglement cross-section\cite{Takayanagi:2017knl,Nguyen:2017yqw,Dutta:2019gen}. The type of entanglement we inject into the tensor network can be tuned by changing the state of the gauge qubit, which one can choose to satisfy known entropy inequalities\cite{Hayden:2011ag,Bao:2015bfa}. The 4-qubit Bacon-Shor code does not support W-like entanglement, because any single qubit has to be maximally entangled with the rest of the system. For 4 qubits, there is no tripartition where all subsystems are larger than one, hence it cannot be supported. However, larger Bacon-Shor codes or the skewed codes may support a greater variety of multi-partite entanglement. It may be worthwhile to investigate if a particular representative of the code admits W-type entanglement in certain tri-partitions.

We will not comment on whether the code models holographic entanglement, but rather to highlight its flexibility. As it is shown from the point of view of entanglement of purification\cite{Nguyen:2017yqw,Takayanagi:2017knl} and reflected entropy\cite{Dutta:2019gen} that because of their presumed relation with minimal cross-section, holographic theories likely require a non-trivial amount of multi-partite entanglement. Constructions like the ones we proposed may be possible in constructing a toy model that reproduces similar features.

\subsection{Einstein Gravity, Entanglement Equilibrium and Constructions Beyond AdS}
Although we claim some versions of ``back-reaction'' in this work, its exact nature and its relation with gravity remain to be understood. In particular, because we expect the entanglement wedge to be consistent with the quantum extremal surface when considering higher order $1/N$ corrections. Back-reactions that are consistent with AdS/CFT would have to reproduce such behaviours, which we hope to explore in future work. 

Furthermore, to actually satisfy the Einstein's equation, matter contribution to entropy and area contribution will likely have to satisfy some form of constraint, which is discussed in \cite{Jacobson:2015hqa,Czech:2016tqr, Cao:2017hrv,Faulkner:2017tkh}. Therefore, one needs to understand what kind of entropy constraint of Einstein gravity (or its generalizations) should such tensor networks satisfy, and whether it can be satisfied with toy models such as these. As we only discuss quantum error correction codes without any \textit{a priori} input from gravity, it will also shed light on whether gravity can be understood as a general phenomenon emerging from quantum mechanics. This is along the same vein of works such as ~\cite{Cao:2016mst,Cao:2017hrv, Giddings:2018cjc}.

Because most of our statements here apply to quantum systems that are constructed by gluing tensor networks, we also hope to understand whether these approaches can be generalized to model other space-time geometries that are not necessarily anti-de Sitter. 

\subsection{Relation to Magic States}
We know that magic scales extensively in CFTs, where it is also non-locally distributed and appears at all scales\cite{White:2020zoz}. Here the generalized (approximate) code construction precisely incorporates magic on each tensor by skewing and taking the superposition of codes. Therefore, the overall logical computational basis states are no longer stabilizer states. These skewed codes can be used to inject magic locally into the bulk, such that when traced together as a tensor network, it can interlace non-stabilizer states extensively at different scales. Such is a straightforward modification of the known stabilizer holographic QEC models where one uses exact stabilizer codes but does not produce the correct CFT entanglement spectrum. By tracing together the same approximate code blocks at each scale, the network structure itself has built-in symmetries that may help construct a promising tensor network for critical systems and producing the correct entanglement spectrum for the CFTs.

\subsection{Connection with Gauge theory}
We briefly touched on two different notions of gauge in this work. First is related to subsystem codes, which are also called gauge codes, where the logical information is preserved in an equivalence class of states that are related to each other through actions of the ``gauge operators'', which preserves the logical information. This is different from stabilizer codes where the logical information is encoded in a single state. In the same vein, the hybrid gauge code also admits gauge operators derived from operator pushing of a bulk gauge operator to the boundary, which shifts one representation of the logical state to another. It is therefore natural to ask, whether these gauge operators that perform ``local'' gauge transformations in the bulk can lead to a structure similar to that of a gauge theory. 

The same type of gauge transformation also does not necessarily preserve the entanglement structure. As we have seen, depending on the state of the gauge qubit, a 4-qubit Bacon-Shor code can be dominated by GHZ type entanglement, which is multi-partite, or Bell-like entanglement, which is bipartite. This is only more complex in the hybrid code. Therefore, it is crucial to understand the meaning of this gauge dependence and its potential connection of entanglement entropy in gauge theories~\cite{Ghosh:2015iwa,Donnelly:2011hn,Casini:2013rba}.

The second type of gauge we discuss is that of the conventional gauge theory, such as gravity. We know in AdS/CFT that a gauge invariant bulk massive operator must be a dressed operator because of any operator that creates mass-energy is coupled to gravity. This dressed operator will have Wilson line that end on either the charge or the boundary. In Section~\ref{sec:6apphybrid}, we discussed a construction of the bulk operator, where the bare operator is analogous to the reference code $\tilde{X}_0$ operator, while the dressed operator is similar to the actual logical $\tilde{X}$ operator. Although restricting to the bare operator allows subregion reconstruction, it will incur a small error and does not fully preserve the logical information (i.e. not ``gauge invariant''). The error-free operator can be constructed when one considers the ``fully-dressed'' operator that is global. Whether this is a superficial similarity or there is a deeper connection between gauge theory\cite{Mintun:2015qda} and approximate quantum error correction in general remains to be explored.

\subsection{Qudit generalization and General \texorpdfstring{$\alpha$}{alpha}-block Implementation}
Given the need for a non-trivial center in the code algebra, a qudit code\cite{GKP2001} may be better suited in producing certain salient features of holography such that each $\alpha$-block can contain a variety of subspaces of different dimensions. This is because a non-trivial subalgebra of the Pauli algebra is ultimately limited by its dimension. However, it may also be able to expand the $\alpha$ subspace by considering a tensor that encodes multiple logical qubits or using the multi-copy code we discussed. Such variants can be explored, but qudit generalizations seem beneficial not only to holography but also to constructing quantum error correction codes at large, which is important for quantum computing on qudits.

\subsection{Connection to Dynamics}
By the (approximate) Eastin-Knill theorem\cite{Faist:2019ahr,Woods2019,Kubica2020}, it is known that the (non-trivial) continuous time evolution in AdS/CFT and perfect erasure correction on the boundary appear incompatible\cite{HaydenNezami}. It will be informative to further elaborate on the relationship between AQEC and continuous time evolution using our explicit construction.

While speculative, a dynamical process may be essential in selecting the preferred decomposition of the skewed code into superpositions of other quantum codes. For instance, in the setting of decoherence, each term in the superposition may be understood as different branches of the wavefunction, each corresponding to a semi-classical spacetime that is stable under time evolution. 

\subsection{Code properties and the Tensor Network Approach}
Tensor network provides a graphically intuitive way to generate and understand different quantum error correction codes as well as their generalizations and characterizations.  In particular, they provide a convenient way to visualize entanglement structure, to understand the interdependence of multiple encoded qubits, and to produce local decoding circuits~(sec~\ref{subsec:distdecode}). Some of these connections were explored by \cite{FerrisPoulin14}, but have mostly materialized in the forms of holographic codes\cite{Pastawski:2015qua,Hayden:2016cfa,Harris2018}. More recently, it was also claimed \cite{Harris2020,Farrelly:2020mxf} that the holographic codes also have an optimistic error threshold comparable to that of the surface code. Therefore, it is worthwhile to explore whether such approaches can be used to generate codes that have special properties in encoding/decoding, error detection and correction. It is also interesting to establish connections with existing codes that are somewhat graphical in nature, such as low-density parity-check codes and variants~\cite{Camara2005, Bohdanowicz, qexpandc}. 

More specifically, we want to characterize the type of holographic codes proposed here and in \cite{Pastawski:2015qua}, such that code properties like distance, optimal syndrome measurements, and error correction can be fully understood. 
Even with restrictions to holographic codes on hyperbolic geometries, each individual tensor, type of tiling and hence network geometry may be tuned to produce better encoding rate, code distance and decoding properties. In particular, we would like to implement efficient error detection and correction in these codes using the spatial locality and geometry of the tensor network. As spatially local gates are typically easier and less noisy in certain architectures, we also would like to see if such holographic codes can be lifted to a code on the 2-dimensional bulk spatial geometry, where the application of gates is directly related to spatially local tensor moves. In this way, the log-depth $2$-local decoding circuit may also be made spatially local.

It is also important to understand the property of different approximate or skewed codes. This includes a more complete classification of skewed codes based on their properties as well as a more careful study of it being an AQECC. While the reference code decoding map is explicit and manifestly local, it need not be the decoding channel that produces the smallest possible recovery error. On the other hand, the Petz recovery map and generalizations\cite{ohya1993quantum,Junge2015} are used as a recovery channel in the study of approximate quantum codes. Even though it is difficult to implement in practice for our code, it can be used as a theoretical tool to bound optimal recovery errors. Such tools have also been used in holography, which may help establish a connection with existing literature in entanglement wedge reconstruction\cite{Cotler:2017erl}. 

\subsection{Classification of the Skewed Codes}
Although we provided a few examples of skewed Bacon-Shor codes and the perfect state, the construction is generally applicable for other codes as well. On the one hand, we would like to classify how different types of skewing can lead to different code properties. For example, some of these codes which may correct certain errors approximately, some may correct certain errors exactly but are non-additive. In particular, we also like to understand what type of quantum noise can lead to a skewed code, as opposed to the reference stabilizer or gauge code. One can also construct AQECC that protects against asymmetric errors, which is relevant for some near-term devices. 

On the other hand, by constructing and examining different AQECC tensors and inserting them into the holographic tensor network like the one we construct, we can also help disentangle the source of these gravitational features in the AQECC. Specifically, do these effects universally descend from all classes of AQECCs? Or are there AQECCs that manifest some but not all of these features? If it is the latter case, then what type of AQECC is necessary and sufficient in reproducing gravity?

\section*{Acknowledgements}
We thank Chris Akers, Ning Bao, Aidan Chatwin-Davies, Daniel Harlow, Spiros Michalakis, Jason Pollack, John Preskill, Pratik Rath, Ashmeet Singh, Vincent Su, Brian Swingle, Eugene Tang, and Yixu Wang for comments and discussions. C.C. acknowledges the support by the U.S. Department of Defense and NIST through the Hartree Postdoctoral Fellowship at QuICS, by the Simons Foundation as part of the It From Qubit Collaboration, and by the DOE Office of Science, Office of High Energy Physics, through the grant DE-SC0019380.

\appendix

\section{Code Properties of the Hybrid Bacon-Shor holographic code}
\label{app:codeprop}
\subsection{Encoding Rate}

To compute the encoding rate $r=k/n$, we simply count the number of square tiles and the number of boundary legs given a tessellation. Here we will perform the computation for a single copy construction. For multi-copy, each boundary leg corresponds to $N$ qubits while each square tile (and thus bulk leg) corresponds to one qubit. Hence the rate will be $r_{\rm N-copy}=r/N$. 

To start, we identify two types of square and hexagonal tiles. Let $\bar{h}_k, \tilde{h}_k$ be the number of hexagons that have one and two legs pointing inwards respectively at the $k$-th hexagonal tile layer. Similarly, $\bar{s}_k, \tilde{s}_k$ are the same but for squares.  Given the tessellation in Figure~\ref{fig:46TN}, we have the following recursion relations

\begin{equation}
\begin{pmatrix}
\tilde{s}_{k+1}\\
\bar{s}_{k+1}
\end{pmatrix}
=
\begin{pmatrix}
1 & 1\\
2 & 3
\end{pmatrix}
\begin{pmatrix}
\tilde{h}_k\\
\bar{h}_k
\end{pmatrix}
\end{equation}

and 
\begin{equation}
\begin{pmatrix}
\tilde{h}_{k}\\
\bar{h}_{k}
\end{pmatrix}
=
\begin{pmatrix}
1 & 1\\
0 & 1
\end{pmatrix}
\begin{pmatrix}
\tilde{s}_k\\
\bar{s}_k
\end{pmatrix}
\label{eqn:hrecur}
\end{equation}

One can read off these recursion relation as follows. Without loss of generality, assume we are at a layer that terminates on hexagonal tiles. We first add a layer of square tiles that are filling the gaps of the hexagonal tiles. Because each such square tile is adjacent to two hexagonal tiles, they must be tensors of the second type where they have two legs pointing inwards. They are equal to the total number of hexagons at this layer, which is $\bar{h}_k+\tilde{h}_k$. One can then attach square tensors to the remaining ``valent'' legs of the hexagons. For a hexagon that fills the gap between two squares, it has $6-2-2=2$ remaining valent legs after contracting with tensors from the previous layer and the gap square tensors that were just added. For the rest, there are $6-1-2=3$ remaining valent legs. A similar argument applies if one starts with a tile that terminates on squares, except the valency for subsequent hexagonal tiles are given by $4-2-2=0$ and $4-1-2=1$ respectively.

We can choose the layer of 4 hexagons that surrounds the central square tile as the initial layer, which yields $\vec{h}_0 = (0,~4)^t$.

First we count the encoding rate of a tensor network that terminates on square tiles only. The number of squares at the $n$-th layer is given by

\begin{equation}
    \vec{s}_n = N^{n-1}\vec{s}_1
\end{equation}
where
\[
N=
\begin{pmatrix}
1& 2\\
2 & 5
\end{pmatrix},
\]
$\vec{s}_0 = (4,12)$, and 
\[N^k=\frac {\eta^k} {4}
\begin{pmatrix}
2-\sqrt{2} & \sqrt{2}\\
\sqrt{2} & 2+\sqrt{2}
\end{pmatrix}
+\frac{\epsilon^k}{4}
\begin{pmatrix}
2+\sqrt{2} & -\sqrt{2}\\
-\sqrt{2} & 2-\sqrt{2}
\end{pmatrix}=\frac{\eta^k}{4}A+\frac{\epsilon^k}{4}B.
\] for $k>0$.

Here $\eta=3+2\sqrt{2}\approx5.83 , \epsilon =3-2\sqrt{2}\approx 0.17$. For large $k$, the first term dominates. In the limit of large $n$,

\begin{equation}
    \vec{s}_n \approx \frac{\eta^{n-1}}{4} A \vec{s}_0
    =\eta^{n-1}\begin{pmatrix}
    2+2\sqrt{2}\\
    \sqrt{2}+3(2+\sqrt{2})
    \end{pmatrix}
    \approx
    \eta^n
    \begin{pmatrix}
    0.83\\
    2.00
    \end{pmatrix}
\end{equation}

Thus, the total number of bulk degrees of freedom in this tessellation up to the $n$-th square tile layer is

\begin{equation}
    N_{\rm bulk}= 1+\sum_{k=1}^{n} |s_k|_1 \approx 3.41 \eta^n,
\end{equation}
where $+1$ accounts for the central square tile. 

We compute the boundary degrees of freedom by
\begin{equation}
    N_{\rm boundary}^{(S)} = 3\bar{s}_n+2\tilde{s}_n \approx 7.66 \eta^n,
\end{equation}
because each gap-filling square tile has $4-2$ legs facing outward, and the other ones have $4-1$ legs facing outward. 

The encoding rate is 
\begin{equation}
    r=\frac k n = \frac{N_{\rm bulk}}{N_{\rm boundary}^{(S)}}\approx 0.445
\end{equation}
in the gauge code. If we treat each Bacon-Shor tensor as a $[[4,2,2]]$ stabilizer code by promoting the gauge qubit to a logical qubit, then the encoding rate is $\sim 0.890$ in the large layer $n\gg 1$ limit.

For a tensor network that terminates on a purely hexagonal layer, we simply use (\ref{eqn:hrecur}).

Using \ref{eqn:hrecur}, we find
\begin{equation}
    \vec{h}_n  = 
    \begin{pmatrix}
    1 & 1\\
    0 & 1
    \end{pmatrix} \vec{s}_n \approx \eta^n
    \begin{pmatrix}
    2.83\\
    2
    \end{pmatrix}
\end{equation}
Therefore, the number of boundary qubits is given by 

\begin{equation}
    N_{\rm boundary}^{(H)} = 5\bar{h}_n+4\tilde{h}_n \approx 21.32 \eta^n
\end{equation}

Then, the overall encoding rate is 
\begin{equation}
    r=k/n = \frac{N_{\rm bulk}}{N_{\rm boundary}^{(H)}} \approx 0.160
\end{equation}
in the gauge code, and $0.320$ in the stabilizer code where we take each square tensor to encode two qubits. In Figure \ref{fig:46TN} where we terminate on mostly hexagons but mixed with a few square tiles on the boundary, the gauge code encoding rate is $21/124\approx 0.169$. 

\subsection{Code Distance}
Here we set bounds on the code distance for the central bulk qubit. These bounds are not tight, but are sufficient for our purposes, especially in establishing connections with features in holography. 

The lower bound of the distance is set by first demoting all logical qubits other than the central one to gauge qubits. Then we assume each Bacon-Shor block takes one input to a single output. This is true for inputs such as $X,Z$ type operators using gauge operator pushing, however it is at least weight 2 for a $Y$ type input, thus our assumption results in a lower bound. The central tile has two outputs, each feeding into the trapezoidal region we defined previously. For a perfect tensor, it takes a single non-identity input to at least 3 outputs. This is because the stabilizers have a minimal of weight 4. For two inputs in a perfect tensor, however, the output has a minimal weight 2, if the inputs are arranged to have to correct operators. Then it is clear that for the 3 outputs from the Bacon-Shor blocks, two of them are fed into a duo-input hexagon, whereas the other one is fed into a single-input hexagon. The hexagon with two inputs produces two outputs, which again can be fed into a single hexagon, establishing a repeating pattern. This part of the operator pushing only contributes a constant weight of 2 all the way to the boundary. 

For the remaining single input hexagon, it again generates 3 outputs; which produces a duo-input hexagon and a single-input hexagon. Again, we see a self-similar pattern, where at each layer, the minimal weight operator picks up an additional weight 2 from the duo-input hexagon. Hence the lower bound grows linearly with the number of layers,
\begin{equation}
    d_{\downarrow} = 2(2n+3)\leq d
\end{equation}
where $n\geq 0$ denotes the layer number starting from the inner-most hexagons that are next to the central Bacon-Shor tile. Each layer is defined as one layer of squares followed by their adjacent hexagons. Therefore, the code distance for the central qubit is not a constant, which is the case in the Bravyi-Bacon-Shor construction, but instead grows at least logarithmically with the number of physical qubits.

For the upper bound, we again demote all but the central logical qubits to gauge qubits. Then it is clear that any single input of the Bacon-Shor isometry can be pushed on to any two of the three outputs. The remaining output remains an identity operator. A logical operator of the central tile is at least weight 2 but at most weight 3, where each of the outputs feeds into the trapezoidal region we discussed in the previous section. Because each perfect tensor can clean at most 3 legs, we will assume that a hexagon has 3 non-identity outputs.  

By observation, it is clear that the 3 outputs from the first layer hexagons feed into 3 Bacon-Shor squares, which in turn, can be cleaned and feed into 5 hexagons. From then on, the procedure is self-similar: each hexagon has 3 outputs, which feed into 3 squares, and then 5 hexagons. Hence, an upper bound on code distance is given by
\begin{equation}
    d_{\uparrow} = 3 (5^n) >d,
\end{equation}
assuming we terminate on a layer of hexagons. Because $5< \eta \approx 5.83$, 
\begin{equation}
    \frac{d_{\uparrow}}{N_{\rm boundary}} \xrightarrow{n\rightarrow \infty}  0,
\end{equation}
where a logical operator can have support over a measure zero set of the boundary qubits. This is precisely the observation of uberholography, which is first observed in the context of AdS/CFT for configurations that are near the vacuum state. 

\section{Gauge Operator Pushing in the Double-Copy}
\label{app:dcpushing}
The code has a rotational symmetry up to swapping the two layers. Because for any single copy of the Bacon-Shor code, a single site operator can always be pushed to at most two sites in the adjacent edges, it suffices to show that any single site operator on the double-copy tensor can be pushed to one site that is diagonally across from the input, and another adjacent to it.

\begin{figure}[H]
    \centering
    \includegraphics[width=0.9\textwidth]{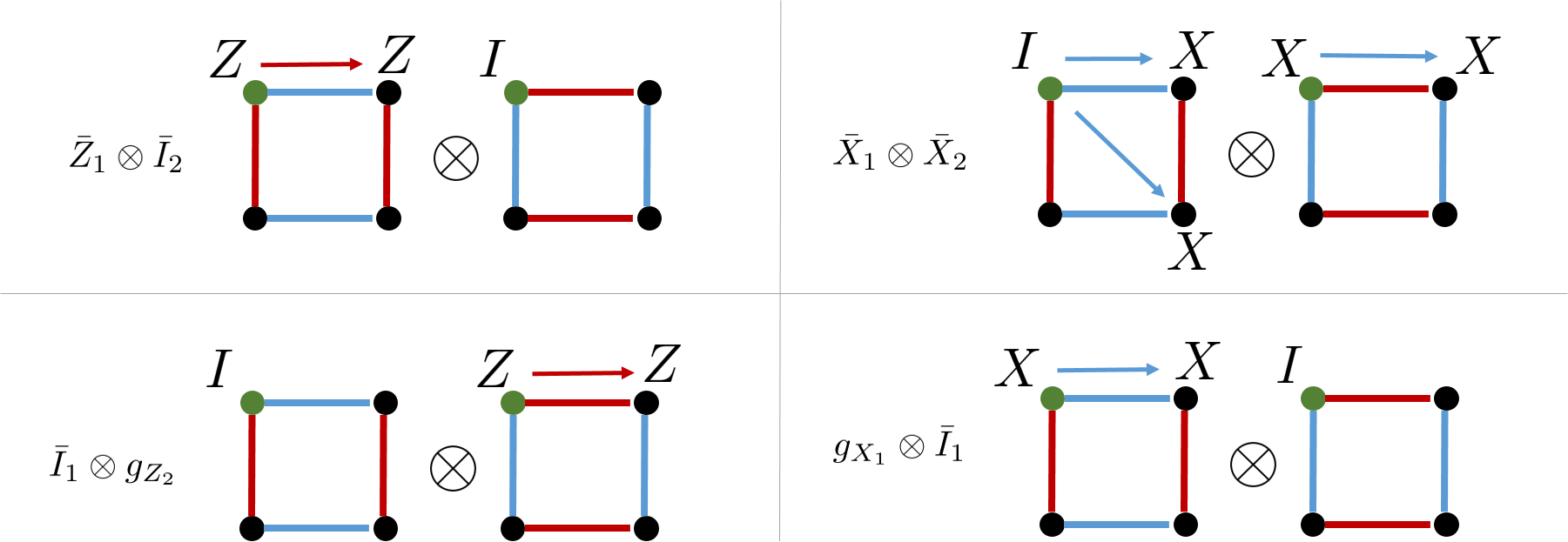}
    \caption{We assume, without loss of generality, there is a single site input in the double-copy tensor in the top left corner (green) and will find the equivalent operator to the right column up to a gauge. }
    \label{fig:double_copy_gauge_push}
\end{figure}

Because any single site Pauli input can be generated by the input of $XI, IX, ZI,IZ$, it suffices to show that these operators can be pushed to two legs one of which is diagonally across from the input. We show in Figure \ref{fig:double_copy_gauge_push} that this is indeed possible. The top left corner labels the input Pauli at each layer, and pushing can be done by multipling the operators we listed to the left of the diagrams. Arrows denote the operator we multiply on each copy for pushing. Blue indicates $X$ type gauge or logical operator, and red arrow indicates $Z$ type gauge or logical operators. $I$ are dropped on the outputs.

However, this kind of restricted pushing is not always doable, even using gauge operators. For example, in the six-copy code we take $XIIIII$ as input. 

\begin{figure}[H]
    \centering
    \includegraphics[width=0.4\textwidth]{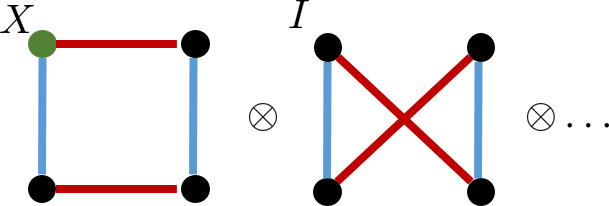}
    \caption{For six copies, we find that there are certain configurations where the gauge pushing does not allow an equivalent operator on arbitrary two sites without leaving the code subspace. }
    \label{fig:six_copy_gauge_push}
\end{figure}

In this case, we show only two of the cases, but we can already see the pushing rule breaks down. As before, we push an operator with single site support on the top left corner to two sites on the right column of the six-copy-tensor. For all copies, we only want non-identity elements on the right column and identities on the bottom left corner. For the first diagram, the $X$ operator cannot be pushed to the right column without using the logical $\bar{X}_1$ operator. However, to preserve the code subspace, we have to apply logical $\bar{X}_i$ operators to all other copies as well. We see no $\bar{X}_2$ is supported on the right column in the second diagram. The logical operators are only supported on the diagonals or on the rows, which is either not completely contained in the second column, or it has support on the input qubit, which is supposed to be identity, instead of $X$. This completes the counterexample.

\section{Details on Tensor Contraction}
\label{app:tensorcontract}
In this section, we show that SBST we constructed in section ~\ref{subsec:powerlaw} is an isometry when we split the physical legs as 3-1. Beginning with the simplest example we used in the earlier sections, we fix the logical state to $|\tilde{0}\rangle= |\bar{0}\bar{0}\rangle$. Recall that $XX$ acting on each copy of the 4 qubit Bacon-Shor code is either a logical $\bar{X}_L$ operator or a gauge $\bar{X}_G$ operator; they flip the logical and gauge qubits from 0 to 1, when we fix the $ZZ=+1$ gauge. Because the code also detects 1 error, the subspaces span by the remaining 3 qubits are mutually orthogonal. Explicitly, we see that for 

\begin{equation}
    W_0(0) = |000\rangle\langle 0|+|111\rangle \langle 1|
\end{equation}
the $XX$ operator yields 
\begin{align}
    W_1(0) &=X_1X_2W_0(0)= |110\rangle\langle 0|+|001\rangle\langle 1|\\
    W_2(0) &= X_2X_3W_0(0) = |011\rangle\langle 0|+|100\rangle\langle 1|\\
    W_3(0) &=X_1X_3W_0(0)= |101\rangle\langle 0|+|010\rangle\langle 1|.
\end{align}

\begin{figure}[H]
    \centering
    \includegraphics[width=0.7\textwidth]{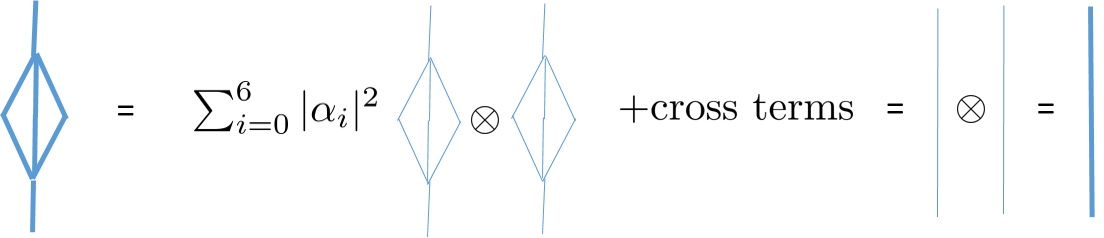}
    \caption{Explicit tensor contraction yields the above terms, which consist of identity, as $\sum |\alpha_i|^2=1$, and cross terms.}
    \label{fig:ABST_detail_contr}
\end{figure}

Indeed, $W_i^{\dagger}W_j =\delta_{ij} I$, as the 3 qubit states are all orthogonal. Therefore, as claimed in section~\ref{subsec:powerlaw}, the (double-copy) SBST~(Figure \ref{fig:abst_powerlaw}) is indeed an isometry when the bulk state fixed to $|\tilde{0}\rangle$. This can be easily verified through direct tensor contraction, where diagonal terms add up to identity~(Figure~\ref{fig:ABST_detail_contr}). 

\begin{figure}[H]
    \centering
    \includegraphics[width=0.3\textwidth]{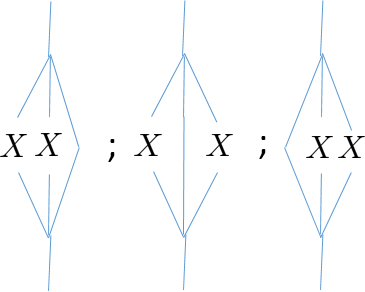}
    \caption{Each cross term contains contributions like one of the above three in a tensor product.}
    \label{fig:xterm_ABSTcontr}
\end{figure}

The cross terms, on the other hand, consist of tensor product of terms shown in Figure~\ref{fig:xterm_ABSTcontr},
which vanish by orthogonality we showed above. 

\begin{figure}[H]
    \centering
    \includegraphics[width=0.3\textwidth]{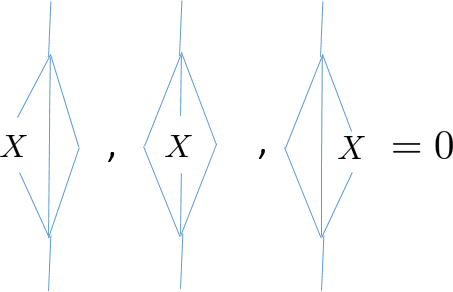}
    \caption{Single X insertion}
    \label{fig:single_insertABST}
\end{figure}

\begin{figure}[H]
    \centering
    \includegraphics[width=0.8\textwidth]{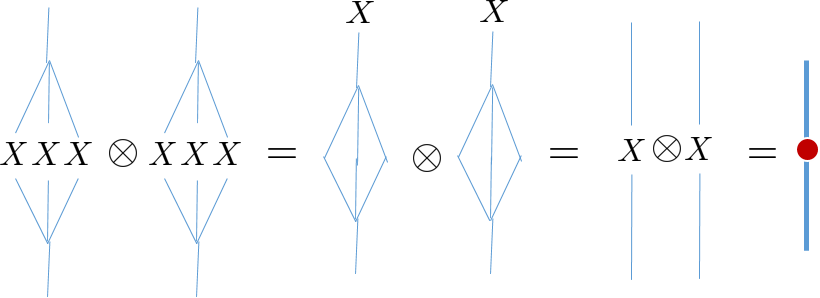}
    \caption{Inserting $X\otimes X$ on the same leg of the double copy code only yield non-zero terms as shown in the figure. }
    \label{fig:abst_xinsert}
\end{figure}

Therefore the entire SBST contracts to identity when the bulk state is $|\tilde{0}\rangle$. Note that single insertion of $X$ also vanishes for a single copy (Figure~\ref{fig:single_insertABST}).
For any insertion of $X\otimes X$ operator, we expand the SBST into the superposition in the definition with the appropriate operator insertions. Then contraction gives the following types of non-vanishing terms (Figure~\ref{fig:abst_xinsert}).

The specific term that the inner product picks up depends on where $X$ is inserted. For example, for $X\otimes X$ inserted on the middle leg, the only nonzero term is like the one above, but with coefficient $2\Re(\alpha_1\alpha_2)$. See Figure~\ref{fig:abst_powerlaw}.

If the bulk state is fixed to 1, then 
\begin{align}
    W_0(1)=&|110\rangle\langle 0|+|001\rangle\langle 1|\\
    W_1(1) =& |000\rangle\langle 0|+|111\rangle\langle 1|\\
    W_2(1) =& |101\rangle\langle 0|+|010\rangle\langle 1|\\
    W_3(1) =& |011\rangle\langle 0|+|100\rangle\langle 1|.
\end{align}
which is a permuted version of $W_i(0)$ with $(01)(23)$. Because all the orthogonality conditions above remain valid,  the tensor contraction properties we derived still hold when the bulk is $|\tilde{1}\rangle = |\bar{1}\bar{1}\rangle$.

For a state that is not in the logical computational state, the situation is slightly more complicated, because this encoding map of 4 physical qubits into 1 logical is non-isometric. Specifically, we mean that the physical state representing $|\tilde{0}\rangle$ is not orthogonal to that representing $|\tilde{1}\rangle$. This can be easily read off from the encoding map, where both $|\tilde{0}\rangle, |\tilde{1}\rangle$ have non-zero support over $|\bar{1}\bar{0}\rangle$ and $|\bar{0}\bar{1}\rangle$. Because of this, defining a tensor for an arbitrary bulk state is somewhat ambiguous in that either the bulk state or the boundary state will not have unit norm. 

However, let us normalize an arbitrary superposition of $a|\tilde{0}\rangle+b|\tilde{1}\rangle$ in the physical representation. For whatever logical output the encoding map defines, one can show that these tensors are still isometries when we contract 3 in-plane legs and $X$ is still a scaling operator under such contractions. An easy way to see this is to expand such tensor in terms of the reference isometry with X insertions, where there exists an mutually orthogonal basis under the 3 leg contraction. 

\begin{figure}[H]
    \centering
    \includegraphics[width=\textwidth]{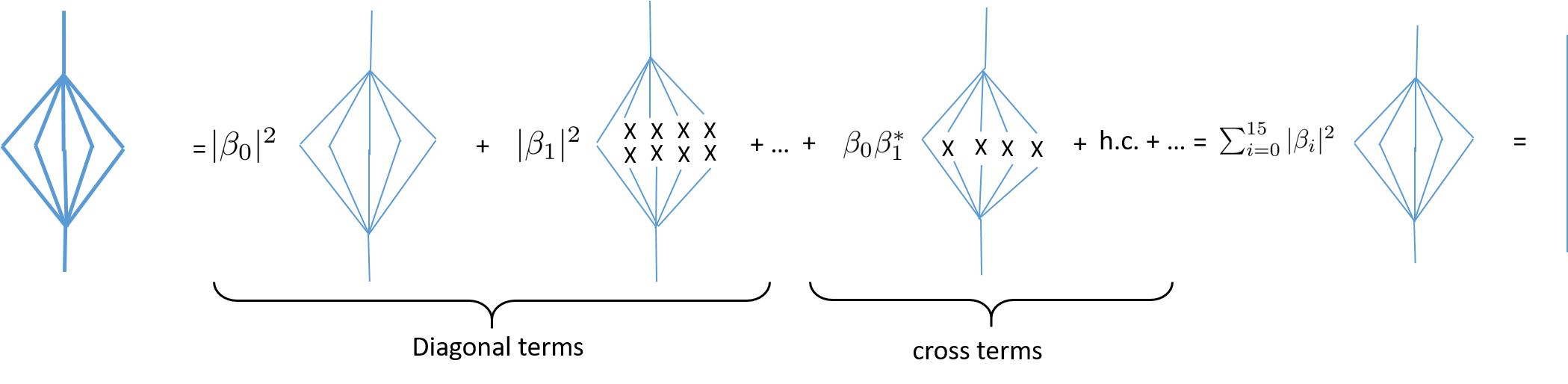}
    \caption{The imperfect tensor (bolded line) contraction can be computed term by term. For diagonal terms, they are simply the perfect tensor contraction, which reduce to identity. The cross terms vanish by orthogonality.}
    \label{fig:ipt_detail_contract}
\end{figure}

We can prove similar properties for the imperfect tensor, which is a superposition of the perfect (reference) tensor and all weight 2 X errors applied to the perfect tensor. 
Again, the isometric properties can be shown through explicit tensor contraction by expanding the imperfect tensor into a superposition in the definition. This reduces to tensor contractions of the perfect tensor 5-1 isometries and cross terms that corresponding to contracting such isometries with operator insertions. The diagonal terms that do not have operator insertions add up to identity while the cross terms vanish (Figure~\ref{fig:ipt_detail_contract}). To see this, let $W_{PT}$ be the 5-1 isometry constructed from the perfect tensor. Because for each such weight 2 or weight 4 X operator, one can find a stabilizer $S\in \mathcal{S}$ of the $[[5,1,3]]$ code such that it acts trivially on the uncontracted leg (which we can take to the be logical subspace) but anticommutes with the inserted operator, it follows that

\begin{equation}
    W_{PT}^{\dagger}O_{X}W_{PT} = W_{PT}^{\dagger} O_{X} SW_{PT} = - W_{PT}^{\dagger}S O_X W_{PT} = - W_{PT} O_X W_{PT}=0.
\end{equation}

This is because all such X insertions are detectable errors. Thus they take the encoded state, as represented by the subspace span by the 5 legs to be contracted, to an orthogonal error subspace as represented by the same 5 legs but with X error insertions.
Therefore, the imperfect tensor we constructed is also an isometry with 5 contracting legs. 

\begin{figure}[H]
    \centering
    \includegraphics[width=0.7\textwidth]{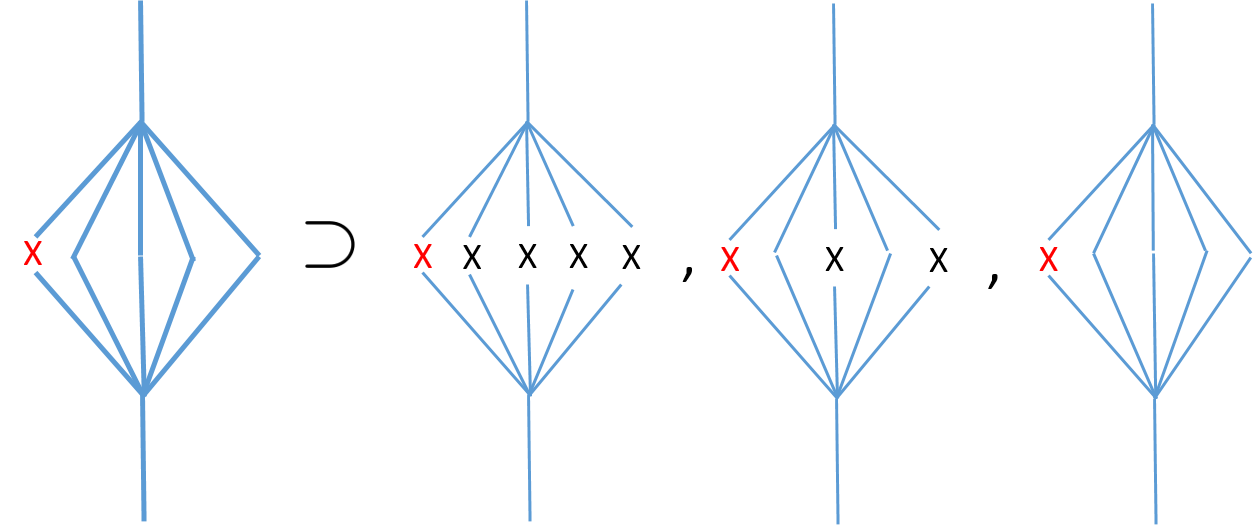}
    \caption{A contraction of the imperfect tensor with a single X insertion give rise to 3 types of terms. The inserted X operator is marked red, and the rest of the X operators are from the definition. Only the first one contributes by applying a stabilizer and pushing the weight 5 X operator to a weight 1 operator to the top or bottom. The rest vanish by orthogonality.}
    \label{fig:single_x_insert}
\end{figure}

Now we derive the pushing rule for a single X operator insertion (marked in red) (Figure \ref{fig:single_x_insert}). Again, by expanding the SBST with the operator insertion, we end up with three types of terms in the superposition, which correspond to the contraction of 5 legs of perfect tensor isometry with weight 1, weight 3 or weight 5 X operator insertions. Again, by finding stabilizers of the perfect code that anti-commute with the inserted operators, only one of them does not reduce to zero. It is precisely corresponds to the logical X operation of the 5 qubit code. Then we can push the weight 5 X operator up to the single leg using $XXXXXX$, which stabilizes the perfect tensor. One can also determine the corresponding amplitude of such terms, which gives the multiplicative factor that depends on $\beta_i$. We thus recover the pushing rule in  Figure~\ref{fig:imperfect_tensor_contract}. Note that the multiplicative factor is proportional to the size of the noise terms we add to the perfect tensor.

\section{Mechanics of tracing and entanglement structure}
\label{app:tracing}

Tools from the stabilizer formalism can often aid in our understanding of tracing and tensor contraction and its resulting entanglement structure. An illuminating example is the use of graph states to visualize the entanglement structure of the $2\times 2$ Bacon-Shor code as analytically treated in section~\ref{subsec:bs4}. As there we begin with the code in the $Z_1Z_2 = +1$ gauge, which has check matrix
$$\left(\begin{array}{cccc|cccc} 
1 & 1 & 1 & 1 & 0 & 0 & 0 & 0 \\
0 & 0 & 0 & 0 & 1 & 1 & 0 & 0 \\
0 & 0 & 0 & 0 & 0 & 0 & 1 & 1
\end{array}\right).$$

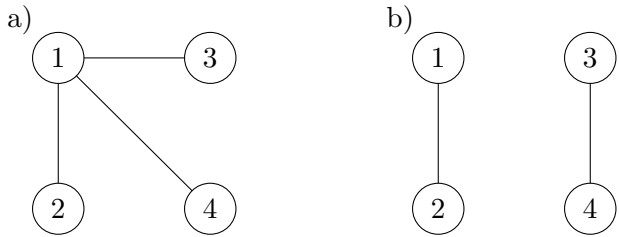
\begin{figure}[t]
\begin{center}\begin{tikzpicture}
\draw (-0.5,2.5) node {a)};
\draw (0,2) -- (0,0);
\draw (0,2) -- (2,2);
\draw (0,2) -- (2,0);
\draw (0,0) node[circle, draw, fill=white] {$2$};
\draw (0,2) node[circle, draw, fill=white] {$1$};
\draw (2,0) node[circle, draw, fill=white] {$4$};
\draw (2,2) node[circle, draw, fill=white] {$3$};
\draw (4.5,2.5) node {b)};
\draw (7,0) -- (7,2);
\draw (5,2) -- (5,0);
\draw (5,0) node[circle, draw, fill=white] {$2$};
\draw (5,2) node[circle, draw, fill=white] {$1$};
\draw (7,0) node[circle, draw, fill=white] {$4$};
\draw (7,2) node[circle, draw, fill=white] {$3$};
\end{tikzpicture}\end{center}
\caption{Graphs associated to (a) $|\overline{0}\rangle$ and (b) $|\overline{+x}\rangle$ as graph states.}
\label{fig:graph-states}
\end{figure}

Consider the logical basis $\{|\bar{0}\rangle, |\bar{1}\rangle\}$, which are the eigenstates of $\bar{Z}$. As these two states are locally unitarily equivalent (namely under $\bar{X}$) they present the same entanglement structure. Adding $\bar{Z}$ to our check matrix yields
$$\left(\begin{array}{cccc|cccc} 
1 & 1 & 1 & 1 & 0 & 0 & 0 & 0 \\
0 & 0 & 0 & 0 & 1 & 1 & 0 & 0 \\
0 & 0 & 0 & 0 & 0 & 0 & 1 & 1 \\
0 & 0 & 0 & 0 & 0 & 1 & 0 & 1 
\end{array}\right).$$
As a local transformation does not change the entanglement structure we may apply $H_2 H_3 H_4$ to get a locally-equivalent code with check matrix
$$\left(\begin{array}{cccc|cccc} 
1 & 0 & 0 & 0 & 0 & 1 & 1 & 1 \\
0 & 1 & 0 & 0 & 1 & 0 & 0 & 0 \\
0 & 0 & 1 & 1 & 0 & 0 & 0 & 0 \\
0 & 1 & 0 & 1 & 0 & 0 & 0 & 0
\end{array}\right) \sim
\left(\begin{array}{cccc|cccc} 
1 & 0 & 0 & 0 & 0 & 1 & 1 & 1 \\
0 & 1 & 0 & 0 & 1 & 0 & 0 & 0 \\
0 & 0 & 1 & 0 & 1 & 0 & 0 & 0 \\
0 & 0 & 0 & 1 & 1 & 0 & 0 & 0
\end{array}\right).$$
This check matrix describes the graph state with graph as seen in Figure \ref{fig:graph-states}a. Namely the adjacency matrix of the graph is the $Z$-block of the check matrix in this reduced form.

If on the other hand we examine the entanglement structure of either $\{|\overline{+x}\rangle, |\overline{-x}\rangle\}$, we augment our check matrix with $\bar{X}$ to produce.
$$\left(\begin{array}{cccc|cccc} 
1 & 1 & 1 & 1 & 0 & 0 & 0 & 0 \\
0 & 0 & 0 & 0 & 1 & 1 & 0 & 0 \\
0 & 0 & 0 & 0 & 0 & 0 & 1 & 1 \\
0 & 0 & 1 & 1 & 0 & 0 & 0 & 0 
\end{array}\right).$$
Now we use $H_2 H_4$ to obtain a locally-equivalent code with check matrix
$$\left(\begin{array}{cccc|cccc} 
1 & 0 & 1 & 0 & 0 & 1 & 0 & 1 \\
0 & 1 & 0 & 0 & 1 & 0 & 0 & 0 \\
0 & 0 & 0 & 1 & 0 & 0 & 1 & 0 \\
0 & 0 & 1 & 0 & 0 & 0 & 0 & 1
\end{array}\right) \sim
\left(\begin{array}{cccc|cccc} 
1 & 0 & 0 & 0 & 0 & 1 & 0 & 0 \\
0 & 1 & 0 & 0 & 1 & 0 & 0 & 0 \\
0 & 0 & 1 & 0 & 0 & 0 & 0 & 1 \\
0 & 0 & 0 & 1 & 0 & 0 & 1 & 0 
\end{array}\right).$$
This provides a different graph as seen in Figure \ref{fig:graph-states}b.

There is a good deal of information about a graph state encoded in its associated graph. For example, a Bell pair is represented as a graph with two nodes connected by an edge; and so we see that $|\overline{+x}\rangle$ is formed of two Bell pairs (after locally changing basis in qubits 2 and 4). The graph of $|\bar{0}\rangle$ represents a graph state not locally equivalent $|\overline{+x}\rangle$ (it appears as graph No. 3 of Table V of \cite{hein2006entanglement}), and so has a distinct entanglement structure. One rational for expecting this behavior is that the logical Hadamard gate does not act transversally on this code, and so transforming $|\bar{0}\rangle \mapsto |\overline{+x}\rangle$ will involve entangling gates on the physical qubits, changing the entanglement structure of the state.

\subsection{Single trace of two codes}
One can generate new stabilizer codes by ``tracing'' two codes. As we have previously discussed in the main text, this corresponds to gluing together legs of the tensor network.
Such gluing procedure through projection is equivalent to contracting individual tensor legs in the conventional tensor network notation, which we briefly review.

For example, consider two quantum states

\begin{align}
\begin{split}
    |T^1\rangle = \sum_{i_l} T^1_{i_1,i_2,\dots, i_n}|i_1, i_2,\dots, i_n\rangle,~~~~ 
     |T^2\rangle = \sum_{j_l} T^2_{j_1,j_2,\dots, j_m}|j_1, j_2,\dots, j_n\rangle.
 \end{split}
\end{align}

Their coefficients are $T^1_{i_1,i_2,\dots, i_n}, T^2_{j_1,j_2,\dots, i_m}$, which can be represented as an $n$-legged tensor and an $m$-legged tensor respectively. The tensors can be glued together through the contraction of sites $i_1,\dots, i_h$ and $j_1,\dots, j_h$ without loss of generality,

\begin{equation}
    \sum_{k_l, l=1,\dots,h} T^1_{k_1, \dots, k_h,i_{h+1},\dots, i_n} T^2_{k_1, \dots,k_h,j_{h+1},\dots, j_m} = T'_{i_{h+1},\dots, i_n, j_{h+1},\dots, j_m}.
\end{equation}
This is precisely the coefficient of the ``traced'' state one generates through projections up to a constant normalization factor.

 Graphically, if the codes can correct erasure errors on the legs that are traced over, then the new tensor generated from gluing is a code whose stabilizers and logical operators can be obtained from stabilizer matching. That is, an operator $P^{(1)}\otimes P^{(2)}$ is a logical operator or a stabilizer of the new code if there exists logical or stabilizer operator $P^{(1)}\otimes Q^{(1)}_i$ from the first code and $P^{(2)}\otimes Q^{(2)}_j$ from the second code, such that $Q_i^{(1)}=Q_j^{(2)}$ are matching Pauli strings over qubits $i$, $j$ that are being glued. Here we will recast these graphical operations on the tensor network as operations on check matrices. 

For tracing together two codes by gluing together one leg, we first consider the scenario in which both codes detects at least one error, or corrects one located error. 
Check matrices of such codes can always be organized into the following form

\begin{equation}
H=
\left(\begin{array}{cc|cc} 
1 & v^t & 0 & u^t \\
0 & w^t & 1 & r^t\\
0 & A & 0 & B 
\end{array}\right).
\label{eqn:gen_check}
\end{equation}
Here we have singled out a particular qubit, which we will glue to another code. Because it corrects an erasure error on this qubit, the code must contain at least two stabilizer elements that respectively have $X$ and $Z$ on this qubit. We can then perform Gaussian elimination to clear up the remaining entries on the same column. In the above notation, $v,u$ are vectors, whereas $A,B$ are matrices. 

We obtain the check matrix for the tensor product of two such codes by direct summing their check matrices. 

\begin{equation}
H=H_1\oplus H_2=
\left(\begin{array}{cccc|cccc} 
1 & v_1^t & 0 & 0 & 0 & u_1^t & 0 & 0\\
0 & w_1^t & 0 & 0 & 1 & r_1^t & 0 & 0\\
0 & A_1 & 0 & 0 & 0 & B_1 & 0 & 0\\ 
 0 & 0 & 1 & v_2^t  & 0 & 0 & 0 & u_2^t\\
 0 & 0 & 0 & w_2^t & 0 & 0 & 1 & r_2^t\\
 0 & 0 & 0 & A_2  & 0 & 0 & 0 & B_2
\end{array}\right).
\end{equation}

Without loss of generality, tracing the first qubit of these two codes correspond to the following operation
\begin{equation}
    H\rightarrow
    \left(\begin{array}{cc|cc} 
v_1^t & v_2^t & u_1^t & u_2^t\\
w_1^t & w_2^t & r_1^t & r_2^t\\
A_1 & 0 & B_1 & 0\\ 
 0 & A_2 & 0 & B_2 \\
\end{array}\right)\xrightarrow{\mathrm{Gauss. Elmin.}} H_{\mathrm{tr}}
\end{equation}

We can derive this from the stabilizer pushing condition, where the Pauli operators on the qubits that are traced over have to match. This means that for any row that have non-zero and matching first entry for both X and Z sections of the check matrix, we retain their rows sans the first columns in the X and Z sections. This procedure gives the first two rows in $H$, where stabilizers that have $X$ or $Z$ in the qubits being traced are matched. If the first entries in the $X$ or $Z$ portion of the check matrix are both zero, then it means the stabilizer elements have $I$ on these qubits. Then the stabilizer for the traced codes are simply the tensor product of the respective Pauli strings in $A_i, B_i$, which correspond to the direct sum of these block matrices, as shown above. We then perform row elimination to obtain the simplified form of the check matrix as the rows may not be linearly independent.

When the code does not correct a located error, then it is possible that the check matrices cannot be organized into the form in (\ref{eqn:gen_check}). In particular, this implies one can only find stabilizers such that only one of the first two rows are present in (\ref{eqn:gen_check}). Or alternatively, if the stabilizer only has $Y$ on the first qubit, the first $(1~ v^t | 1~ u^t)$.
In such cases, we remove the rows whose first column entries (on either X or Z part of the check matrix) do not match.

For instance, consider gluing check matrices of the form 

\begin{equation}
H_1=
\left(\begin{array}{cc|cc} 
1 & v_1^t & 0 & u_1^t \\
0 & A_1 & 1 & B_1 \\
0 & A'_1 & 0 & B'_1
\end{array}\right),
~~
H_2=
\left(\begin{array}{cc|cc} 
1 & w_2^t & 1 & r_2^t \\
1 & A_2 & 0 & B_2 \\
0 & A'_2 & 0 & B'_2
\end{array}\right).
\end{equation}
We attain the check matrix $H$ by tracing the first qubit, where we drop any row whose first X or Z entry does not match. More concretely, the second row in $H_1$ does not match anything in $H_2$, and neither does the first row in $H_2$. Therefore, information from these rows do not enter the new check matrix, which reads

\begin{equation}
H=
    \left(\begin{array}{cc|cc} 
    V_1 & A_2 & U_1 & B_2\\
A'_1 & 0 & B'_1 & 0 \\
0 & A'_2 & 0 & B'_2
\end{array}\right),
\end{equation}

where
\begin{equation}
    V_1 = 
    \begin{pmatrix}
    v^t_1\\
    \vdots\\
    v^t_1
    \end{pmatrix},~~
        U_1 = 
    \begin{pmatrix}
    u^t_1\\
    \vdots\\
    u^t_1
    \end{pmatrix}
\end{equation}
and $row(V_1)=row(U_1)=row(A_2)=row(B_2)$. We then perform row elimination as before.

\subsubsection{Example: tracing together two 4-qubit Bacon-Shor codes in the \texorpdfstring{$ZZ=1$}{zz1} gauge}
\label{appsub:ex_sing_tr}
Recall that the check matrix of one such code is 

\begin{equation}
H_1= H_2=
\left(\begin{array}{cccc|cccc} 
1 & 1 & 1 & 1 & 0 & 0 & 0 & 0\\
0 & 0 & 0 & 0 & 1 & 0 & 1 & 0\\ 
 0 & 0 & 0 & 0  & 0 & 1 & 0 & 1
\end{array}\right).
\end{equation}

Suppose we contract the tensors by gluing the first leg of both codes, it is easy to identify $v^t = (1, 1, 1), u^t = (0,0,0), w^t = (0,0,0), r^t=(0, 1,0), A=(0,0,0), B=(1,0,1)$.
Then resulting check matrix is 

\begin{equation}
\left(\begin{array}{cccccc|cccccc} 
1 & 1 & 1 & 1 & 1 & 1 & 0 & 0 & 0 & 0 & 0 & 0\\
0 & 0 & 0 & 0 & 0 & 0 & 0 & 1 & 0 & 0 & 1 & 0\\ 
0 & 0 & 0 & 0 & 0 & 0 & 1 & 0 & 1 & 0 & 0 & 0\\
0 & 0 & 0 & 0 & 0 & 0 & 0 & 0 & 0 & 1 & 0 & 1
\end{array}\right).
\end{equation}
We recognize this is nothing but the $[[6,2,2]]$ code by Bravyi \cite{Bravyi2011}, which is consistent with the code generated by gluing together tensor networks and identify the corresponding gauge and stabilizer operators using operator pushing.

\subsection{Multi-trace and self-trace}
Instead of gluing one leg, we can also glue multiple legs between two tensors. However, this is equivalent to performing a self-trace of a single code, i.e. gluing two legs of the same tensor, after performing a single trace on two codes. Therefore, we focus on the latter scenario. Performing self-trace following the same stabilizer matching algorithm, such that if $S'$ is a stabilizer element of the self-traced code, it must be obtained from a parent code that has stabilizer $S'\otimes P_i\otimes P_j$ with $i,j$ being the two legs that are glued, $P_{i,j} \in \{I, X,Y,Z\}$ and $P_i=P_j$. We arrive at this rule because when performing a multi-trace on two tensors, the stabilizers on all the legs that are glued have to match. If two stabilizers do not match, we eliminate it from the stabilizer group of the self-traced code.

Again, for a generic check matrix, we organize it into the canonical form for the qubits to be traced over, where the first two columns in the X and Z sections are arranged to match. If this is not possible, then just like the single trace of two codes, the incompatible rows have to be eliminated. Consider a check matrix that can be arranged in the following form
\begin{equation}
H=
\left(\begin{array}{ccc|ccc} 
1 & 0 & v_1^t & 0 & 0 & u_1^t \\
0 & 1 &  v_2^t & 0 & 0 & u_2^t\\
0 & 0 & v_3^t & 1 & 0 & u_3^t\\
0 & 0 & v_4^t & 0 & 1 & u_4^t\\
0 & 0 & A & 0 & 0 & B
\end{array}\right)
\end{equation}
We then perform row operations such that the columns corresponding to the first two qubits match as much as possible. Self-trace can be easily performed by removing the first two columns if they match. If certain rows in these columns do not match, we also eliminate the corresponding rows. We also removed the manifestly linearly-dependent rows during the self-trace step.
\begin{equation}
H\xrightarrow{\mathrm{row~opt.}}
\left(\begin{array}{ccc|ccc} 
1 & 1 & v_1^t+v_2^t & 0 & 0 & u_1^t+u_2^t \\
1 & 1 & v_1^t+v_2^t & 0 & 0 & u_1^t+u_2^t \\
0 & 0 & v_3^t+v_4^t & 1 & 1 & u_3^t+u_4^t\\
0 & 0 & v_3^t+v_4^t & 1 & 1 & u_3^t+u_4^t\\
0 & 0 & A & 0 & 0 & B
\end{array}\right)
\xrightarrow{\mathrm{self-trace}}
\left(\begin{array}{c|c} 
 v^t_1+v_2^t &  u_1^t+u_2^t \\
 v_3^t+v_4^4 & u_3^t+u_4^t\\
 A & B
\end{array}\right)\xrightarrow{\mathrm{row~ elim.}} H_{\rm tr}.
\end{equation}

Self-tracing can be trivially generalized to other matrices that have lower ranks. 

\subsubsection{Example: self-trace in Bravyi's \texorpdfstring{$[[6,2,2]]$}{622} code}
Let us continue the example in section \ref{appsub:ex_sing_tr}, where we obtained Bravyi's $[[6,2,2]]$ code. We can perform a self-trace, which corresponds to double tracing two 4-qubit Bacon-Shor tensors. Without loss of generality, let us perform a self-trace on the first two qubits. Performing the row operators such that the first two columns are matching, we have the following check matrix operations

\begin{align}
H=
&\left(\begin{array}{cccccc|cccccc} 
1 & 1 & 1 & 1 & 1 & 1 & 0 & 0 & 0 & 0 & 0 & 0\\
0 & 0 & 0 & 0 & 0 & 0 & 1 & 1 & 1 & 0 & 1 & 0\\ 
0 & 0 & 0 & 0 & 0 & 0 & 1 & 1 & 1 & 0 & 1 & 0\\
0 & 0 & 0 & 0 & 0 & 0 & 0 & 0 & 0 & 1 & 0 & 1
\end{array}\right)\\
\xrightarrow{\mathrm{self-tr}}
&\left(\begin{array}{cccc|cccc} 
1 & 1 & 1 & 1  & 0 & 0 & 0 & 0\\
0 & 0 & 0 & 0  & 1 & 0 & 1 & 0\\ 
0 & 0 & 0 & 0  & 1 & 0 & 1 & 0\\
0 & 0 & 0 & 0  & 0 & 1 & 0 & 1
\end{array}\right)\\
\xrightarrow{\mathrm{row~reduce}}
&\left(\begin{array}{cccc|cccc} 
1 & 1 & 1 & 1  & 0 & 0 & 0 & 0\\
0 & 0 & 0 & 0  & 1 & 0 & 1 & 0\\ 
0 & 0 & 0 & 0  & 0 & 1 & 0 & 1
\end{array}\right),
\end{align}

which reduces to a single 4-qubit code. This is indeed consistent with the code generated by gluing together tensor networks.

\bibliographystyle{unsrt}
\bibliography{all}

\end{document}